\RequirePackage{fix-cm}
\documentclass[epjc3]{svjour3}  
\journalname{Eur. Jour. of Phys. C}

\usepackage{graphics}
\usepackage{amsmath}
\usepackage{mathrsfs}

\usepackage{amsfonts}
\usepackage{hyperref}
\usepackage{amsthm}
\usepackage{xcolor}
\usepackage{mathrsfs}
\usepackage{subfigure}
\usepackage[demo]{graphicx}
\usepackage{tocbibind}
\usepackage[utf8]{inputenc}
\usepackage[english]{babel}
\usepackage{microtype}
\usepackage{cite}
\newcommand{\cB}{\mathcal{B}}
\newcommand{\cC}{\mathcal{C}}

\newcommand{\cE}{\mathcal{E}}

\newcommand{\cH}{\mathcal{H}}
\newcommand{\cI}{\mathcal{I}}
\newcommand{\cM}{\mathcal{M}}
\newcommand{\cN}{\mathcal{N}}
\newcommand{\cK}{\mathcal{K}}
\newcommand{\cO}{\mathcal{O}}
\newcommand{\cP}{\mathcal{P}}

\newcommand{\cR}{\mathcal{R}}

\newcommand{\cV}{\mathcal{V}}

\newcommand{\tr}{\mathrm{Tr}}

\begin{document}

\title{Holographic spacetime, black holes and quantum error correcting codes: A review}
\author{Tanay Kibe\and Prabha Mandayam\and Ayan Mukhopadhyay\\
\normalfont{\textit{Center for Quantum Information Theory of Matter and Spacetime, and Center for Strings, Gravitation and Cosmology, Department of Physics, Indian Institute of Technology Madras, Chennai 600036, India}\\}
\email{tanayk@smail.iitm.ac.in, prabhamd@physics.iitm.ac.in, ayan@physics.iitm.ac.in}
}                     
\institute{}
\date{Received: date / Revised version: date}
%
\maketitle
\begin{abstract}
This article reviews the progress in our understanding of the reconstruction of the bulk spacetime in the holographic correspondence from the dual field theory including an account of how these developments have led to the reproduction of the Page curve of the Hawking radiation from black holes. We review quantum error correction and relevant recovery maps with toy examples based on tensor networks, and discuss how it provides the desired framework for bulk reconstruction in which apparent inconsistencies with properties of the operator algebra in the dual field theory are naturally resolved. The importance of understanding the modular flow in the dual field theory has been emphasized. We discuss how the state-dependence of reconstruction of black hole microstates can be formulated in the framework of quantum error correction with inputs from extremal surfaces along with a quantification of the complexity of encoding of bulk operators. Finally, we motivate and discuss a class of tractable microstate models of black holes which can illuminate how the black hole complementarity principle can emerge operationally without encountering information paradoxes, and provide new insights into generation of desirable features of encoding into the Hawking radiation. 
%
%
\end{abstract}

\tableofcontents
\section{Introduction}
\label{intro}
The AdS/CFT correspondence \cite{Maldacena:1997re,Gubser:1998bc,Witten:1998qj} is the most well understood example of the holographic emergence of spacetime and gravity. The heuristic reasoning for the holographic principle of gravity is simply that if we stuff in enough matter in a box, then eventually it will collapse to form a black hole whose maximum possible size would be that of the box \cite{Susskind:1994vu}. The maximal entropy of a theory of gravity inside the box would then be the Bekenstein-Hawking entropy \cite{Bekenstein:1973ur,Bardeen:1973gs,Bekenstein:1974ax,Bekenstein_1981} of the black hole whose horizon is of the size of the box,  which explicitly is $A/4G$, i.e. the quarter of the area $A$ of the horizon measured in Planck units (we set $\hbar = c =1$). This heuristic argument relies on a semi-classical description of gravity and should be approximately correct if the size of the box is very large so that gravity is weak even at the black hole horizon when the black hole is of the same size as the box. More precise versions of this argument incorporating covariance under diffeomorphisms and inputs from quantum information theory \cite{Bousso:1999cb,Bousso:1999xy,Casini:2008cr} have been instrumental in providing a concrete ground for the holographic principle which states that a quantum theory of gravity in a spacetime with appropriate asymptotic boundary conditions can be described in terms of a (non-gravitational) quantum many-body system living at the boundary. The AdS/CFT correspondence is a concrete instance of a holographic duality between quantum (super)gravity with asymptotically anti-de Sitter ($AdS$) boundary conditions (i.e. with constant negative curvature near the boundary) and a (super)conformal gauge theory. 

\begin{figure}
\centering
\subfigure[A decoupling limit isolates the massless (red) open string excitations in presence of a stack of D-branes. These can be described exactly by a non-Abelian gauge theory.]{\includegraphics[width = 2in]{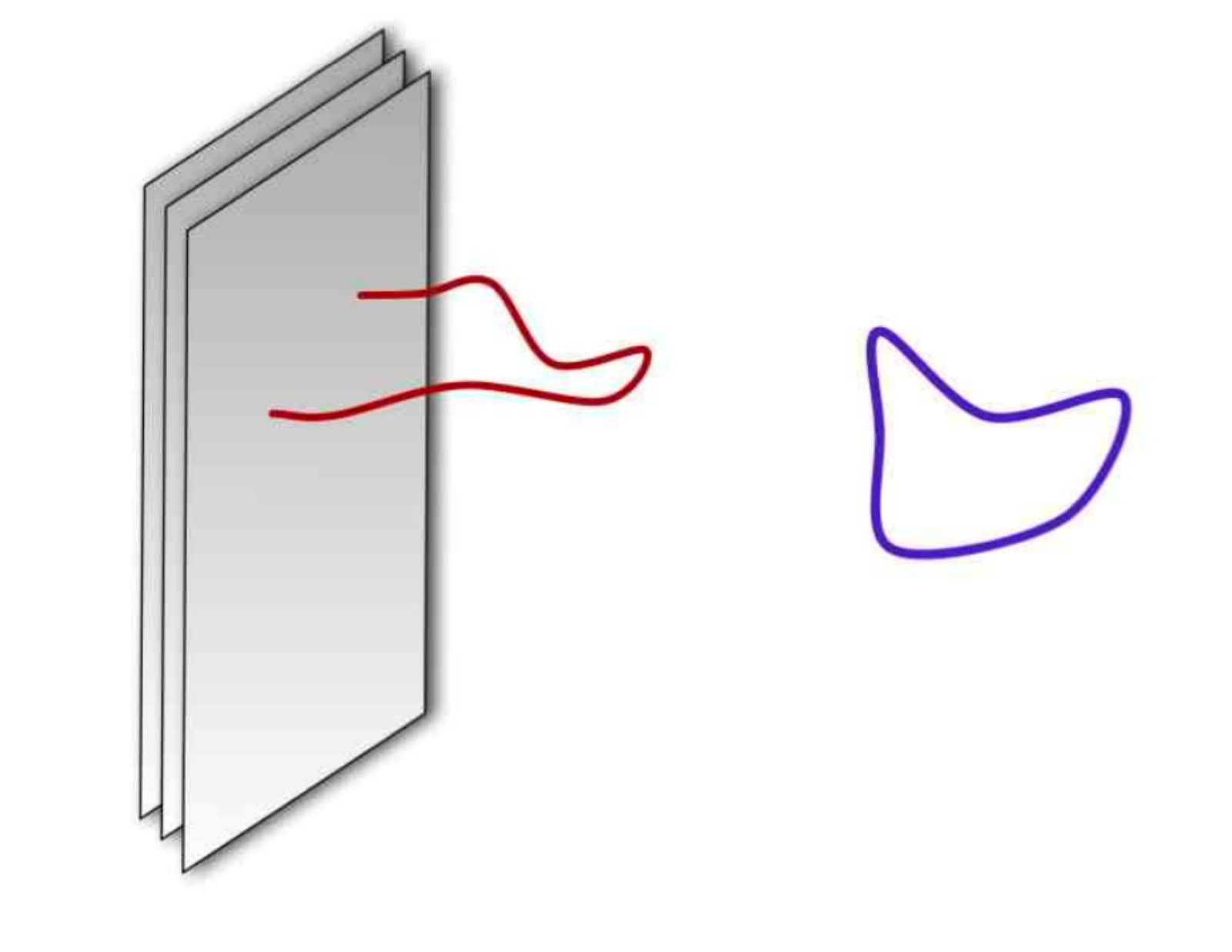}}\,\,\,\,
\subfigure[The same decoupling limit isolates the closed string excitations of the near-horizon throat geometry in the supergravity solution describing the same stack of D-branes in closed string theory due to an infinite redshift factor. The exact description  in this limit is thus the closed string theory in the throat which is asymptotically $AdS_5\times X$.]{\includegraphics[width = 2.2in]{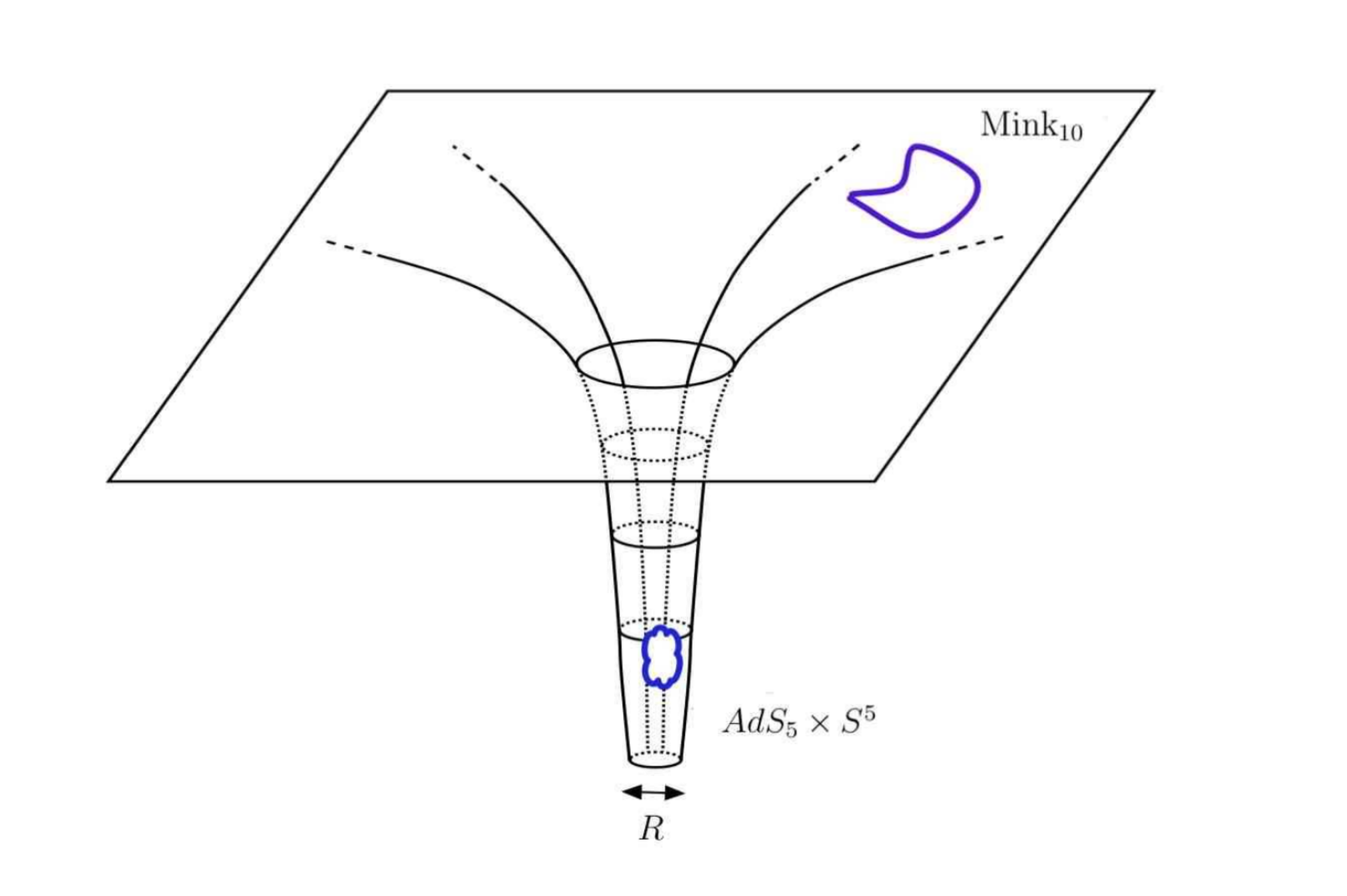}}
\caption{ An illustration of how the AdS/CFT correspondence emerges from two dual descriptions of coincident D-branes in open and closed string theory. These figures are from \cite{Mateos:2007ay}. }\label{Fig:Decoupling}
\end{figure}

The origin of the AdS/CFT correspondence is from the description of D-branes in open and closed string theory, and is a consequence of open-closed string duality \cite{Maldacena:1997re} (see \cite{Pinaki} for a very accessible account). In closed string theory, coincident $Dp$ branes are solitonic solutions of ten (or eleven) dimensional supergravity with $AdS_{p+2}\times X$ throats, where $X$ is a compact space of $8-p$ (or $9-p$) dimensions. In open string theory, these coincident branes are $p+1$ spacetime dimensional defects (extended over $p$ spatial dimensions), whose low energy descriptions are given by non-Abelian gauge theories living on the worldvolume. A decoupling limit (see Fig. \ref{Fig:Decoupling}) isolates the gauge theory from the remaining stringy degrees of freedom in the open string description, whereas in the closed string description, the excitations living in the $AdS_{p+2}\times X$ near-horizon geometry decouples from those in the remaining spacetime. It follows then that the closed string theory (quantum gravity) in $AdS_{p+2}\times X$ can be described by a precise $p+1$-dimensional gauge theory\footnote{The compact space $X$ is related to the global symmetries of the gauge theory}. One can obtain more examples of such holographic (a.k.a. gauge/gravity) duality in various dimensions via such string-theoretic setups including those where the gauge theories are non-conformal (and the asymptotic boundary conditions of gravity are non-AdS). We also obtain a precise mapping between the gauge coupling $g_{YM}$ and the rank of the gauge group ($N$) with the parameters specifying the boundary conditions of the quantum gravity theory, e.g. $L$, the asymptotic curvature radius of $AdS$ and the size of the internal manifold $X$. The AdS/CFT correspondence and its generalizations are often called gauge/gravity dualities \cite{Aharony:1999ti}. 

The most remarkable aspect of these dualities follow from two features of the holographic dictionary. Firstly, the 't Hooft coupling $\lambda = g_{YM}^2 N$ is related to a positive power of $L/l_s$ ($l_s$ is the string length) that controls the corrections to a two-derivative Einsteinian gravity theory arising from the finite length of the string. Furthermore, $N$, the rank of the gauge group, is related to a positive power of $L/l_p$ ($l_p$ is the Planck length) that controls the quantum corrections to classical gravity. As a result, when both $N$ and $\lambda$ are large, both $l_s$ and $l_p$ are small in units where the asymptotic curvature radius $L$ is set to unity, so that both stringy and quantum effects are suppressed. Thus the dual description is just a classical gravity theory to a very good approximation. The duality therefore implies that a many-body system which evades a quasi-particle description can be described by a classical Einsteinian theory of gravity in one higher dimension with a negative cosmological constant and minimally coupled to a few fields. This has led to an enormous impact on our understanding of the collective description of many strongly interacting systems, including strongly correlated quantum materials \cite{Hartnoll:2016apf}, non-perturbative dynamics of QCD \cite{Kim:2012ey,Rebhan:2014rxa} and its various phases such as the  quark-gluon plasma \cite{Casalderrey-Solana:2011dxg}. 

Over more than two decades, the correspondence has also been subjected to stringent tests, where non-trivial dual quantities such as the anomalous dimensions of single-trace gauge-invariant operators and the spectrum of strings in anti-de Sitter space have been matched using techniques like integrability \cite{Beisert:2010jr} and localization \cite{Zarembo:2016bbk}. Recently a derivation of the correspondence has been achieved when the dual gauge theory is free (zero 't Hooft coupling) while the string worldsheet sees a quantum spacetime and gets \textit{localized} at the boundary \cite{Eberhardt:2019ywk,Gaberdiel:2021qbb}.

The fundamental aspects of how the bulk spacetime and its gravitational dynamics emerge from the dual gauge theory are still shrouded in many mysteries. Nevertheless, there has been remarkable progress in this direction via the tools of quantum information theory. A path breaking proposal by Ryu and Takanayagi \cite{RT} and its further refinements \cite{HRT,EngelhardtWall} that an appropriate codimension two bulk extremal surface anchored to the boundary $\partial R$ of a boundary spatial subregion $R$ captures the entanglement entropy of that region $R$ in the dual field theory, have been at the heart of these developments. To be specific, we need to consider the causal domain of dependence $D_R$ of a region $R$ as the set of points where the values of measurements can necessarily be influenced by or influence the data on $R$. Then the bulk operators in the entanglement wedge, which is the causal domain of dependence of (any) Cauchy slice bounded by the bulk extremal surface anchored to $\partial R$ and the subregion $R$ at the boundary, can be decoded from the algebra of operators in the dual CFT in $D_R$ as illustrated in Fig. \ref{Fig:shadow} \cite{Czech_2012,Wall:2012uf,FLM,Headrick_2014,Jafferis_2016,Faulkner_2017}. 

Subsequently, quantum information theory has played a fundamental role in understanding how this entanglement wedge of the emergent spacetime can be reconstructed in the dual conformal field theory (CFT) without encountering inconsistencies. It has been shown that the correct framework which achieves bulk reconstruction in a consistent way can be obtained by reformulating the AdS/CFT correspondence as a quantum error correcting code in which the bulk spacetime is encoded in a redundant way in the Hilbert space of the CFT. Furthermore, the encoding is protected against deletion \textit{errors}, i.e. allowing for the \textit{entanglement wedge} to be (approximately) reconstructed from its corresponding boundary subregion even after the complement of the latter is traced out \cite{Almheiri_2015,Harlow:2016vwg,dong_2016_2,cotler2019_univR}.

The connection between the holographic principle of gravity and quantum information theory is currently a topic of fundamental interest to researchers in diverse fields. In fact, this interdisciplinary area of research has been instrumental in developing new perspectives in quantum error correction itself, and has produced novel connections between quantum fields (many-body systems) and quantum information theory as well. One example of such a connection is the postulate of the quantum null-energy condition which states that the expectation value of a null projection of the energy-momentum tensor is bounded from below by a specific null variation of the entanglement entropy \cite{Bousso_2016}. This postulate has not only been proven in holographic field theories \cite{Koeller_2016}, but also in generic two-dimensional CFTs \cite{Balakrishnan:2017bjg} and free field theories \cite{Bousso_2016_2, Malik:2019dpg}, and is also expected to hold generally. Furthermore, such developments have led to new understanding of connections between entanglement and the renormalization group (RG) flow especially with respect to the existence of quantities which evolve monotonically under the flow \cite{Freedman:1999gp,Myers:2010xs,Casini:2012ei}.

The most fundamental test of our understanding of the AdS/CFT correspondence is whether we can find explicit mechanisms for the resolution of black hole information paradoxes which are further refinements of Hawking's original result that a semi-classical black hole should lose its mass to thermal (Hawking) radiation violating unitarity \cite{Hawking:1975vcx,Hawking:1976ra}. The understanding of the AdS/CFT correspondence in the information theory framework has driven remarkable progress in this frontier as well. In particular, it has been shown that the AdS/CFT correspondence itself leads us to the correct way to compute the fine grained entanglement entropy of the Hawking radiation\footnote{Following \cite{Ghosh:2021axl}, we should call it the "not-so-fine-grained entropy" actually. It is assumed that some averaging has been done due to which an approximate notion of factorization of the Hilbert space into black hole interior and radiation is valid for observables of an effective field theory. Only in such an operational context, a Page curve can be defined suitably. In this review, we will discuss a class of microstate models in Section \ref{sec:microstate} which illustrates some aspects of how such factorizations can emerge. In \cite{Ghosh:2021axl}, there is a discussion on this issue within the semi-classical approximation in the context of braneworld cosmology.}. Surprisingly, these computations can be done in the semi-classical evaporating black hole geometry itself and produce results which are consistent with unitarity. In setups where the holographic system is connected to a bath which collects the Hawking quanta of an evaporating black hole, the bath develops an entanglement wedge, called the island which contains portions of the interior of the black hole, once the black hole is past the Page time (approximately the time when the black hole and the Hawking radiation have the same number of degrees of freedom). The inclusion of this spatially disconnected island leads to reproduction the Page curve \cite{Page_1993,Page_2013} for the time-dependence of the von Neumann entropy of the Hawking radiation in consistency with unitarity without invalidating the effective semiclassical description of bulk physics \cite{AEMM,PeningtonQES,AlmheiriQES,Penington:2019kki,Almheiri2020Islands}. The further understanding of how information of the black hole interior is encoded without encountering fundamental inconsistencies is probably the most exciting topic at the intersection of quantum information theory and gravity.

The present review is aimed to provide an accessible account for researchers in diverse fields to follow the developments connecting the holographic emergence of spacetime and gravity with quantum information theory. We also give a special emphasis on recent developments in connection with black holes. Our account is somewhat complementary to existing reviews in literature, and is also self-contained.  As instances of focused reviews on subtopics covered here, we would especially like to mention \cite{Rangamani:2016dms} which reviews the holographic entanglement entropy proposal, \cite{Mathur:2009hf,Harlow:2014yka,Raju:2020smc} which review the black hole information puzzles and their possible resolutions, \cite{harlow2018tasi,Jahn:2021uqr,Chen:2021lnq} which review aspects of bulk reconstruction, and \cite{Almheiri:2020cfm} (see \cite{mahajan} for a very accessible summary) which reviews recent progress in reproduction of the Page curve (the time-dependence of the entanglement entropy of Hawking radiation) from the AdS/CFT correspondence. We also update the content of these reviews, and describe the central concepts and proposals along with essences of their derivations or arguments supporting them in a sufficiently detailed manner. Furthermore, we present perspectives on the open questions and some promising directions for research in the near future.

The plan of the review is as follows. In section \ref{sec:QES}, we introduce the Ryu-Takanayagi surface and its generalization the quantum extremal surface which is the key to compute the entanglement entropy of a subregion in the dual field theory. As mentioned before, this is at the heart of the connection between quantum information theory and the holographic correspondence. We also review the proof of these proposals for bulk extremal surfaces at leading and subleading orders. We furthermore discuss the consistency checks for the proposal for the quantum extremal surface. We especially outline the proof that the entanglement wedge contains the bulk causal wedge which is the key to the understanding of the non-triviality of bulk reconstruction. Furthermore, we discuss how the holographic presciptions for computing entanglement entropy reproduce entanglement inequalities (especially the strong subadditivity) and the alternative maximin construction.

In section \ref{sec:bulkreconstruction}, we introduce the entanglement wedge reconstruction hypothesis and the key postulate of equality of bulk and boundary relative entropies in the bulk semi-classical approximation. We then discuss various implications of the latter postulate especially how it implies the emergence of gravitational field equations at linearized order, and furthermore how the bulk canonical energy gets connected to Fisher information at the boundary. We also discuss progress in understanding of the emergence of bulk from modular flow at the boundary, and the basic reasons for reformulating bulk reconstruction as a quantum error correcting code. Afterwards, we introduce the appearance of islands which are disconnected portions of the entanglement wedge especially in the context of double holography, and introduce and justify the island rule for computing the entanglement entropy of a subregion of a bath in contact with a holographic system described by a semiclassical evaporating black hole while sketching how this reproduces the Page curve of Hawking radiation.

In section \ref{sec:QECandholo}, we review quantum error correction especially in relation to operator algebras, and discuss how this framework together with the postulate of equality of bulk and boundary relative entropies imply reconstruction of local bulk operators in the entanglement wedge in terms of the operators of the dual field theory in the region of interest. We discuss how apparent inconsistencies of bulk reconstruction are mitigated by formulating the bulk reconstruction in AdS/CFT correspondence as a quantum error correcting code that protects against deletion of complementary boundary subregions. Toy models of AdS/CFT correspondence based on tensor networks which provide examples of perfect recovery maps for the entanglement wedge in the dual subregion of the field theory are then presented. We focus particularly on the Petz map, and discuss how a specific variation could provide the desired approximate recovery even at sub-leading order, and relate to the modular flow at the boundary. We furthermore discuss the connection between the bulk radial coordinate and renormalization group flow in this context with a perspective on issues which may spur further universal understanding of the holographic correspondence.

In section \ref{sec:bhinterior}, we review the exciting recent progress in the reproduction of the Page curve from Hawking radiation. In particular, we focus on how replica wormhole saddles imply the quantum extremal surfaces and islands responsible for restoring behavior of the Page curve that is consistent with unitarity although these wormholes provide an intrinsically averaged description of the black hole microstates. We then discuss the state-dependence of the encoding of the interior in the framework of universal (approximate) subsystem recovery and also how the Python's lunch mechanism generates exponential complexity of the state-dependent encoding.

Furthermore, we present a perspective on issues that are crucial to fully understand how the black hole complementarity principle can emerge operationally without encountering information paradoxes such as the AMPS paradox while emphasizing the need for a complex encoding of the interior. We motivate the need for tractable microstate models of black holes which can demonstrate the realization of all desirable features of the encoding in the Hawking radiation especially information mirroring (with decoding possible without full knowledge of interior) and exponentially complex encoding of the black hole interior excitations simultaneously. Furthermore, it should explain the origin of self-averaging in real time to be consistent with the implications of replica wormholes. We proceed to discuss a class of tractable microstate models explicitly, their promising results in this direction along with implications and some open questions.

In section \ref{sec:discussion}, we conclude with a discussion on some of the topics of significance which are not covered in this review and some promising directions for further research.

\section{Quantum extremal surfaces}
    \label{sec:QES}
\subsection{Partial proof of the quantum extremal surface proposal}
    \label{subsec:QESprop}
    The entanglement entropy of a boundary subregion provides the most fundamental link between holography and quantum information. 
    
\begin{figure}
        \centering
        \resizebox{0.6\textwidth}{!}{%
        \includegraphics{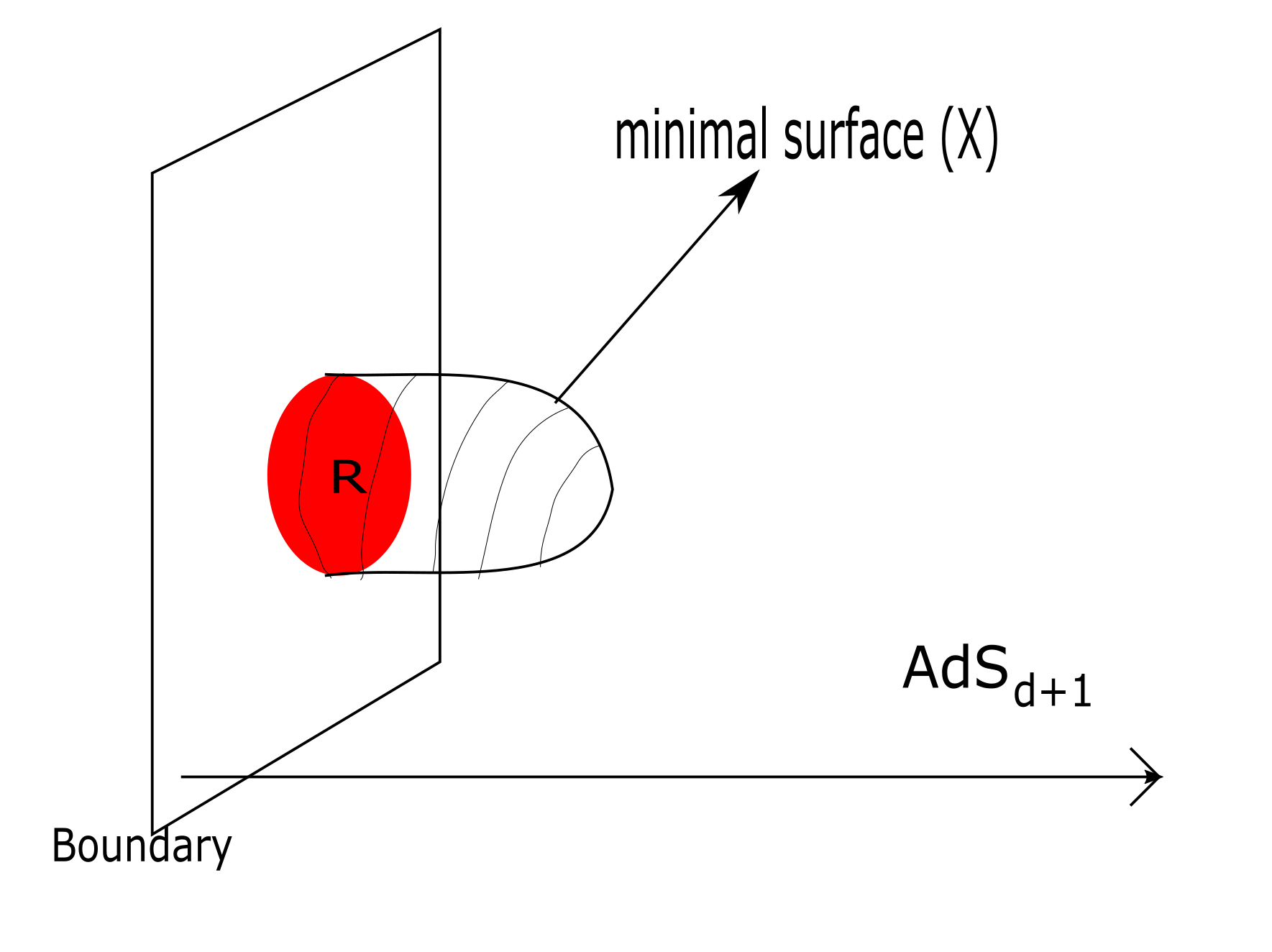}
        }
        \caption{The red region is the boundary sub-region of interest (R). The classical extremal surface anchored to the boundary of R is shown.}
        \label{Fig:RT}       
    \end{figure}

    Ryu and Takayanagi (RT) \cite{RT} were the first to describe how entanglement entropy could be computed holographically in a static semi-classical spacetime (dual to a large $N$ quantum field theory). Then Hubeny, Rangamani and Takayanagi (HRT) \cite{HRT} extended this idea to time dependent geometries. The RT/HRT formula states that the entanglement entropy of a subregion R of the boundary theory is proportional to the area of the classical, bulk co-dimension  two \footnote{If $N$ is a submanifold of $M$ then the codimension of $N$ in $M$ is defined to be: $codim(N)= dim(M)-dim(N)$.} extremal surface that is anchored to the boundary of R and is homologous to R (see Fig. \ref{Fig:RT}). The RT proposal was shown to be correct at the leading order in $\hbar$ by Lewkowycz and Maldacena (LM)\cite{LM}. A generalization of this proof for the HRT proposal using a bulk version of the Schwinger-Keldysh contour was described in \cite{Dong_2016}. The RT/HRT prescription has led to a much simpler proof of the strong sub-additivity of entanglement entropy \cite{Headrick:2007km,Wall:2012uf} and also to new inequalities that holographic entanglement entropy should satisfy \cite{Hayden:2011ag,Bao:2015bfa,Hubeny:2018ijt,Hubeny:2018trv}. The next to leading order correction to the holographic entanglement entropy for semi-classical static situations was computed by Faulkner, Lewkowycz and Maldacena (FLM) \cite{FLM}. They found that the entanglement entropy correct to $\mathcal{O}(\hbar^{0})$ is given by 

\begin{figure}
    \centering
    \resizebox{0.6\textwidth}{!}{%
    \includegraphics{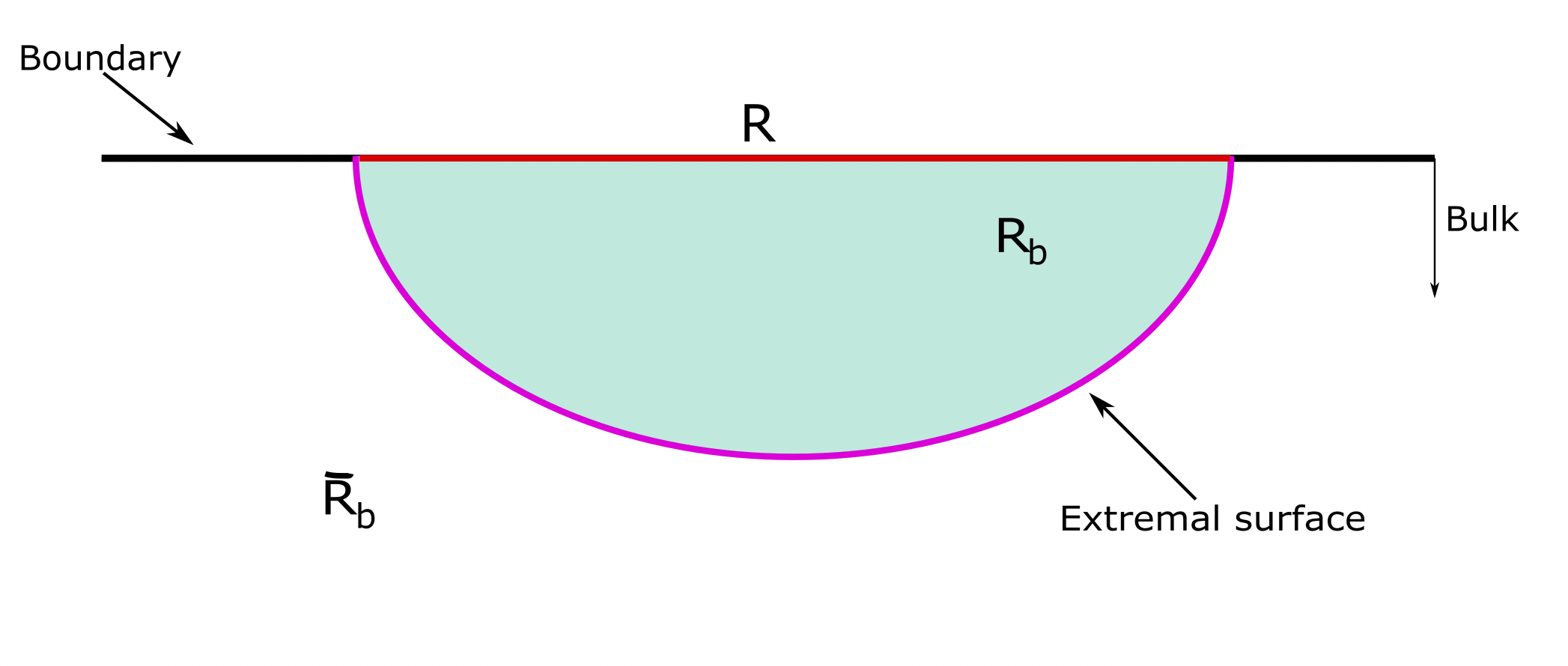}
    }
    \caption{The red region is the boundary sub-region of interest. The magenta extremal surface divides the bulk into two regions. $R_{b}$ is the bulk sub-region connected to the boundary region of interest.}
    \label{Fig:FLM} 
    \end{figure}

    \begin{equation}
	        S(R) = \frac{\mathcal{A}(X_{R})}{4 G \hbar} + S_{\text{ent-bulk}} = S_{\text{gen}}(X_{R}),
        \label{Eq:SGen}
        \end{equation}
    where $\mathcal{A}(X_{R})$ is the area of the classical bulk extremal HRT surface ($X_{R}$) and the leading order quantum corrections are given by the bulk entanglement entropy ($S_{\text{ent-bulk}}$) which is as follows. As shown in Fig. \ref{Fig:FLM}, the classical extremal surface divides the bulk into two sub-regions $R_{b}$ and $\bar{R}_{b}$; $S_{\text{ent-bulk}}$ is the entanglement entropy of the reduced density matrix (in the bulk effective field theory) on the bulk sub-region $R_{b}$ that is connected to the boundary region $R$. The bulk quantum fields are assumed to be described by an effective field theory (EFT) living on a fixed background and the entanglement entropy of the bulk sub-region connected to the boundary region $R$ is computed using standard quantum field theory techniques. The quantity defined in Eq. \eqref{Eq:SGen} is called the generalized entropy. The quantity $S_{\text{ent-bulk}}$ suffers from divergences, which can be absorbed into the renormalization of the Newton's constant leading to a well defined generalized entropy \cite{Srednicki:1993im,Susskind:1994sm,Kabat:1995eq,Larsen:1995ax,Jacobson:1994iw}.
    It is not clear if the LM and FLM proofs (reviewed below) can be extended to include higher order corrections involving multiple loops of quantized gravitons. However, the Engelhardt-Wall proposal is expected to work for all orders in $\hbar$ in which we consider the full quantum theory of the bulk matter on a fixed (but backreacted) semiclassical gravitational background as discussed below.
    
    Engelhardt and Wall (EW) \cite{EngelhardtWall} conjectured that the holographic entanglement entropy of a boundary sub-region $R$ is given by the generalized entropy of the quantum extremal surface (QES) $\chi_{R}$ anchored to ($\partial R$), so that  
    \begin{equation}
        S(R) = S_{\text{gen}}(\chi_{R}),
    \label{eq:EWproposal}
    \end{equation}
    where the quantum extremal surface is the surface that extremizes the generalized entropy and is homologous to $R$. If there are multiple such extremal surfaces, the one with the smallest generalized entropy satisfying the homology constraint is picked.\footnote{Beyond the semiclassical gravity limit, the area of the quantum extremal surface should be promoted to an area operator $\hat{A}$ which satisfies the identity $$\frac{\delta\hat{A} }{\delta X^a} \rho =0 = \rho \frac{\delta\hat{A} }{\delta X^a},$$ where $\rho$ is a state of the Hilbert space of the bulk matter theory as discussed in \cite{EngelhardtWall}. Furthermore, we will need to consider $\langle A\rangle + S_{\rm ent-bulk}+ \langle{\rm counterterms}\rangle$ to define the holographic entanglement entropy (where counterterms remove the ultraviolet divergences of the area operator). The validity of such an approach has been examined in \cite{Belin:2018juv,Belin:2019mlt,Belin:2021htw} perturbatively in the $1/N$ expansion.} This \textit{QES proposal} is different from the FLM formula, which is the generalized entropy of the classical RT/HRT surface.
    The QES proposal hasn't been proven, however it is believed to be correct since it satisfies some non trivial consistency checks which we will review in section \ref{subsec:QESProofs}. This proposal has been checked order by order in $\hbar$ for a few sub-leading corrections by comparing computations in the bulk with those in the dual boundary CFT \cite{Belin:2018juv,Belin:2019mlt,Belin:2021htw}. In the rest of this review, we will use units where $\hbar=1$ unless explicitly mentioned.
     
    Before describing the checks on the QES proposal, we will first review the classical argument from \cite{LM} that proves the RT formula. The entanglement entropy of a boundary sub-region $R$ can be computed using the replica trick. This consists of going to Euclidean time and considering an angular direction in the boundary field theory with origin at the boundary of $R$. This is labelled by $\tau$, with $\tau = \tau +2 \pi$. The boundary quantum field theory (QFT) is then considered in a sequence of spaces ($\widetilde{M}_{n}$) with $\tau = \tau + 2 \pi n$, for positive integers n. This sequence is holographically dual to a sequence of bulk geometries labelled by $\widetilde{B}_{n}$ with the asymptotic boundary $\widetilde{M}_{n}$. One then computes the R\`{e}nyi  entropy ($S_{n}$) as follows:
    
    \begin{equation}
        S_{n} = \frac{1}{1-n} (\ln Z[\widetilde{M}_{n}] -n \ln Z[\widetilde{M}_{1} ]).
    \end{equation}
    Analytic continuation to non-integer $n$ is well defined due to the Carlson theorem \cite{carlson}. The $n \to 1$ limit gives the von Neumann entropy. Here $Z[\widetilde{M}_{n}]$ and $Z[\widetilde{M}_{1}]$ are the QFT partition functions for the spacetime $\widetilde{M_{n}}$ and the original spacetime $M_{1}$. The holographic dictionary in the large $N$ limit tells us the following:
    
    \begin{equation}
        Z[\widetilde{M}_{n}] = e^{-I[\widetilde{B}_{n}]},
    \end{equation}
    where $I[\widetilde{B}_{n}]$ is the on-shell classical bulk action for the bulk geometry $\widetilde{B}_{n}$. The bulk geometries $\widetilde{B}_{n}$ have a $\mathbb{Z}_{n}$ symmetry that corresponds to cyclic permutations of the n replicas. Taking a quotient with this, one can define $B_{n} = \widetilde{B}_{n}/ \mathbb{Z}_{n}$. Due to the quotient, these bulk geometries have the same boundary conditions as the original geometry ($\widetilde{B}_{1}$), that is $\tau = \tau + 2\pi$. These geometries typically have a conical defect with opening angle $\frac{2 \pi}{n}$ at the fixed points of the $\mathbb{Z}_{n}$ symmetry. The classical bulk action is a $\tau$ integral of a local Lagrangian density, therefore it follows that $I[\widetilde{B}_{n}] = n I[B_{n}]$. Thus the R\`{e}nyi  entropy can be written as follows:
    \begin{equation}
        S_{n} = \frac{n}{1-n}(I[B_{n}] - I[B_{1}]).
        \label{eq:Renyi}
    \end{equation}
    Note that when evaluating $I[B_{n}]$ one excludes any contributions from the conical singularity. We can now analytically continue to non-integer n and take the $n \to 1$ limit. The von Neumann entropy is then:
    \begin{equation}
        S = -\partial_{n} I[B_{n}]
        \label{eq:Renyi2}
    \end{equation}
    Varying $n$ corresponds to changing the opening angle of the conical defect. The metric and other fields also have to change due to this change in $n$. However since the geometry $B_{n}$ is a solution of the equations of motion, the first order variations of the bulk action away from these solutions should vanish. Thus the only change in the action comes from a boundary term (at the conical singularity). 
    Therefore, the von Neumann entropy is essentially a boundary term at the conical singularity, which is a co-dimension 2 hypersurface in the bulk. This boundary term was calculated in \cite{LM} and was shown to reproduce the RT formula.
    
    The above argument was extended by FLM \cite{FLM} to include quantum corrections (to $\mathcal{O}(\hbar^{0})$) by considering quantum fluctuations in the bulk EFT. The partition function of bulk quantum fields is then given by the following:
    
    \begin{equation}
        Z_{bulk}^{(n)}= Tr[\rho^{n}_{n}], 
    \end{equation}
    where $\rho_{n}$ is a state of the bulk quantum fields in the bulk geometry $B_{n}$ . The R\`{e}nyi entropy can now be written as follows:
    
    \begin{equation}
        S_{n} = \frac{1}{1-n}(\ln Tr[\rho^{n}_{n}] - n \ln Tr[\rho_{1}]).
    \end{equation}
    The von Neumann entropy is then:
    \begin{equation}
        S =-\partial_{n}(\ln Tr[\rho^{n}_{n}] - n \ln Tr[\rho_{1}])_{n=1}.
    \label{eq:split1}
    \end{equation}
    We can add and subtract the term $-\partial_{n}(\ln Tr[\rho_{1}^{n}])$ to Eq. \eqref{eq:split1} and after some algebra we get:
    \begin{equation}
        S= S_{\text{ent-bulk}} + S_{\text{area}},
    \end{equation}
    with
    \begin{equation}
        S_{\text{ent-bulk}} = -\partial_{n}(\ln Tr[\rho_{1}^{n}] - n \ln Tr[\rho_{1}])_{n=1}
    \end{equation}
    and
    \begin{equation}
        S_{\text{area}} = -\frac{Tr [\partial_{n} \rho_{n}]_{n=1}}{Tr[\rho_{1}]}.
    \end{equation}
    
    The $S_{\text{ent-bulk}}$ term involves $\rho_{1}$ which is the density matrix of bulk quantum fields in the original geometry. This therefore computes the bulk entanglement entropy. The $S_{\text{area}}$ term can be expressed as a variation of a local Lagrangian, which for the usual 2 derivative gravity action gives the area term as seen before for the classical case. This concludes a heuristic review of the arguments of FLM for the quantum correction to the RT formula. 
    
\begin{figure}
   \centering
    \resizebox{0.6\textwidth}{!}{%
    \includegraphics{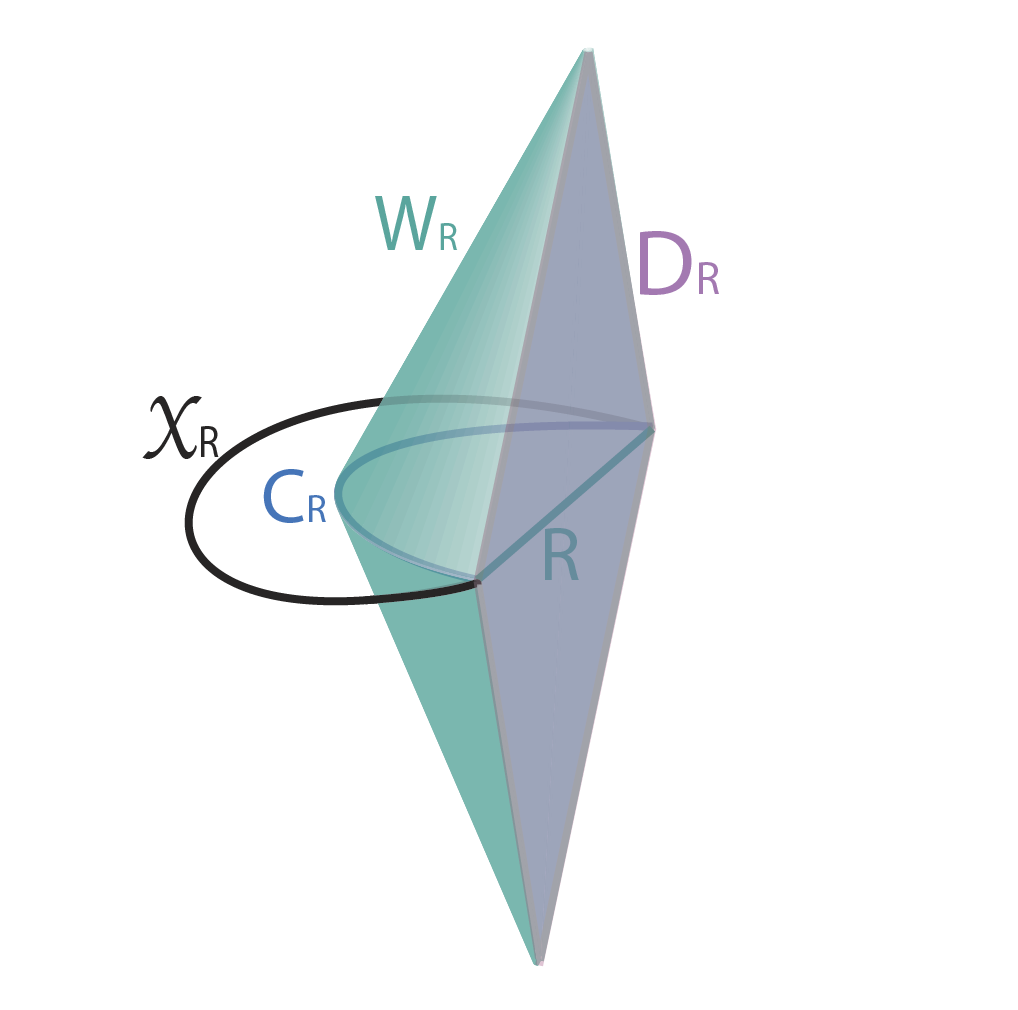}
    }
    \caption{The boundary domain of dependence is coloured purple, the causal wedge is the green bulk region and the causal surface is coloured blue. The quantum extremal surface is marked as $\chi_{R}$ and is shown to lie deeper into the bulk than the causal surface. Figure from \cite{EngelhardtWall}.}
    \label{Fig:CausalWedge} 
    \end{figure}

    We will now define the causal wedge and the entanglement wedge. These are important for bulk reconstruction, which is the major focus of this review. For a boundary sub-region $R$, the boundary domain of dependence of $R$, labelled by $D_{R}$ is defined to be the set of all boundary points such that any in-extensible timelike curve that passes through any point in $D_{R}$ necessarily intersects $R$ 
    (i.e. $D(R)$ is the set of points where the values of measurements can necessarily be influenced by or influence the data on $R$). The bulk casual wedge ($W_{R}$) is then defined to be the intersection of the causal past and future of ${R}$. $W_{R} = \mathcal{J^{+}}(D_{R}) \cap \mathcal{J^{-}}(D_{R})$ The boundary of $W_{R}$ is called the causal surface and is labelled by $C_{R}$. See Fig.\ref{Fig:CausalWedge} for a pictorial depiction of these definitions.
   
    Classically it should be possible to reconstruct any bulk operator within $W_{R}$ in terms of boundary operators on $D_{R}$ since they are in causal contact with each other. Entanglement wedge reconstruction however states that any operator in the entanglement wedge of $R$ can be reconstructed from operators in $D_R$, where the entanglement wedge is defined to be the bulk domain of dependence of the Cauchy surface that interpolates between $R$ and the extremal surface anchored to $\partial R$. This has led to the notion of \textit{sub-region duality} \cite{Bousso:2012mh,Czech:2012bh,Bousso:2012sj}, which states that sub-regions of the boundary are dual to sub-regions of the bulk. Bulk reconstruction will be described in more detail in section \ref{sec:bulkreconstruction}. We will now use the definitions from above to review certain checks on the QES proposal.
    \paragraph{Sanity checks}:Any proposal for the holographic entanglement entropy that claims to be correct to all orders in $\hbar$ must pass the following preliminary checks.

        \begin{enumerate}
	        \item It must agree with the RT/HRT formula at leading order in $\hbar$
	        \item It must agree with the FLM formula at next to leading order in $\hbar$
        \end{enumerate}
         Engelhardt and Wall \cite{EngelhardtWall}  argued that their proposal indeed passes these two checks. We review these arguments below.
         
         Since the FLM proposal is valid only upto $\mathcal{O}(\hbar^{0})$, it is enough to show that the following is true:

        \begin{equation}
	        S_{gen}(X_{R}) = S_{gen}(\chi_{R}) + \mathcal{O}(\hbar).
        \end{equation}
        The generalized entropy consists of two terms $S_{gen}(X) = \frac{\mathcal{A}(X)}{4 G \hbar} + S_{ent} $ .
        In a semi-classical bulk the classical and quantum extremal surfaces are expected to be a distance $\mathcal{O}(\hbar)$ apart. Thus the entanglement entropies of bulk fields ($S_{ent}$) for the two surfaces are expected to differ only at $\mathcal{O}(\hbar)$. That is $S_{ent}(X_{R}) - S_{ent}(\chi_{R}) = \mathcal{O}(\hbar)$. Now we can look at the area term. Since the classical extremal surface extremizes the area, first order variations of the classical surface do not affect the area. That is since $X_{R}$ and $\chi_{R}$ are a distance $\hbar$ apart, the leading order difference in the two areas is at most $\hbar^{2}$, that is $A(X_{R}) - A(\chi_{R}) = \mathcal{O}(\hbar^{2})$. This therefore proves that the holographic entanglement entropy proposal agrees with the RT/HRT and FLM formulas at the appropriate orders.
        The FLM formula and the QES proposal will not agree at higher orders in $\hbar$ and one can perform computations at higher orders to determine if the QES proposal is correct \cite{Belin:2018juv,Belin:2019mlt,Belin:2021htw}. The QES proposal passes some essential consistency checks and is therefore believed to be correct. These checks will be the focus of the rest of this section.

    \subsection{The entanglement wedge contains the causal wedge}
    \label{subsec:QESProofs}
    If the sub-region duality is consistent then the QES must lie deeper in the bulk than causal surfaces. This is an important consistency check that the QES proposal passes. In this section we reason why this consistency condition should hold and review a proof of why the QES proposal passes this check.
    
\begin{figure}
    \centering
    \resizebox{0.6\textwidth}{!}{%
    \includegraphics{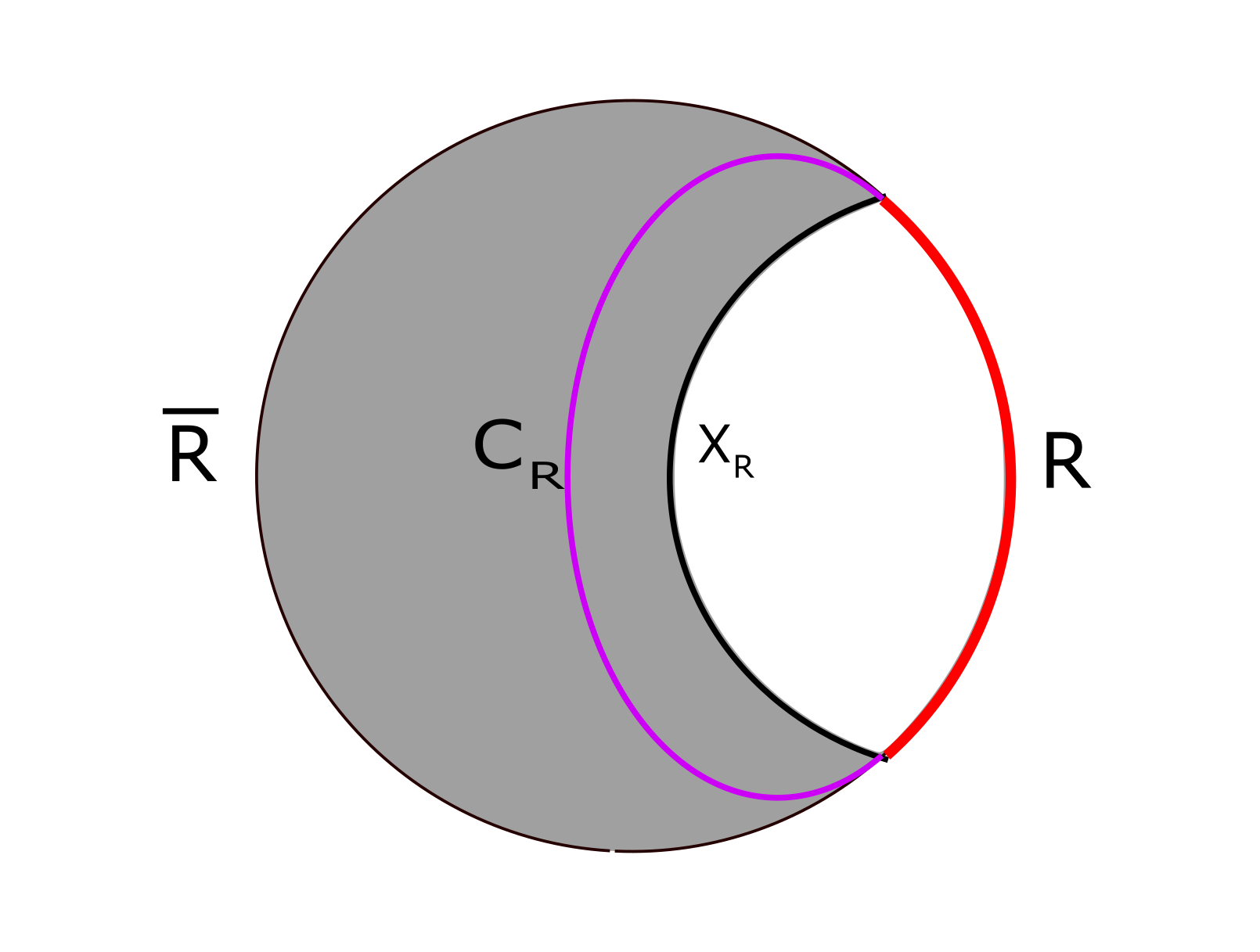}
    }
    \caption{This is a spatial slice of the geometry. The boundary is the circumference of the circle. $R$ (red) is the boundary subregion of interest and $\bar{R}$ is its complement. $C_{R}$ (magenta) is the causal surface for $R$ and $X_{R} = X_{\bar{R}}$ (black) is the classical extremal surface, which is shown to lie within the causal wedge of R. The grey region is the entanglement wedge of $\bar{R}$. Anything in the grey region can be reconstructed from $\bar{R}$. A signal from R can propagate upto $C_{R}$ and therefore into the entanglement wedge of $\bar{R}$. This violates micro-causality of the CFT.}
    \label{Fig:Signalling1} 
    \end{figure}

    Let us assume that the extremal surface can lie closer to the boundary than the causal surface. Let us consider a pure state in the dual theory at the boundary. The entanglement entropy of its reduced density matrices on $R$ and its complement ($\rho_{R}$ and $\rho_{\bar{R}}$) must therefore be the same. This can be reproduced by the QES proposal if $X_{R} = X_{\bar{R}}$ since the bulk fields are also in a pure state. As shown in Fig. \ref{Fig:Signalling1} if $X_{R}$ lies within the causal wedge $W_{R}$ then the region between $C_{R}$ and $X_{R}$ would be in the entanglement wedge of $\bar{R}$ and can therefore be reconstructed on $\bar{R}$. However this region is in causal contact with $D_R$ and therefore can be affected by a signal propagating from $D_R$. This leads to a contradiction in the dual boundary theory since $R$ and $\bar{R}$ are causally disconnected. This can be avoided only if  $X_{R}$ lies deeper in the bulk than $C_{R}$ and is spacelike to it.
    
    It was shown in \cite{HubenyRangamani} that if the classical null energy condition (NEC) holds then the classical extremal surface lies deeper in the bulk than the causal surface. The NEC states that $T_{kk}\geq 0$ for any future directed null vector $k^{\mu}$. This can be violated if there is quantum matter in the bulk and the classical extremal surface can therefore be closer to the boundary than $C_{R}$ or it could be timelike separated from it. The way to avoid this inconsistency is to use the QES instead of the classical extremal surface. Below we review the argument from \cite{EngelhardtWall} which shows that the QES $\chi_{R}$ lies deeper in the bulk than $C_{R}$ and is spacelike to it.
    
    We will first state the generalized second law (GSL). The usual second law of thermodynamics states that the thermodynamic entropy of any closed system is nondecreasing in time. The GSL is a statment about the monotonicity of the generalized entropy \cite{Wall_2013}. The generalized entropy is computed on a Cauchy slice (at some "time"), the GSL then states that the variation of this generalized entropy along any future directed normal to the Cauchy slice is non negative. The generalized entropy can be defined for any causal horizon ($H^{+}$), which is defined to be the boundary of the past of any future directed timelike or null worldline. Define $H = H^{+} \cap \Sigma$, where $\Sigma$ is a Cauchy slice and $H^{+}$ is a future causal horizon. Then the GSL states the following \cite{EngelhardtWall,Wall_2013}
    
    \begin{equation}
        \frac{\delta S_{gen} (H)}{\delta H^{\mu}} k^{\mu} \geq 0,
    \end{equation}
    
    where $\delta H^{\mu}$ is a normal to H and $k^{\mu}$ is any future directed null vector. We will also require the following theorem from \cite{Wall_2013}, see also \cite{EngelhardtWall}.
    
\begin{figure}
   \centering
    \resizebox{0.6\textwidth}{!}{%
    \includegraphics{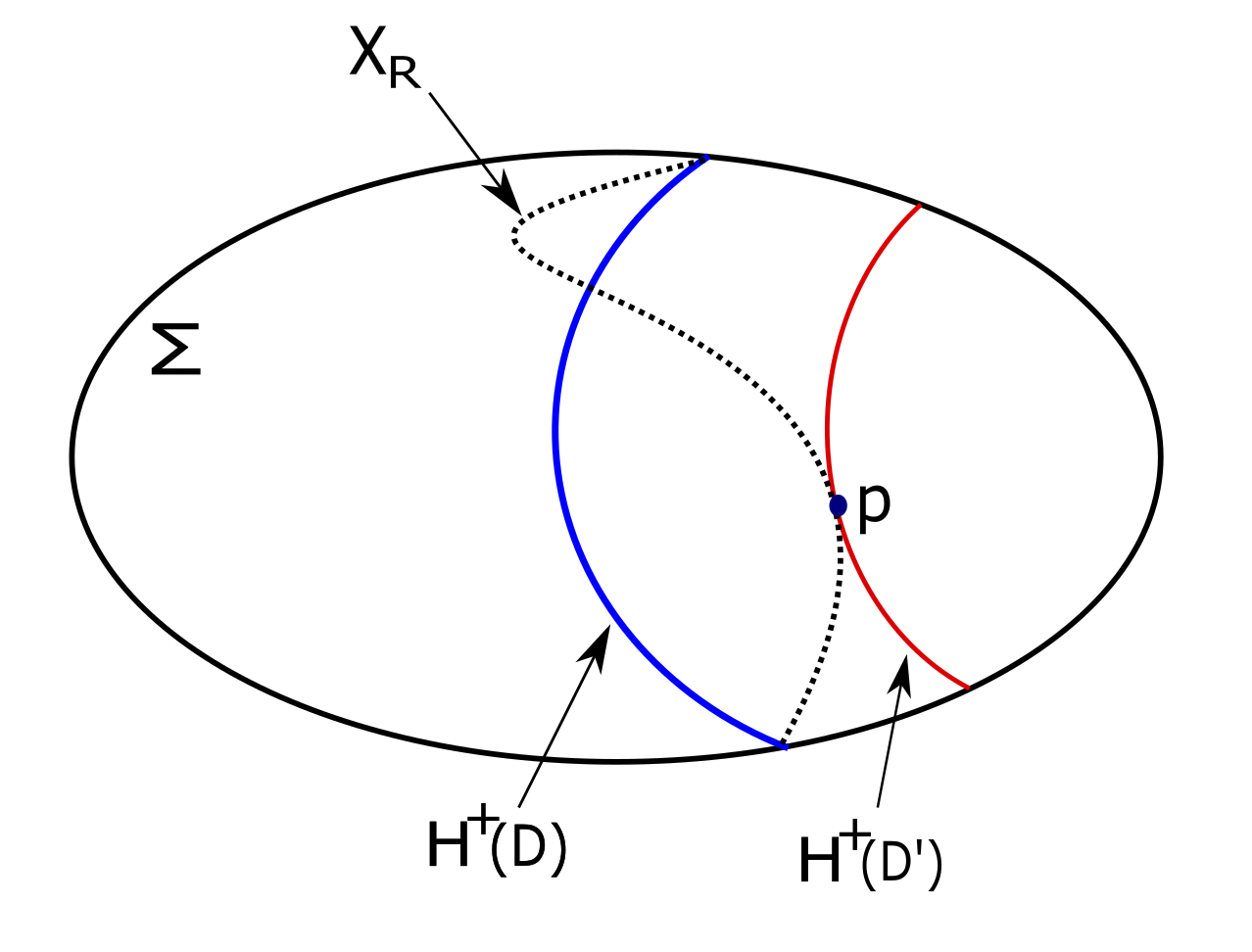}
    }
    \caption{$\Sigma$ is a spatial slice that contains the QES $\chi_{R}$. The projections of $H^{+}(D)$ (blue) and $H^{+}(D')$ (red) are shown. The dotted curve is the QES that intersects and is tangent to $H^{+}(D')$ at point p. The region to the right of the QES is $Ext(\chi_{R})$. Figure reproduced from \cite{EngelhardtWall}}
    \label{Fig:proof} 
    \end{figure}

    \begin{theorem}
        Let M and N be co-dimension one null surfaces that split the spacetime into two parts, an interior (Int) and exterior (Ext), where Ext is defined to be the region containing the boundary subregion $D_{R}$ which is of interest. Let $M \cap Ext(N)$ be empty. Also assume that M and N coincide at some point p and that M and N are smooth in the neighbourhood of p. Let $\Sigma$ be a spatial slice that passes through p. Then there exists a normal ($\delta \Sigma^{\mu}$) to $\Sigma \cap M$ in the neighbourhood of $p$ such that
        
        \begin{equation}
            \frac{\delta S_{gen}(M)}{\delta \Sigma^{\mu}} k^{\mu} \geq \frac{\delta S_{gen}(N)}{\delta \Sigma^{\mu}} k^{\mu},
            \label{Eqn:thm1}
        \end{equation}
    where $k^{\mu}$ is a future directed null normal to $M$ and $N$.
    \end{theorem}
   To use this theorem and show that the QES lies deeper than the causal surface we identify the null splitting surface $M$ with the boundary of the entanglement wedge. This can be generated by shooting null rays from $\chi_{R}$ towards $R$. Let us choose the Cauchy slice $\Sigma$ such that it intersects the boundary of the entanglement wedge at the QES. Let us also assume that $\Sigma$ intersects the future causal horizon of $D_{R}$ at $H^{+}(D)$.
   The statement that the QES lies deeper than the causal surface can be proved by contradiction. Assume that $\chi_{R} \cap {\rm Int} H^{+}(D_{R})$ is non empty as shown in Fig. \ref{Fig:proof}. We can continuously shrink the boundary domain of dependence $D_{R}$ to a new region $D'$ such that the new causal horizon intersects $\Sigma$ at $H^{+}(D')$, which is contained entirely in ${\rm Ext}(\chi_{R})$ (see Fig.\ref{Fig:proof}). Since we are shrinking the region continuously we can choose a $D'$ such that its future causal horizon $H^{+}(D')$ intersects $\chi_{R}$ at $p$, is tangent to it at $p$ and is in ${\rm Ext}(\chi_{R})$ everywhere else. We can now identify $H^{+}(D')$ as $N$ from Eq. \eqref{Eqn:thm1} to obtain the following:
        
    \begin{equation}
        \frac{\delta S_{gen}(EW(\chi_{R}))}{\delta \Sigma^{\mu}} k^{\mu} \geq \frac{\delta S_{gen}(H^{+}(D'))}{\delta \Sigma^{\mu}} k^{\mu}
        \label{eqn:proof1}
    \end{equation}
    where $EW(\chi_R)$ (the boundary of the entanglement wedge) is the surface generated by shooting null rays from $\chi_{R}$ towards $R$. Since the QES extremizes the generalized entropy, the left hand side of Eq. \eqref{eqn:proof1} is zero. Therefore we have
        
    \begin{equation}
        \frac{\delta S_{gen}(H^{+}(D'))}{\delta \Sigma^{\mu}} k^{\mu} \leq 0,
    \end{equation}
    with equality only if $\chi_{R}$ lies on $H^{+}(D')$ in a neighbourhood of p.
    However the generalized second law states that 
    \begin{equation}
        \frac{\delta S_{gen}(H^{+}(D'))}{\delta \Sigma^{\mu}} k^{\mu} \geq 0.
    \end{equation}
    The equality holds only in non-generic spacetimes. Therefore for generic spacetimes we have a contradiction. A similar argument by using the time reversed GSL establishes a contradiction for the past causal horizon. The proof can be extended to the non generic case, see \cite{EngelhardtWall} for details. Therefore the QES is spacelike or null separated from the causal surface and is deeper in the bulk.
    
    This statement leads to an important conclusion. The von Neumann entropy is invariant under unitary transformations. Suppose we perturb the boundary with some unitary operator localized to $R$, this should leave the boundary von Neumann entropy unchanged. This boundary perturbation leads to sources for the bulk fields. However only the bulk fields within $W_{R}$ can be affected due to the boundary unitary. Since $\chi_{R}$ lies deeper in the bulk than $C_{R}$ and is spacelike or null separated from it, the QES is unchanged due to the boundary unitary perturbation. The bulk entanglement entropy $S_{\text{ent-bulk}}$ is also unaffected since the von Neumann entropy is unchanged under unitary transformations. Thus the generalized entropy is unaffected by such boundary unitary perturbations. The classical extremal surface $X_{R}$ can lie inside the causal wedge in spacetimes that violate the classical null energy condition. Thus the FLM holographic entropy would be changed under boundary unitaries, whereas the QES proposal is consistent with entanglement wedge reconstruction and the invariance of the boundary von Neumann entropy under unitary transformations of the state.
    
    \subsection{Maximin vs extremal: strong sub-additivity and entanglement wedge nesting}
    \label{subsec:maxmin}
\begin{figure}
   \centering
    \resizebox{0.6\textwidth}{!}{%
    \includegraphics{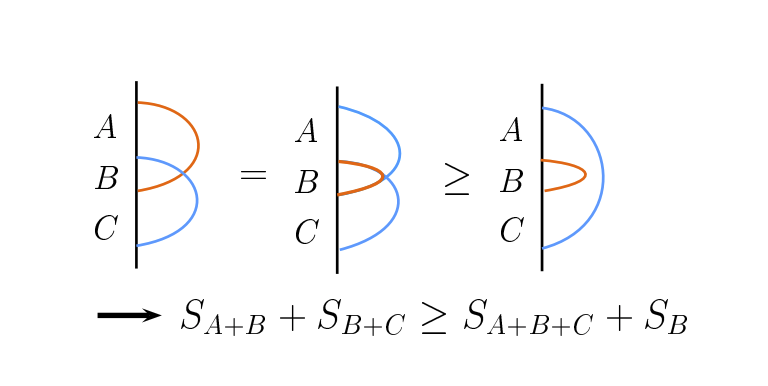}
    }
    \caption{The left most figure computes $S(A\cup B) + S(B \cup C)$ via the RT prescription. The figure in the middle is simply a re-colouring of the different RT surfaces and must have the same entropy as the left most figure. The red and blue curves in the middle figure are not the extremal surfaces for the sub-regions $B$ and $A\cup B \cup C$ respectively, and therefore have a larger area than $S(A \cup B \cup C) + S(B)$. Figure from \cite{Nishioka:2009un}.}
    \label{Fig:RTSSA} 
    \end{figure}

    In the previous subsection we have described how the QES proposal is consistent with the invariance of the boundary von Neumann entropy under unitary transformations. There are two other conditions that any proposal for holographic entanglement entropy should satisfy: (1) strong sub-additivity of the von Neumann entropy and (2) entanglement wedge nesting. The second condition states that if we consider a boundary subregion $R' \subset R$ then the entanglement wedge of $R'$ should lie within the entanglement wedge of $R$. The subalgebra of operators localized to $R'$ must be a subset of the subalgebra of operators localized to $R$ and subregion duality implies the same must be true for the dual operators localized to the corresponding entanglement wedges. Thus entanglement wedge nesting is a consequence of subregion duality.
    The strong sub-additivity condition states that \cite{PhysRevLett.30.434,Lieb:1973cp}
    \begin{equation}\label{eq:SSA-def}
        S(A \cup B) + S(B \cup C) \geq S(B) + S(A\cup B \cup C),
    \end{equation}
    where $A$, $B$ and $C$ are three boundary sub-regions. A simple geometric proof of the strong sub-additivity for the RT prescription was given in \cite{Headrick:2007km} (see Fig. \ref{Fig:RTSSA}). 
    
    Strong sub-additivity and entanglement wedge nesting was shown for the covariant HRT prescription in \cite{Wall:2012uf} using the \textit{maximin} surfaces defined as follows. For the boundary subregion $R$ we consider all possible Cauchy surfaces $\Sigma$ that contain $R$ and find the minimal area surface that is homologous to $R$ ($X_{R}(\Sigma)$) on each of these slices. Then we maximize over all such surfaces $X_{R}(\Sigma)$ to obtain the maximin surface. This surface is more convenient for proofs since it corresponds to a minimal surface on some Cauchy slice just like the RT surface. The maximin surface was shown to be equivalent to the HRT surface if the null curvature condition (NCC) holds \cite{Wall:2012uf}. The NCC states that $R_{\mu \nu} k^{\mu} k^{\nu} \geq 0$ for any null vector $k^{\mu}$. Wall \cite{Wall:2012uf} proved that these maximin surfaces exist in spacetimes without horizons and on spacetimes with Kasner like singularities. Thus in spacetimes satisfying the NCC the existence of HRT surfaces is guaranteed. This existence proof was then extended to generic blackholes in $AdS$ with singularities that are not Kasner like \cite{Marolf:2019bgj}. The quantum generalization of the maximin surface was defined in \cite{Akers:2019lzs} as follows. For a boundary subregion $R$ we consider all possible Cauchy surfaces $\Sigma$ containing $R$ and find the surface that is homologous to $R$ and minimizes $S_{gen}$, then we maximize over all the Cauchy slices. In \cite{Akers:2019lzs} it was proved that the quantum maximin surfaces exist, are identical to the QES and obey strong sub-additivity as well as entanglement wedge nesting.
    
    The RT/HRT and maximin prescription has led to stronger inequalities on the von Neumann entropy that do not hold for non holographic systems \cite{Hayden:2011ag,Bao:2015bfa,Hubeny:2018ijt,Hubeny:2018trv}. These inequalities haven't been shown to hold when we include quantum corrections via the QES prescription, however exploration in this direction was initiated in \cite{Akers:2021lms} where it was shown that if the bulk entropies obey the monogamy of mutual information \cite{Hayden:2011ag} then the dual boundary entropies also obey the same.
    
\section{Bulk reconstruction}
\label{sec:bulkreconstruction}
\subsection{The entanglement wedge reconstruction hypothesis}\label{sec:EWrecons}
    The AdS/CFT dictionary relates observables in the large N strongly coupled QFT at the boundary to the observables in the semi-classical bulk spacetime. The Euclidean partition function ($Z$) of the boundary theory is related to the on-shell bulk gravitational action ($I$) as follows \cite{Witten:1998qj,Gubser:1998bc}:
    \begin{equation}\label{Eq:GKPW}
        Z[\phi_{0}] = e^{- I[\phi_{0}]},
    \end{equation}
    where  $\phi_{0}$ is the boundary value of a bulk field $\phi$ and is identified with the source of the dual boundary operator $\mathcal{O}$. Thus the bulk gravitational action is the generating functional of all connected correlation functions in the boundary theory. This is called the Gubser-Klebanov-Polyakov-Witten (GKPW) prescription in literature. This prescription implies that the connected correlation functions of the field theory can then be obtained by functionally differentiating the on-shell dual bulk gravitational action $I$ with respect to the sources. Divergences in the on-shell bulk gravitational action $I$ arise due to the infinite volume of the $AdS$ spacetime near the boundary and these mimic the local ultraviolet divergences in the dual field theory. These divergences can be systematically removed by first regularizing with a radial cut-off $r =\epsilon$ (the boundary is at $r =0$) and subtracting them with diffeomorphism-invariant local counterterms on the cut-off surface. This procedure is called holographic renormalization  \cite{Henningson:1998gx,BalKraus,deHaro:2000vlm,Skenderis1} (see \cite{Kanitscheider:2008kd} for implementation in more general cases). The radial cutoff thus mimics an energy-scale cutoff in the dual field theory. We discuss more on this issue in section \ref{sec:herg}. The Lorentzian generalization of \eqref{Eq:GKPW} has been discussed in \cite{Son:2002sd,Skenderis:2008dg,Herzog:2002pc,Glorioso:2018mmw}.
    
\begin{figure}
   \centering
    \resizebox{0.6\textwidth}{!}{%
    \includegraphics{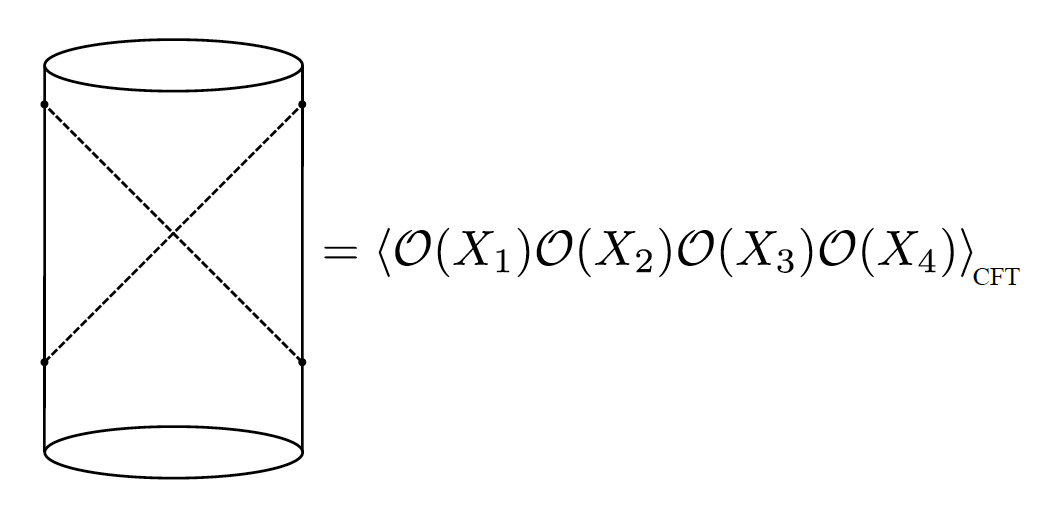}
    }
    \caption{A scattering experiment in the bulk that is equivalent to a 4 point function in the boundary. Figure from \cite{harlow2018tasi}}
    \label{Fig:scattering} 
    \end{figure}

     An equivalent and useful way to state the correspondence which generalizes readily to the Lorentzian signature is as follows. Corresponding to any state $\rho_B$ in the boundary theory there exists an asymptotically $AdS_{d+1}$ solution $B$ in the dual gravity theory which satisfies appropriate smoothness conditions such as absence of naked singularities (unless explicitly mentioned we will assume that $B$ has no horizon). The generic on-shell asymptotic boundary behavior of a scalar field $\phi$ dual to an operator $\mathcal{O}$ with scaling dimension $\Delta$ in such a geometry is:
    \begin{equation}
        \phi(r,t,\mathbf{x})_B = r^{d-\Delta} \phi_{0}(t,\mathbf{x}) (1+ \mathcal{O}(r^2))+ r^{\Delta} \langle O(t,\mathbf{x}) \rangle_{\rho_B}(1+ \mathcal{O}(r^2)).
    \end{equation}
    The source $\phi_0$ which couples to $\mathcal{O}$ is identified with the leading term (non-normalizable mode) in the asymptotic expansion as mentioned before. The coefficient of the sub-leading term (normalizable mode) gets identified with the expectation value of $\mathcal{O}$ in the dual state $\rho_B$ as indicated above. The mass of the field $m$ is related to the scaling dimension $\Delta$ via $$\Delta = \frac{d}{2} +\sqrt{\frac{d^{2}}{4} +m^{2} l^{2}}$$with $l$ the AdS radius.  An \textit{extrapolate} dictionary stated in \cite{Susskind:1998dq,Banks:1998dd} relates correlation functions of the boundary theory in the state $\rho_B$ to scattering $S$-matrices of the semiclassical bulk fields in the geometry $B$ as follows:\footnote{The right hand side of \eqref{Eq:extrapolatecorrln} is more precisely the AdS analogue of the Lehmann-Symanzik-Zimmermann (LSZ) reduction \cite{1955NCimS...1..205L} of $S$-matrices in flat space.}
    \begin{equation}\label{Eq:extrapolatecorrln}
        \left<\mathcal{O}(x_1)\hdots \mathcal{O}(x_n) \right>_{\rho_{B}} = \lim_{r \to0} r^{-n \Delta} \langle\phi(r,x_{1}) \hdots \phi(r,x_n)\rangle_{B}.
    \end{equation}
    For instance a four point function $ \left<O(x_1)O(x_2)O(x_3)O(x_4) \right>$ can be obtained from a $2\rightarrow2$ bulk scattering experiment shown in Fig. \ref{Fig:scattering}.  This extrapolate dictionary has been shown to be equivalent to the GKPW prescription \eqref{Eq:GKPW} in \cite{Giddings:1999qu,Harlow:2011ke}.
    
    The extrapolate dictionary reproduces boundary observables as boundary limits of bulk observables. However, this is not the goal of bulk reconstruction which aims to do the opposite, namely describe (reconstruct) bulk observables in terms of boundary observables. A naive way to do this is to solve the bulk equations of motion with boundary conditions determined by the data of the boundary CFT (expectation values of the dual operators etc) and then use the extrapolate dictionary. For a free bulk scalar field this gives (see \cite{Banks:1998dd} for details of the computation):
    \begin{equation}
        \label{eqn:naiveHKLL}
        \phi(X) = \int d^{d}x K(X,x) \mathcal{O}(x),
    \end{equation}
    where the integration is over all boundary points ($x$) that are spacelike separated from the bulk point $X$ and $K$ is referred to as the smearing function (it is the inverse of the bulk-to-boundary propagator). This expression is correct at the leading order in $N$ and the $1/N$ corrections can be obtained by perturbatively solving the bulk equations of motion including the bulk vertices.
    
    This naive procedure suffers from a major issue. Eq \eqref{eqn:naiveHKLL} says that a bulk local operator $\phi(X)$ depends on all CFT operators localized to a region spacelike to $X$. This non-locality persists even when the bulk operator is pushed to the boundary. Therefore Eq \eqref{eqn:naiveHKLL} doesn't smoothly reduce to the extrapolate dictionary. In order to recover the extrapolate dictionary the smearing function $K(X,x)$ must become more and more local as $X$ is pushed to the boundary. 
    
    Hamilton, Kabat, Lifschytz, and Lowe (HKLL) addressed this issue in the context of the AdS Rindler wedge which is the bulk causal wedge $W_R$ of a ball-shaped region $R$ at the boundary. Recall that the boundary limit of $W_R$, i.e. $\partial W_R$ is $D_R$, the boundary domain of dependence of $R$. HKLL showed that the reconstruction of the bulk operator in a Rindler wedge can be made manifestly consistent with the extrapolate dictionary if we work in Rindler coordinates (which covers $W_R$) instead of the global coordinates of AdS \cite{HKLL,HKLL2}. This is referred to as AdS-Rindler reconstruction. For example, we can choose the CFT state to be the vacuum. The dual geometry is then pure AdS. The HKLL reconstruction procedure is explicitly
    \begin{equation}
    \label{eq:rindler}
        \phi(X) = \int_{D_{R}}{\rm d}^{d-1}x {\rm d}\tau\,\, K^{\text{Rindler}}(X;x,\tau) \mathcal{O}(x,\tau),
    \end{equation}
    where $\tau$ is the Rindler time. Note that the integration is restricted to $D_R$ the boundary domain of dependence of $R$ and $\mathcal{O}(x,\tau)$ is the Heisenberg picture operator evolved with the Rindler Hamiltonian which generates translation in $\tau$, i.e. boosts. The smearing function $K^{\text{Rindler}}$ is known explicitly in terms of the mode functions obtained from the semi-classical quantization of the bulk field in Rindler wedge (see \cite{harlow2018tasi}). This leads to the causal wedge reconstruction conjecture, which states that a bulk field within the causal wedge $W_R$ of a boundary subregion $R$ can be reconstructed on the boundary domain of dependence $D_{R}$, i.e. $\phi(X)$ can be represented using boundary operators within $D_{R}$ provided $X \in W_{R}$. As we move $X$ closer to the boundary, a smaller $D_{R}$ is required to reconstruct $\phi$ on the boundary. This is manifestly consistent with the extrapolate dictionary and therefore solves the issue that occurred in the global reconstruction. It is important to note that the explicit smearing function is known only for the ball-shaped boundary subregions in the vacuum state.
    
    The entanglement wedge reconstruction conjecture states that bulk operators within the entanglement wedge of some boundary region $R$ can be reconstructed from operators on $D_{R}$ at the boundary. This is called the entanglement wedge reconstruction conjecture \cite{Czech_2012,Wall:2012uf,FLM,Headrick_2014,Jafferis_2016}. This conjecture has now been proven using methods of operator algebra (quantum) error correction as will be discussed in section \ref{sec:holoqec}. This automatically implies the causal wedge reconstruction since the causal wedge is contained within the entanglement wedge as shown earlier. Note that in the case of the AdS-Rindler wedge reconstruction, the causal and entanglement wedges coincide. 
    
\begin{figure}
    \centering
    \resizebox{0.4\textwidth}{!}{%
    \includegraphics{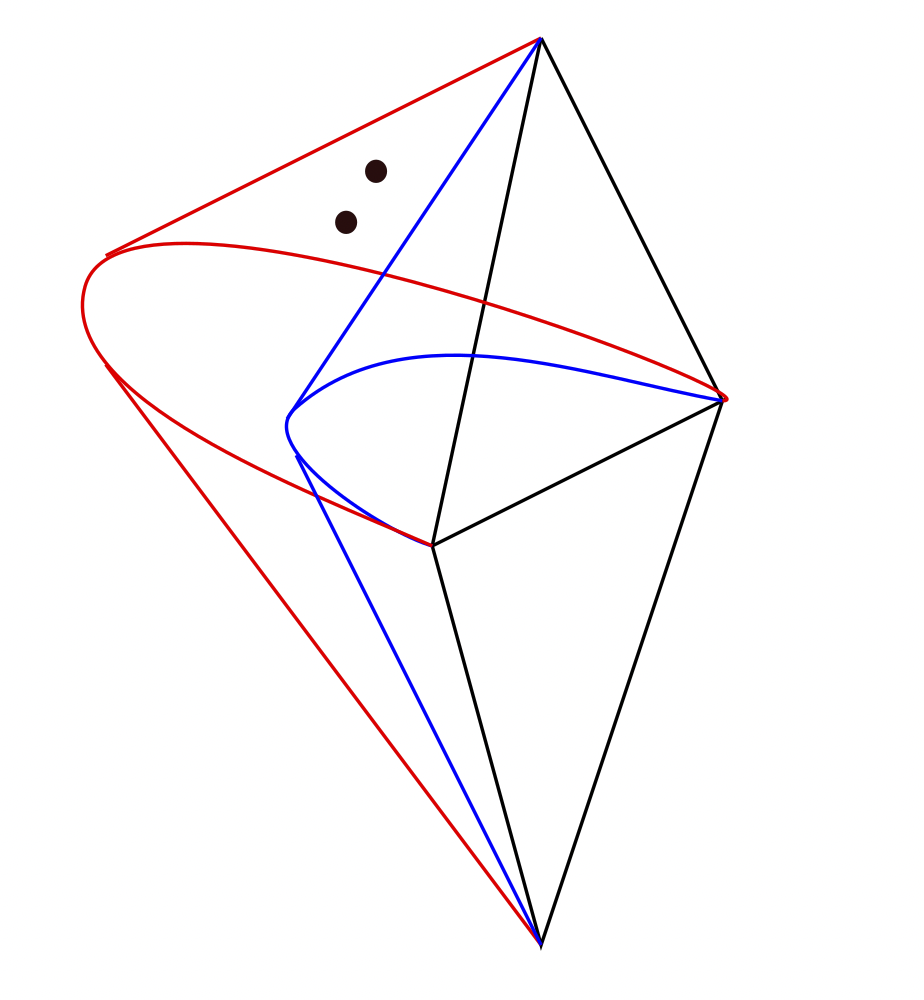}
    }
    \caption{The solid black lines are the edge of the boundary domain of dependence $D_{R}$. Solid blue lines are the edge of the causal wedge and the solid red lines are the edge of the entanglement wedge. The black dots are two bulk operators in the causal shadow region (beyond the causal wedge and within the entanglement wedge). These two operators need not be spacelike separated and need not commute, however a bulk reconstruction formula analogous to Eq.\eqref{eq:rindler} would imply that they commute with each other.}
    \label{Fig:shadow} 
    \end{figure}

    However, the prescription given by Eq. \eqref{eq:rindler} cannot be correct when we consider entanglement wedge reconstruction. Generically the entanglement wedge contains a region which is spacelike separated from the boundary domain of dependence $D_{R}$. This is referred to as the \textit{causal shadow} \cite{Headrick_2014} (see Fig. \ref{Fig:shadow}).\footnote{In \cite{Headrick_2014}, it was shown that a causal shadow is generated for an interval slightly larger than half the boundary (a circle) in the asymptotically $AdS_3$ metric $${\rm d}s^2 =\frac{1}{\cos^2\rho}\left(-f(\rho){\rm d}t^2+\frac{{\rm d}\rho^2}{f(\rho)} +\sin^2 \rho {\rm d}\phi^2\right), \quad f(\rho) = 1 - \frac{1}{2}\sin^2(2\rho).$$The metric is supported by bulk matter satisfying null energy condition.} The causal shadow is spacelike to $D_{R}$, therefore all bulk operators in the causal shadow would commute with operators in $D_{R}$. Thus if an equation analogous to \eqref{eq:rindler} is correct for entanglement wedge reconstruction, then all bulk operators in the causal shadow must commute with each other, which leads to an inconsistency since they are not necessarily mutually spacelike separated. See Fig. \ref{Fig:shadow} for an illustration.
    
    Jafferis, Lewkowycz, Maldacena and Suh (JLMS) \cite{Jafferis_2016} proposed that this inconsistency can be resolved if the bulk reconstruction equation takes the form:
    \begin{align}
    \label{eq:JLMS}
        \phi(X) &= \int_{R} d^{d-1}x \int ds K(X;x,s) \mathcal{O}_{s}(x), \nonumber\\
        \mathcal{O}_{s}(x) &= \rho_{R}^{-i s}  \mathcal{O}(x) \rho_{R}^{is} = e^{i H_\rho s}  \mathcal{O}(x) \rho_{R}^{-i H_\rho s},
    \end{align}
    where $\rho_{R}$ is the reduced density matrix on the boundary subregion $R$ and $H_\rho = - \log\rho$ is the modular Hamiltonian and $K$ is an appropriate smearing function. The conjugation of the operator $\mathcal{O}$ by the density matrix is called modular flow \cite{Haag:1992hx} and $s$ is the modular flow parameter. Since the modular Hamiltonian $H_\rho$ is typically non-local, the modular flowed operators $\mathcal{O}_{s}$ are non-local also, and therefore they do not commute with $\phi$. This resolves the inconsistency due to the causal shadow described before. The JLMS proposal reduces to the HKLL prescription \eqref{eq:rindler} for ball-shaped regions in the vacuum as the modular Hamiltonian is then exactly the Rindler Hamiltonian which generates boosts \cite{Bisognano:1976za,Hislop1981,Casini:2011kv} implying that $\mathcal{O}_{s}(x) = \mathcal{O}(\tau,x)$. Generically, the modular Hamiltonian is local only for boundary regions with sufficient symmetry and for vacuum states. 
    
    The entanglement wedge reconstruction hypothesis states that an appropriate smearing function $K$ should always exist for \eqref{eq:JLMS}. Progress towards an explicit construction of this smearing function will be discussed later in this section.
    
    \subsection{The equivalence of bulk and boundary relative entropies and its consequences}\label{sec:modH_relEnt}
        
    We proceed to first show the key result of \cite{Jafferis_2016} which establishes the equivalence of the boundary and bulk relative entropies as a consequence of the Engelhardt-Wall prescription \cite{EngelhardtWall} for the holographic entanglement entropy in the semi-classical approximation. This will be the fundamental input in the proof of entanglement wedge reconstruction to be discussed in section \ref{sec:recons-proof} in the framework of operator algebra error correction. We will also study some of the striking consequences which follow from this relation.
    
    The relative entropy between two states $\rho$ and $\sigma$ is a measure of their distinguishability (divergence) and is defined as
\begin{equation}
    S(\rho | \sigma) = {\rm Tr} [\rho (\log \rho - \log \sigma) ] 
\end{equation}
Its classical analogue is the Kullback-Leibler divergence between two probability distributions. It could be helpful to see how the relative entropy arises as a measure of distinguishability. Consider a positive-operator-valued-measure (POVM) which is a set of positive semi-definite operators $A_i$ such that $\sum_i A_i = \hat{1}$. We then define the classical probability distributions $p$ and $q$ obtained via $p_i = {\rm Tr}(A_i\sigma)$ and $q_i = {\rm Tr}(A_i\rho)$, and use the classical Kullback-Liebler divergence between $p$ and $q$ to define
$$S_1 := S_1(\rho | \sigma) =  {\rm sup}_{A_i}\left(\sum_i p_i (\log p_i - \log q_i) \right)$$with the supremum taken over all possible POVMs. $S_1$ is thus a measure of distinguishabilty between the two states for a single measurement. We can similarly consider $n$ copies of both $\rho$ and $\sigma$ along with all POVMs acting on these $n$-copies, and define $S_n :=S_n(\rho | \sigma)$. The result of Hiai and Petz is that \cite{HiaiPetz}: $$S(\rho | \sigma)=\lim_{n\rightarrow \infty}S_n.$$We can paraphrase this as the statement that the probability that we can confuse between $\rho$ and $\sigma$ after we perform a large number ($n$) of measurements on $\rho$ decreases as $\exp (-n S(\rho | \sigma))$ as $n\rightarrow\infty$. In quantum field theory, the relative entropy is a measure of how well we can distinguish two states based on the algebra of observables in a subregion $R$. See \cite{RevModPhys.90.045003} for a detailed and illuminating discussion. The relative entropy is invariant under simultaneous unitary transformations of the two states, and therefore like the von-Neumann entropy, it is an observable that depends only on $D_R$ and not the specific choice of $R$.

The first crucial property of the relative entropy is that it is non-negative and vanishes if and only if the two states are identical \cite{nielsen}. It follows also from the similar feature of Kullback-Leibler divergence as should be clear from the above discussion. Furthermore, the relative entropy is related to mutual information. Consider the union of two subregions $A$ and $B$, a joint state $\rho_{A\cup B}$ and the uncorrelated state $\rho_A \otimes \rho_B$ with each density matrix obtained by tracing out the complement of the corresponding subregion. The mutual information between $A$ and $B$ subregions in the joint state is defined as
\begin{equation}\label{Eq:mutin}
    I (A, B) = S(A) + S(B) - S(A\cup B)
\end{equation}
with $S(A)$, $S(B)$ and $S(A\cup B)$ referring to the von Neumann entropies of $\rho_A$, $\rho_B$ and $\rho_{A\cup B}$ respectively. One can readily see that
\begin{equation}\label{Eq:mutinrel} 
   I(A,B) = S(\rho_{A\cup B} | \rho_A \otimes \rho_B).
\end{equation}
The non-negativity of the relative entropy then implies that $I(A,B)$ is positive and vanishes only when the two intervals are fully uncorrelated.

The second crucial property of the relative entropy is its monotonicity under completely positive\footnote{Note that a map $\cN:\cB(\cH)\rightarrow \cB(\cK)$ between bounded linear operators in two Hilbert spaces $\cH$ and $\cK$ is said to be \emph{positive} if it maps positive operators on $\cH$ to positive operators on $\cK$. The map $\cN$ is said to be \emph{completely positive} if any extension of the map is also a positive map. In other words, suppose the map $\cN$ acts on a subsystem $\cH_{A}$ of a composite system $\cH_A \otimes \cH_{B}$, then complete positivity ensures that $(\cN \otimes \cI)(\rho_{AB}) \geq 0$, for all (positive) $\rho_{AB} \geq 0$.} trace preserving (CPTP) maps. A density matrix maps to another density matrix under a CPTP map and will characterize an arbitrary noise channel in the context of quantum error correction. It has been shown that \cite{uhlmann1977relative}
\begin{equation}\label{Eq:relentmon}
    S(\mathcal{N}(\rho) | \mathcal{N}(\sigma)) \leq S(\rho | \sigma)
\end{equation}
for an arbitrary CPTP map $\mathcal{N}$. Considering $\mathcal{N}$ to be the tracing out of a subregion $C$, we can readily see that the monotonicity of the relative entropy implies the strong subadditivity property \eqref{eq:SSA-def} of the entanglement entropy as follows. Under this trace operation we should have the inequality
\begin{equation}\label{Eq:SSA2}
    S(\rho_{A\cup B\cup C} | \rho_A \otimes \rho_{B\cup C}) \leq S(\rho_{A\cup B} | \rho_A \otimes \rho_{B}),
\end{equation}
from the monotonicity property. We easily obtain \eqref{eq:SSA-def} from the above inequality using \eqref{Eq:mutinrel} and \eqref{Eq:mutin}. The strong-subadditivity of the entanglement entropy is saturated for quantum Markov chain states in which $A$ and $C$ are independently conditioned by $B$ as will be discussed in section \ref{sec:recons-proof}. This will have implications for toy models of holography.

    To proceed further, we rewrite the relative entropy in the following form
        \begin{eqnarray}\label{Eq:relent21}
            S(\rho | \sigma) 
                               &=& Tr[\rho \log \rho - \sigma \log \sigma] +Tr[(\sigma  - \rho) \log \sigma] \nonumber\\
                              &=& -\Delta S + \Delta \left< H_{\sigma}\right>, \nonumber
        \end{eqnarray}
        where $\Delta S$ is the difference between the von Neumann entropies of the states $\rho$ and $\sigma$, and $\Delta\left< H_{\sigma}\right>$ denotes the difference between the expectation value of the modular Hamiltonian of $H_\sigma = - \log \sigma$ in these two states. Note these differences are exact and not infinitesimal. Since the relative entropy is non-negative and reaches its extremal vanishing value when the two states are identical, the first order change in the relative entropy must vanish for an infinitesimal difference between the two states  i.e. when $\rho = \sigma +\delta \sigma$. This implies the first law of entanglement entropy \cite{Blanco:2013joa}
        \begin{equation}\label{Eq:FirstLaw}
            \delta S = \delta \left<H_{\sigma} \right>
        \end{equation}
        for any infinitesimal variation of the state $\sigma$.
        
        The Engelhardt-Wall prescription states that for any boundary subregion $R$
        \begin{equation}
        \label{eq:QES}
             S(R)_{\text{bdy}} = \frac{\mathcal{A}(\chi_{R})}{4 G} + S_{\text{ent-bulk}}
        \end{equation}
        where $\chi_R$ is the quantum extremal surface which extremizes the generalized entropy with $S_{\text{ent-bulk}}$ the entanglement entropy of the bulk matter within the corresponding entanglement wedge. The area term can be viewed as the expectation value of an operator in the bulk effective field theory ($Tr[\rho \frac{\hat{A}_{\chi_{R}}}{4 G }]$). Note that the von-Neumann entropy is simply the expectation value of the modular Hamiltonian: $S(\sigma) = Tr[ \sigma H_{\sigma} ] = \left< H_{\sigma}\right>_{\sigma}$. Therefore Eq \eqref{eq:QES} can be written as an equivalence between the bulk and boundary modular Hamiltonians as \cite{Jafferis:2014lza}:
        \begin{equation}
        \label{eq:blkdbymod}
            H_{\text{bdy}} = \frac{\hat{A}_{\chi_{R}}}{4 G \hbar} + H_{\text{bulk}}
        \end{equation}
        Note that the area operator $\hat{A}_{\chi_{R}}$ will commute with both $H_{\text{bulk}}$ and $H_{\text{bdy}}$ since $X_R$ is spacelike separated with all points in the corresponding bulk and boundary regions. Furthermore, the boundary relative entropy in the form \eqref{Eq:relent21} is
        \begin{equation}
          S(\rho | \sigma)_{\text{bdy}} = -\Delta S_{\text{bdy}} + \Delta \left< H_{\sigma}^{\text{bdy}}\right>.
        \end{equation}
        From \eqref{eq:QES} we readily obtain that
        \begin{equation}
        \label{eq:deltaS}
            \Delta S_{\text{bdy}} = \Delta \left(\frac{\mathcal{A}(X_{R})}{4 G \hbar} \right) + \Delta S_{\text{bulk}} 
        \end{equation}
         and similarly from \eqref{eq:blkdbymod} we obtain
        \begin{equation}
        \label{eq:deltaH}
            \Delta \left< H_{\sigma}^{\text{bdy}}\right> = \Delta\left<\frac{\hat{A}_{\chi_{R}}}{4 G \hbar} \right> + \Delta \left< H_{\sigma}^{\text{bulk}}\right>.
        \end{equation}
        Subtracting Eq. \eqref{eq:deltaS} from Eq \eqref{eq:deltaH} and noticing that the area term cancels, we immediately obtain the desired result \cite{Jafferis_2016}
        \begin{equation}
            S(\rho | \sigma)_{\text{bdy}} =  S(\rho | \sigma)_{\text{bulk}} \label{eq:entropy_eqv}
        \end{equation}
        stating the equivalence between the boundary and bulk relative entropies. It is interesting to note that a similar equivalence between boundary and bulk mutual information (between two subregions) can also be readily proved \cite{FLM}.
        
        Let us consider a bulk field $\phi$ within the entanglement wedge of some boundary region $R$. Eq. \eqref{eq:blkdbymod} implies:
        
        \begin{equation}
        \label{eq:modflow}
            [H_{\text{bdy}},\phi ] = [H_{\text{bulk}},\phi]
        \end{equation}
        The area term in $H_{\text{bdy}}$, which is localized on the extremal surface and is spacelike to the interior of the entanglement wedge, drops out. Therefore bulk causality implies that the bulk and boundary modular flows are identical.
        
        The definition of the modular Hamiltonian contains some ambiguities. For example in lattice gauge theories observables are located on the links of the lattice. Therefore, when a surface splitting space into two parts cuts a link, it is not clear which side of the surface should include the observable on that link. These ambiguities are localized on the boundary of $R$. Nevertheless, the relative modular Hamiltonian $H_{\text{rel-bdy}}= H_{R-\text{bdy}}-H_{\bar{R}-\text{bdy}}$ is free of such ambiguities. Similarly, the bulk modular Hamiltonian has ambiguities localized on the extremal surface, but the bulk relative modular Hamilton is free of these ambiguities. The relative bulk and boundary modular Hamiltonians should be identical, i.e. $H_{\text{rel-bdy}} = H_{\text{rel-bulk}}$, if the area term  cancels out, i.e. if $R$ and its complement $\overline{R}$ share the same extremal surface (and hence the two corresponding entanglement wedges are complements of each other). This happens for a pure state (and horizonless bulk geometries). This ambiguity free modular Hamiltonian can be used to define the modular flow.
        
        To see consequences of the equivalence of bulk and boundary relative entropies, we need the first law of entanglement  entropy \eqref{Eq:FirstLaw} which simply follows from the vanishing of the first order variation of the relative entropy as shown above. We will also need the result for the second order variation of the relative entropy. Let $$\rho = \sigma(\epsilon) = \sigma + \epsilon \delta_1 \sigma + \epsilon^{2}\delta_2 \sigma + \mathcal{O}(\epsilon^3)$$. We readily see that
        \begin{equation}\label{Eq:QFI}
            \delta S(\sigma(\epsilon) | \sigma) = \epsilon^{2} \frac{1}{2} {\rm Tr}\left[\delta_1 \sigma \frac{{\rm d}}{{\rm d} \epsilon}\log(\sigma + \epsilon \delta_1 \sigma)\Big\vert_{\epsilon =0}\right] + \mathcal{O}(\epsilon^3)
        \end{equation}
        where terms containing $\delta_2\sigma$ vanish for the same reason as in the case of the first order variation mentioned before. This implies
        \begin{equation}\label{Eq:QFI1}
            \frac{{\rm d}^2 S(\sigma(\epsilon) | \sigma)}{{\rm d} \epsilon^2}\Big\vert_{\epsilon =0} =  {\rm Tr}\left[\delta_1 \sigma \frac{{\rm d}}{{\rm d} \epsilon}\log(\sigma + \epsilon \delta_1 \sigma)\Big\vert_{\epsilon =0}\right] := \left<\delta_1\sigma, \delta_1\sigma\right>_\sigma
        \end{equation}
        The quantity $\left<\delta_1\sigma, \delta_1\sigma\right>_\sigma$ is called the quantum Fisher information which defines a Riemannian metric on the space of states. Quantum Fisher information is important in the study of quantum metrology and state estimation, where it bounds the amount of information that can be obtained about a state by generalized measurements \cite{Petz2011}. The positivity of relative entropy implies that the quantum Fisher information is positive. 
        
        Let us consider $\sigma$ to be the reduced density operator on a ball shaped subregion $B$ of the vacuum state of a holographic CFT and $\rho$ to be the corresponding reduced density operator of a perturbed state \textit{close} to the CFT vacuum. The vacuum is dual to pure AdS and the perturbed state is dual to a perturbation of pure AdS. It was shown in \cite{Lashkari_2014,Faulkner_2014} using results from \cite{Casini:2011kv,Hollands_2012} that at the leading and subleading orders the variation of the bulk relative entropy can be written in the form (with the bulk metric $g = g_0 + \epsilon \delta g + \mathcal{O}(\epsilon^2)$):\footnote{The key to this result is the map of the bulk Rindler wedge dual to the domain of dependence $D_B$ of the ball shaped region at the boundary $B$ to a hyperbolic black hole using\cite{Casini:2011kv} (the Casini-Huerta-Myers (CHM) map). This map is dual to the statement that the vacuum state in $D_B$ can be mapped conformally to a thermal density matrix in hyperbolic space with radius of curvature $R_H = 1/(2\pi T)$ given by the temperature. The perturbations of the vacuum is dual to gravitational perturbations of the hyperbolic black hole with a bifurcate horizon which can be analyzed by the method of Wald and Hollands in \cite{Hollands_2012}. Note that the QES is mapped to the bifurcate horizon. Then the change in bulk relative entropy can be split using \eqref{Eq:relent21} with the change in von-Neumann entropy given by the change of the black hole entropy and the change in the modular Hamiltonian (which gets identified with the usual Hamiltonian after the CHM map) given by the change in the Arnowitt-Misner-Deser energy. Then the change in the bulk relative entropy can be reproduced essentially from the variation of the gravitational action according to \cite{Hollands_2012} which further develops results in \cite{Iyer:1994ys}.}
        \begin{equation}\label{Eq:first-order}
            \delta S(\rho | \sigma)_{\text{bulk}} = \mathcal{E}(g_0,\delta g,\mathcal{L}_{\xi}g ) - 2 \int_{\Sigma} \xi^{\mu} \delta E_{\mu \nu}(g) d \Sigma^{\nu}
        \end{equation}
        where $\Sigma$ is the Cauchy slice bounded by the boundary subregion and the bulk extremal surface, $\delta g$ is the bulk metric perturbation and$$\mathcal{L}_{\xi} g = \nabla_{\mu}\xi_{\nu}+ \nabla_{\nu}\xi_{\nu}$$is the Lie derivative of the bulk metric $g$ in the direction of $\xi$ which is the timelike Killing vector associated to the bulk Rindler wedge (note that $\xi$ vanishes on the extremal surface). $E_{\mu \nu}$ is proportional to the equations of motion. Furthermore, $\mathcal{E}$ is a symplectic form on $\Sigma$ given by:
        \begin{equation}
            \mathcal{E}(g_0,\delta_{1}g,\delta_{2}g) = -\frac{1}{16 \pi} \int_{\Sigma} \delta_{1} h_{\mu\nu} \delta_{2}p^{\mu\nu} - \delta_{2} h_{\mu\nu} \delta_{1}p^{\mu\nu} 
        \end{equation}
        where $h$ is the induced metric on $\Sigma$ and $p^{\mu \nu} = \sqrt{h}(K^{\mu \nu} -h^{\mu \nu} K ) $ with $K_{\mu\nu}$ denoting the extrinsic curvature of the Cauchy surface (and $K = h^{\mu\nu}K_{\mu\nu}$).
        
        Since $\mathcal{L}_{\xi} g_0  = 0$ as $\xi$ is a Killing vector associated to the unperturbed metric $g_0$, the canonical energy  vanishes, i.e. $\mathcal{E} =0$ at leading order in $\epsilon$. The first order variation of relative entropy must vanish due to the positivity of relative entropy. Since this should hold for any Cauchy surface $\Sigma$, we obtain:
        \begin{equation}
            \delta E_{\mu \nu}(g) = \mathcal{O}(\epsilon^2).
        \end{equation}
        This simply implies that the perturbed metric should satisfy linearized Einstein's equation expanded about the background for an arbitrary perturbation. Therefore positivity of relative entropy is equivalent to linearized Einstein equations.
        
        The second order variation of the relative entropy is then
        \begin{equation}
            \left.\frac{d^{2} S(\rho | \sigma)_{\text{bulk}}}{d \epsilon^{2}}\right\vert_{\epsilon=0} = \mathcal{E}(\delta g,\mathcal{L}_{\xi}\delta g ) - 2 \int_{\Sigma} \xi^{\mu} \left.\frac{\partial^{2} E_{\mu \nu}(g)}{\partial \epsilon^{2}}\right\vert_{\epsilon =0} d \Sigma^{\nu}.
        \end{equation}
        The vanishing of the linearized equations of motion finally implies that \cite{Lashkari_2016}:
        \begin{equation}
            \left.\frac{d^{2} S(\rho | \sigma)_{\text{bulk}}}{d \epsilon^{2}}\right \vert_{\epsilon=0} = \mathcal{E}(g_0,\delta g,\mathcal{L}_{\xi}\delta g ).
        \end{equation}
        The right hand side is the canonical energy of the linearized perturbation \cite{Hollands_2012}. Using the equivalence of the bulk and boundary relative entropies and \eqref{Eq:QFI1}, we obtain that the left hand side is exactly the quantum Fisher information in the boundary. Therefore, the quantum Fisher information of a perturbation of the density matrix in the CFT is dual to the bulk canonical energy of the dual linearized gravitational perturbation \cite{Lashkari_2016}. The positivity of the Fisher information (which follows from that of the relative entropy) then must imply the positivity of the canonical energy as indeed is the case for perturbation about any stable vacuum.
        
    \subsection{Modular flow and bulk reconstruction}
    \subsubsection{The JLMS smearing function}
        In this section we describe the explicit construction of the smearing function \cite{Faulkner_2017} in equation \ref{eq:JLMS}. For this it is useful to first consider the Fourier transform of the modular flowed operators:
        
        \begin{equation}
        \label{eq:fouriermod}
            O_{\omega} = \int_{-\infty}^{\infty} ds e^{-i s \omega} e^{i H_{\sigma} s} O  e^{-i H_{\sigma} s}, \quad [H_{\sigma},O_{\omega}] = \omega O_{\omega},
        \end{equation}
        where $H_{\sigma}$ is the modular Hamiltonian for the state $\sigma$. We describe below an explicit expression for the smearing function from \cite{Faulkner_2017} that can be obtained by looking at the zero modular frequency mode. We can consider a bulk operator $\phi$ in the entanglement wedge and look at its zero mode $\phi_{0}(X)$. Since the bulk modular flow is the same as boundary modular flow as seen in Eq.\eqref{eq:modflow}, it follows that $[\phi_{\omega} , H_{\sigma} ] = \omega \phi_{\omega}$. Therefore the zero mode $\phi_{0}$ commutes with the modular Hamiltonian and this field must be localized on the extremal surface. It was then shown in \cite{Faulkner_2017} that the zero mode of the dual boundary operator is:
        
        \begin{equation}
            O_{0}(x) = \int_{\chi_{R}} dX_{\chi_{R}} \left<\phi(X_{\chi_{R}}) O(x) \right> \phi(X_{\chi_{R}}),
        \end{equation}
        where $\chi_{R}$ is the extremal surface corresponding to a boundary subregion $R$. This generalizes the results from \cite{Czech:2016xec,deBoer:2016pqk,CarneirodaCunha:2016zmi}. Inverting this expression gives
        
        \begin{equation}
            \phi(X_{\chi_{R}}) = \int_{R} dx K_{0}(X_{\chi_{R}}, x) O_{0}(x),
        \end{equation}
        where $K_{0}$ can be obtained by inverting the usual bulk to boundary correlator defined as follows:
        
        \begin{equation}
            \left< \phi(X_{\chi_{R}}), O(x)\right> = \int dy K_{0}(X_{\chi_{R}},y) \left< O(x), O_{0}(y)\right>.
        \end{equation}
        This shows how bulk operators on the QES can be reconstructed. If we foliate the entanglement wedge by extremal surfaces corresponding to smaller and smaller boundary subregions contained within $R$ we can reconstruct operators on the full entanglement wedge. This foliation is well defined due to entanglement wedge nesting which was reviewed in section \ref{subsec:maxmin}. The inversion of the bulk to boundary propagator only needs to be done over the extremal surface making this a much simpler computation compared to the inversion of the full bulk to boundary propagator. (A similar technical computation in a different context was done in \cite{Banerjee:2016mhh}.)

        These modular zero modes have been used to define the modular Berry connection for the boundary CFT \cite{Czech:2017zfq}, which has been related to the Riemann curvature in the bulk \cite{Czech:2019vih}.  
    \subsubsection{A note on modular Hamiltonians for excited states and bulk reconstruction}
    \label{subsec:modularhamexcited}
        We have described the JLMS proposal and an explicit construction of the smearing function. The only missing ingredient for entanglement wedge reconstruction is the modular Hamiltonian, which can be explicitly obtained for sufficiently symmetric boundary subregions of the vacuum state \cite{Bisognano:1976za,Hislop1981,Casini:2011kv,Faulkner:2016mzt,Casini:2017roe}. However in generic situations the modular Hamiltonian is a non-local object that is very hard to compute (for examples see \cite{Klich:2015ina,Lashkari:2015dia,Klich:2017qmt}). Using the Casini, Huerta and Myers map \cite{Casini:2011kv} Sarosi and Ugajin \cite{S_rosi_2018} have described an explicit CFT construction of the modular Hamiltonian for ball shaped subregions in slightly excited states close to the vacuum. Let the excited state be $\rho = \sigma + \delta \sigma$, then the modular Hamiltonian is \cite{S_rosi_2018}:
        
        \begin{align}
        \label{eq:modularham}
            &H_{\rho} = H_{\sigma} + \sum_{n=1}^{\infty} (-1)^{n} \int_{-\infty}^{\infty} ds_{1}\hdots ds_{n} \mathcal{K}_{n}(s_{1}\hdots s_{n})\mathcal{P}\nonumber\\
            &\mathcal{P}= \prod_{i=1}^{n} e^{-\left(i\frac{s_{i}}{2\pi}+\frac{1}{2}\right)H_{0}} \delta \sigma e^{\left(i\frac{s_{i}}{2\pi}+\frac{1}{2}\right)H_{0}}
        \end{align}
        This construction of the modular Hamiltonian doesn't assume that the CFT is holographic. Sarosi and Ugajin \cite{S_rosi_2018} show using the results of \cite{Faulkner_2014} (valid for the Rindler wedge) that their construction of the modular Hamiltonian relates the quantum Fisher information to the canonical energy of an \emph{emergent} bulk \emph{for any CFT} without assuming either the RT/HRT formula in the bulk or the large $N$ limit for the CFT.
        The expressions for the kernel $\mathcal{K}_{n}$ in Eq.\eqref{eq:modularham} can be found in \cite{S_rosi_2018}, here we reproduce the kernels for $n=1,2$:
        
        \begin{align}
            \mathcal{K}_{1}(s_1) &= \frac{1}{(2\cosh{\frac{s_1}{2}})^{2}}\\
            \mathcal{K}_{2}(s_1,s_2)&=\frac{1}{16 \pi} \frac{i}{\cosh{\frac{s_1}{2}}\cosh{\frac{s_2}{2}}\sinh{\frac{s_2-s_1}{2}}}
        \end{align}
        Note that the kernel $K_{1}$ is the same as the kernel seen in the twirled Petz map \cite{Junge_2018,cotler2019_univR} described in section \ref{sec:holoqec}. See \cite{Kabat:2020oic,Kabat:2021akg} for an explicit evaluation of Eq. \eqref{eq:modularham}.
    \subsection{Why bulk reconstruction is quantum error correction}\label{sec:BR_qec}
\begin{figure}
    \centering
    \resizebox{0.5\textwidth}{!}{%
    \includegraphics{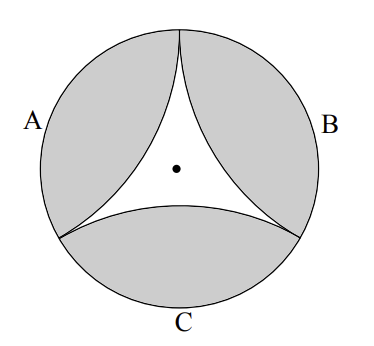}
    }
    \caption{Three boundary subregions with their corresponding entanglement wedges are shown. The dot indicates a bulk operator. Figure from \cite{Almheiri_2015}}
    \label{Fig:redundancy} 
    \end{figure}

        We have already seen hints of a connection between entanglement wedge reconstruction and quantum error correction. For instance Eq.\eqref{eq:blkdbymod} and \eqref{eq:modflow} have been argued to be equivalent to the conditions for quantum error correction \cite{dong_2016_2}. Equation \ref{eq:modularham} has a structure reminiscent of the twirled Petz map \cite{Junge_2018,cotler2019_univR} which will be reviewed in section \ref{sec:holoqec}.
        Moreover entanglement wedge reconstruction leads to an interesting puzzle \cite{Almheiri_2015}. Consider three boundary subregions $A$,$B$ and $C$, as shown in Fig. \ref{Fig:redundancy}. The gray regions are the entanglement wedges for the corresponding boundary subregions. The dot indicates a bulk field $\phi(X)$. This bulk field lies outside the entanglement wedges of each of the three boundary regions. However the field lies in the entanglement wedge of $A\cup B$, $B\cup C$ and $A \cup C$. Thus the bulk operator $\phi$ can be reconstructed on $A\cup B$ and must therefore commute with all operators on $C$ due to causality in the boundary theory. Similarly we can argue that $\phi$ must commute with all operators on $A$ and $B$ by reconstructing it on $B\cup C$ and $A\cup C$ respectively. Therefore the reconstructed bulk operator must be proportional to the identity since it commutes with all CFT operators. This inconsistency can be avoided if the three representations of the bulk operator on the three subregions are not the same. Thus the same bulk operator is \textit{encoded} as different operators on the boundary subregions. Such redundant encoding is essential to quantum error correction. This connection between bulk reconstruction and quantum error correction will be described in detail in section \ref{sec:holoqec}.
    
\subsection{A first look at islands}
\label{sec:islands}
    Following the QES proposal by Engelhardt and Wall \cite{EngelhardtWall}, the Page curve \cite{Page_1993,Page_2013} for an evaporating black hole in $AdS_{2}$ was computed in \cite{AEMM,PeningtonQES}. Similar models for black hole evaporation were studied in \cite{Mertens_2019,Rozali_2020}. The results from these models show that the semi-classical geometry can see features of unitarity (Page curve) in black hole evaporation. The information paradox is of course not resolved since it is still not clear how the information about the black hole interior is encoded in the radiation and how it can be decoded. Nevertheless this was an important step towards the resolution of the paradox.
    
\begin{figure}
    \centering
    \resizebox{0.6\textwidth}{!}{%
    \includegraphics{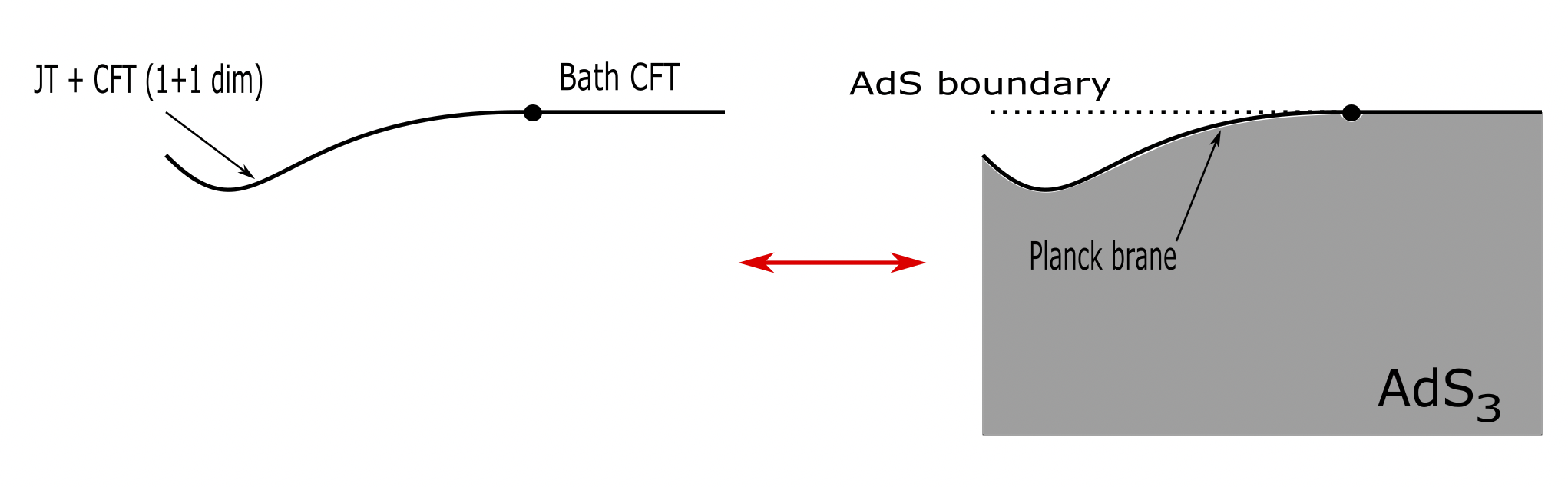}
    }
    \caption{The doubly holographic setup from \cite{AlmheiriQES}. A quantum dot (shown in black) described holographically by JT gravity with holographic bulk matter is brought in contact with the same bath holographic CFT without gravity. The boundary condition is such that the black hole in JT gravity can evaporate. The holographic dual of the full setup is locally $AdS_{3}$ spacetime with a codimension one Planck brane where the JT theory lives.}
    \label{Fig:randall} 
    \end{figure}

    The setup for these computations is Jackiw-Teitelboim \cite{JACKIW1985343,TEITELBOIM198341,almheiri2015models} (JT) gravity with bulk matter described by a $1+1$ dimensional CFT. This is dual to a quantum dot. At some time $t=0$ the bulk boundary conditions are changed by coupling the dual quantum dot to a wire described by the same $1+1$ dimensional CFT as in the bulk but on a flat background without dynamical gravity (see Fig. \ref{Fig:randall}). Therefore quanta can now flow across the boundary and the black hole in $AdS_{2}$ starts evaporating. The QES can be explicitly computed in this setup since the bulk entanglement entropy is that of a two dimensional CFT which can be obtained using the methods of Cardy and Calabrese \cite{Calabrese:2009qy}. It was shown in \cite{AEMM,PeningtonQES} that the QES has a phase transition which leads to the \textit{turning around} of the Page curve of the black hole. At early times, the QES remains close to the bifurcation point of the original black hole horizon (before coupling to the bath) and starts moving outward towards the boundary. The von Neumann entropy of the dual quantum dot increases due to the emitted Hawking quanta. After the Page time, when the black hole and the radiation have same number of degrees of freedom, (a more precise definition of the Page time is in Section \ref{sec:ReplicaWomhole}), a different extremal surface has minimal generalized entropy. This QES is located just inside the event horizon of the black hole and has a decreasing entropy, and thus implying that the the entanglement entropy of the quantum dot decreases. This is the desired feature of the Page curve if the full system has an unitary evolution (to be discussed later). 
    
    Although a Page curve was seen for the entropy of the quantum dot (dual to the evaporating black hole), the entropy of the Hawking radiation computed semi-classically in \cite{AEMM} was shown to grow monotonically as seen in Hawking's original computation. In section \ref{sec:ReplicaWomhole}, we will analyze these models from the point of view of the full gravitational path integral which computes the R\`{e}nyi entropies of the Hawking radiation and explain how the naive semi-classical computation should be refined by including appropriate saddles which automatically reproduce the location of the QES and give results consistent with unitarity.  Such computations however simplify remarkably in a so-called doubly holographic setup when the bulk matter comprising the Hawking quanta is itself holographic. So, we briefly discuss this below.
    
    In the setup of \cite{AlmheiriQES} the full two-dimensional quantum dot plus wire (bath) system  has a three dimensional holographic dual. This can be though of as a locally $AdS_{3}$ geometry with a dynamical boundary where the JT gravity theory is located (see Fig. \ref{Fig:randall}). This is essentially the same as the setup in \cite{Randall_1999,Karch_2001} where the dynamical  boundary was called the \textit{Planck brane}.
    
\begin{figure}
    \centering
    \resizebox{0.6\textwidth}{!}{%
    \includegraphics{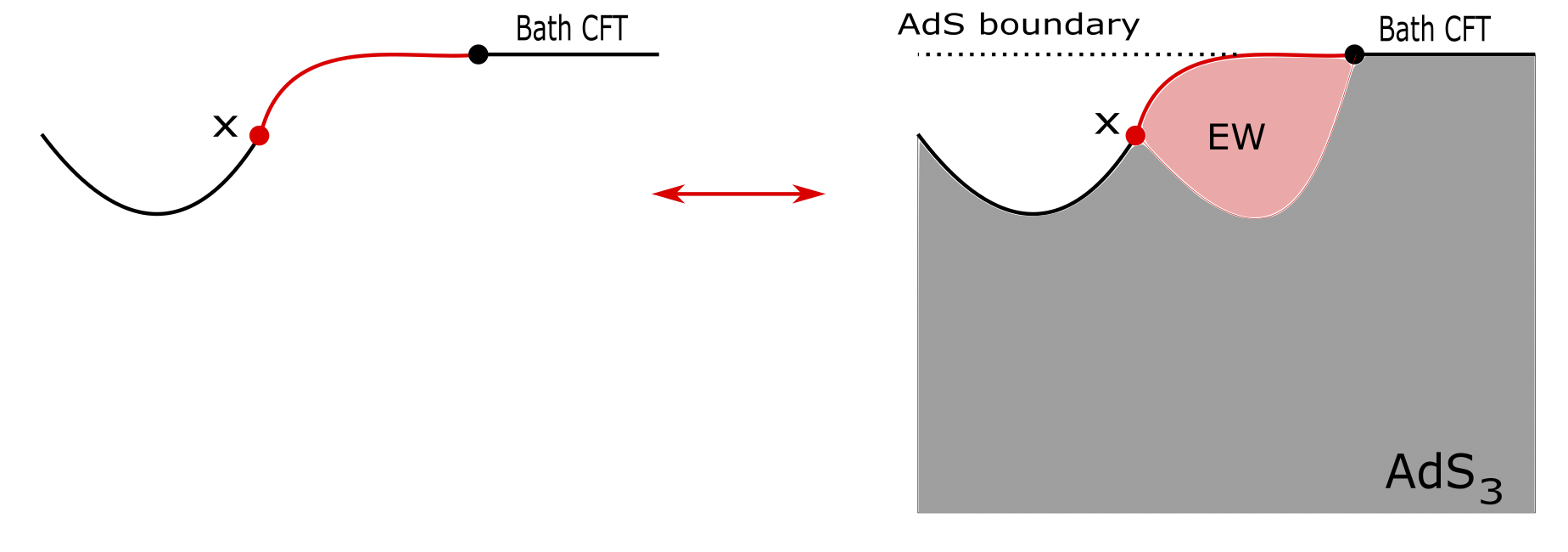}
    }
    \caption{The red segment in the figure on the left indicates where the bulk entanglement entropy must be computed from the point of view of the 2D gravity theory. The red region on the right is the late time entanglement wedge of the black hole in the 3D theory. The entanglement wedge of the bath is the grey region in the right figure, which includes an \textit{island} behind the horizon.}
    \label{Fig:eweg} 
    \end{figure}

    The computation of the generalized entropy in this setup is simple since the bulk entanglement entropy can be computed via the original RT/HRT prescription in the dual three dimensional gravity theory to obtain:
    
    \begin{equation}\label{Eq:GenEntropyJT}
        S_{gen}(x) = \frac{\phi(x)}{4 G^{(2)} \hbar} + \frac{A^{(3)}(X_{x})}{4 G^{(3)} \hbar},
    \end{equation}
    where $x$ is the location of the QES on the Planck brane and $X_{x}$ is the classical extremal surface in three dimensions that is anchored to $x$ at one end and the boundary of the semi-infinite interval in the bath system at the other (see Fig. \ref{Fig:eweg}). The first \textit{area term} in this two dimensional case is simply the value of the dilaton ($\phi$) from the JT theory at the location of the QES. The entanglement wedge for the black hole in this setup is shown in Fig. \ref{Fig:eweg}.
    
    If we try to compute the entanglement entropy of the bath CFT by usual methods, that is by tracing out everything to the left of the black dot in Fig. \ref{Fig:eweg} (left), we would end up with an entropy that increases forever in time. This was seen in \cite{AEMM}. (This naive computation is the coarse-grained entropy as will be defined later.) However in the doubly holographic setup the full CFT (bulk + bath) is dual to a three dimensional geometry. We should therefore use the RT/HRT prescription to compute the entanglement entropy of the bath. This results in an entanglement wedge for the bath that is exactly the complement of the entanglement wedge for the quantum dot (black hole) and hence they have the same entanglement entropies and Page curves as should be the case if the full system is a pure state and evolves unitarily. (The QES has exactly the same phase transition in the doubly holographic setup at Page time as in the computation in \cite{AEMM} which is for a general CFT.) If we look from the two dimensional perspective after the Page time (Fig. \ref{Fig:eweg} (left)) we see two disconnected pieces in the entanglement wedge of the bath (black curves in Fig. \ref{Fig:eweg}) forming \textit{islands}. However if we look at this from the three dimensional perspective, the two islands are connected (Fig. \ref{Fig:eweg} (right)). This is a realization of $ER = EPR $ \cite{Maldacena_2013} paradigm in which the bath should be connected to the interior via a bridge in an extra dimension (wormhole) as a result of large amount of entanglement between them generated by the accumulation of semiclassical Hawking EPR (maximally entangled) pairs.
    
    Based on the above computation the authors of \cite{AlmheiriQES} proposed a new rule for computing entanglement entropies in setups that involve reference systems coupled to gravitational systems. This new rule for $S(R)$ the entanglement entropy of a subregion $R$ in the bath (without gravity) coupled to a gravitating system is as follows:
    
    \begin{equation}\label{Eq:IslandRule}
        S(R) = \min_{I} \underset{I}{\rm Ext} \left[S_{\text{ent-bulk}}(R \cup I) + \frac{A(\partial I)}{4 G}\right].
    \end{equation}
    According to this rule, we should first extremize over the islands $I$ in the gravitating system and then minimize over the extrema. $S_{\text{ent-bulk}}$ is the entanglement obtained from the semiclassical description of the system (in the doubly hologrpahic case it is given by the area of the RT surface). When $R$ is the entire bath (right of the black dot in Fig. \ref{Fig:eweg}), the first term in Eq.\eqref{Eq:IslandRule} equals the entropy of the black hole ($B$) since the full state on $I \cup B \cup R$ is pure. The second term is simply the area of the shared QES. Therefore this new \textit{island rule} gives the same entropy for the bath as that of the black hole leading to the same Page curve for both subsystems.
    
    After Page time the position of the QES in all such setups geometrically realizes the Hayden-Preskill time for information mirroring \cite{Hayden_2007} (more discussion in Section \ref{sec:AMPS}) \cite{AEMM}. More precisely, if some information is thrown into the blackhole post Page time then after the Hayden Preskill time the information crosses the QES. Therefore, the information escapes the entanglement wedge of the boundary and enters the island i.e. the entanglement wedge of the wire (Hawking radiation). It follows that the information thrown into the black hole can be recovered from the Hawking radiation after the Hayden-Preskill time. 
    
    These island computations in two dimensional gravity have been extended to higher dimensions in \cite{Almheiri_higher,chen2020quantum,chen2020quantum2}. It has also been shown that islands can extend outside of event horizons \cite{almheiri2019islands}. The doubly holographic setups have been further analyzed in \cite{Chen_2020} where the Page transition has been studied after excising intervals in the bath CFT . Further studies have been done in \cite{Gautason:2020tmk,Akers:2019nfi,Hartman:2020swn,Hollowood:2020cou,Anegawa:2020ezn,Hashimoto:2020cas,Balasubramanian:2020hfs,Alishahiha:2020qza,Geng:2020qvw}. Islands in the context of de-Sitter and cosmological spacetimes have been studied in \cite{Hartman:2020khs,Chen:2020tes,Krishnan:2020fer,VanRaamsdonk:2020tlr,Balasubramanian:2020xqf,Sybesma:2020fxg,Geng:2021wcq} and also in \cite{Geng:2021iyq} in the context of AdS/BCFT duality. The island rule \eqref{Eq:IslandRule} can be derived generally without invoking the doubly holographic setup as will be reviewed in section \ref{sec:ReplicaWomhole}.

\section{Holography and Quantum Error Correction}
\label{sec:QECandholo}


\subsection{Preliminaries: Quantum Error Correction}\label{sec:qec}

Quantum error correction (QEC) is a mathematical framework that allows for partial or complete recovery of quantum information that is corrupted or lost by \emph{noise} arising due to unwanted interactions of the quantum system with the environment~\cite{gottesman_qec}. Formally, such noise is modelled as a completely positive trace-preserving (CPTP) map $\cN :  \cB(\cH) \rightarrow \cB(\cK)$ from the set of bounded linear operators on Hilbert space $\cH$ to the set of bounded linear operators on  another Hilbert space $\cK$~\cite{nielsen}. Such a CPTP map on the system density operators can be described in terms of a set of \emph{Kraus} operators $\{E_{i}\}$, as, $\cN(\rho) = \sum_{i} E_{i}\rho E_{i}^{\dagger}$. The operators $E_{i}$ are often said to be the \emph{error} operators associated with the noise map $\cN$. 

Since the no-cloning theorem prevents perfect copying of an arbitrary quantum state~\cite{wootters1982}, QEC aims to protect against the effects of noise by \emph{encoding} the information into entangled states of a larger Hilbert space. Specifically, for an $d$-dimensional quantum system with associated Hilbert space $\cH$, an $[[n,k]]$ quantum code protects $k$ qudits by encoding them into a $d^{k}$-dimensional subspace $\mathcal{C}$ of the $n$-qudit space $\mathcal{H}^{\otimes n}$. Throughout this discussion, we will assume that the noise acts identically and independently on each of the $n$ qudits that constitute the encoded space.  

A QEC code $\cC$ is said to correct {\bf perfectly} for the noise $\cN$, iff there exists a CPTP map $\cR: \cB(\cH^{\otimes n}) \rightarrow \cB(\cC)$ -- often called the recovery map -- such that, 
\begin{equation} 
(\cR\circ\cN)(\rho) = \rho, \forall \rho \quad {\rm such} \; {\rm that} \quad \cP \rho \cP = \rho, \label{eq:perfectQEC}
\end{equation}
where $\cP$ is the projection map onto the codespace $\cC$. Note that the noise acting on the encoded state is now a map on the $n$-qudit space, leading to single-qudit as well as multi-qudit errors. A given quantum code can only correct for some subset of these errors on the $n$-qudit space, indicated by a third parameter called the \emph{distance} $t$ of the code. Thus, an $[[n,k,t]]$ quantum code can correct perfectly for the loss of any set of $t \leq n$ qudits, or, equivalently it can correct for arbitrary errors on upto $\frac{t-1}{2}$ qudits. Algebraic and information-theoretic conditions for perfect QEC are known~\cite{knill_laflamme97}. The algebra of the Pauli operators has lead to the rich framework of stabilizer codes and topological QEC. We refer to the comprehensive review by Terhal et al.~\cite{qec_review2015} for further details and references. 

On the other hand, a quantum code $\cC$ is said to correct \emph{approximately} for the noise map $\cN$, iff there exists a (CPTP) recovery map $\cR$ such that,
\begin{equation}
    (\cR\circ\cN)(\rho) \approx \rho, \forall \rho \quad {\rm such} \; {\rm that} \quad \cP \rho \cP = \rho, \label{eq:aqec}
\end{equation} 
where $\cP$ is the projection map onto the codespace $\cC$. The performance of a QEC protocol $(\cC, \cR)$ described by the pair of codespace $\cC$ and recovery $\cR$, is quantified by the \emph{fidelity} function $F$ which is a measure of how close two quantum states are. The fidelity between a pair of states $\rho, \sigma \in \cB(\cH)$ is defined as 
\begin{equation}
    F (\rho, \sigma) = {\rm Tr}\sqrt{\rho^{1/2}\sigma\rho^{1/2}} . \label{eq:fidelity}
\end{equation} 


Moving beyond states, the framework of QEC can be easily extended to operator error correction~\cite{kribs_oqec_2005,operatorqec_nielsen2007}. Indeed, the operator QEC (OQEC) framework is the most relevant one in the context of bulk reconstruction of observables in holography, and it will be useful to describe it in some detail here. OQEC generalises the susbspace structure of standard QEC to a subsystem structure as follows. Suppose the system Hilbert space  $\cH$ has a decomposition of the form $\cH \equiv \cH_{A} \otimes \cH_{B} \oplus \cH_{C}$, for some choice of Hilbert spaces $\cH_{A}$, $\cH_{B}$ and $\cH_{C}$. Then, $\cH_{A}$ is said to be an error-correcting subsystem for a noise map $\cN$ acting on $\cH$, if, for all $\rho$ with support on $\cH_{A}$ and $\sigma$ with support on $\cH_{B}$, there exists a recovery map $\cR$ such that $(\cR\circ\cN)(\rho\otimes\sigma) = \rho\otimes \sigma'$, for some $\sigma' \in \cB(\cH)$ . Physically, this implies that information stored in subsystem $\cH_{A}$ can be recovered from the action of noise $\cN$ by the recovery map $\cR$. Note that we recover the structure of standard QEC codes when the system $\cH_{B}$ is trivial (one-dimensional).

The subsystem QEC structure can then be used to recover for an algebra of observables, via the framework of \emph{operator algebra quantum error correction}~\cite{beny_oqec2007}.  Here, the focus is on identifying subspaces or more generally subsystems of the system Hilbert space $\cH$, such that we can reliably recover observables $X$ that have support on the chosen subspace or subsystem. More generally, we may consider the $C^{*}$-algebra $\mathscr{A}$ of observables on $\cH$\footnote{Note that the operator algebra of observables is closed under addition, multiplication and Hermitian conjugation, thus forming a $C^{*}$-algebra.}. We will first formally state the necessary and sufficient conditions for an algebra of operators to be correctable under a noise map $\cN$.

\begin{theorem}[Operator Algebra QEC Condition]\label{thm:OAQEC}
The algebra of observables $\mathscr{A}$ on a codespace $\cC$ with projector $P$ is correctable against noise $\cN$ with operators $\{E_{i}\}$ if and only if $[E_{i}P, X] = 0$ for all errors $E_{i}$ associated with the noise $\cN$ and all observables $X \in \mathscr{A}$. 
\end{theorem}


\subsection{Holography as QEC}\label{sec:holoqec}

We now proceed to restate the concept of bulk reconstruction in AdS/CFT in the language of quantum error correction. This connection has already been touched upon in Sec.~\ref{sec:bulkreconstruction}, especially in the context of relating operators in the bulk spacetime to operators of the boundary CFT (Sec.~\ref{sec:BR_qec}). Early works in this direction demonstrated -- using certain toy models and specific quantum codes -- that the local operators in the bulk can be interpreted as \emph{encoded} operators on certain subspaces of the states of the CFT at the boundary~\cite{Almheiri_2015,happy_2015}.  The theory of quantum error correction is then invoked to show that these encoded operators are naturally protected against erasures on the boundary by virtue of their entanglement structure. 

\subsubsection{Operator error correction and Bulk Reconstruction}
\label{subsec:OEC_BR}
The first concrete application of QEC was to explain the somewhat counter-intuitive property that emerges in the context of AdS-Rindler bulk reconstruction, namely that the same bulk operator $\phi({\bf x})$ can be reconstructed on the union of different pairs of boundary subregions as shown in Fig.~\ref{Fig:redundancy}. The framework of QEC allows for this to happen in a non-trial manner, via a proposal that the same bulk operator is encoded as different operators on the boundary subregions via a suitable erasure QEC code~\cite{Almheiri_2015}. Note that the erasure noise map is a specific example of a quantum  noise map, wherein the information is either left unaffected or erased with certain probability. The idea of bulk reconstruction using a quantum erasure code can be made concrete via two simple toy examples. The first example is provided the $3$-qutrit code~\cite{qutrit}, defined by the span of the following three qutrit ($3$-dimensional) quantum states.
\begin{eqnarray}
|\tilde{0}\rangle &=& \frac{1}{\sqrt{3}}(|000\rangle + |111\rangle + |222\rangle) \nonumber \\
|\tilde{1}\rangle &=& \frac{1}{\sqrt{3}} (|012\rangle + |120\rangle + |201\rangle) \nonumber \\
|\tilde{2}\rangle &=& \frac{1}{\sqrt{3}}(|021\rangle + |102\rangle + |210\rangle). \label{eq:qutrit}
\end{eqnarray}
The erasure-correcting property of this code ensures that any three-qutrit state within the codespace can be reconstructed even if one of the three qutrits is lost or erased. Given an operator $O$ that acts on the single-qutrit space, one can always find a $3$-qutrit \emph{encoded} operator $\tilde{O}$ which acts in the same way on the $3$-qutrit code subspace. The unique feature of this specific code is that there exist encoded operators (also referred to as \emph{logical} operators in the literature) that have support only on two of the three qutrits that constitute the encoded space. Furthermore, it is then possible to identify sets of encoded operators that have the same action on the codespace, but have nontrivial support on \emph{different} pairs of qutrits. This indeed captures the essence of the bulk reconstruction via the entanglement wedges shown in Fig.~\ref{Fig:redundancy} and demonstrates that such  redundant encoding of operators is indeed possible.

More generally, we may consider the larger encoded space $\cH$ to be partitioned into subsystems $\cH_{E}$ and $\cH_{\bar{E}}$, where $\cH_{E}$ denotes the subsystem whose states get erased due to the noise and $\cH_{\bar{E}}$ denotes the subsystem whose states are unaffected by the erasure noise. Codespaces are then subspaces of the form  $\cC \subset \cH_{E}\otimes\cH_{\bar{E}}$, such that any state $|\tilde{\psi}\rangle \in \cC$ can be recovered after erasure of the $\cH_{E}$ subsytem. In the context of operator error correction, this translates to the statement that corresponding to an operator $O$ that acts on the codespace, there exists an operator $O_{\bar{E}}$ acting only on subsytem $\cH_{\bar{E}}$ such that
\begin{equation}
    O_{\bar{E}}|\tilde{\psi}\rangle = O|\tilde{\psi}\rangle, \; \forall |\tilde{\psi}\rangle \in \cC. \label{eq:oqec}
\end{equation}
It is now easy to identify the set of operators $O$ which can be corrected by a given codespace $\cC$, using the operator algebra QEC condition in Thm.~\ref{thm:OAQEC}. Formally, the operator error correction property in Eq.~\ref{eq:oqec} holds if and only if $O$ commutes with any operator $X_{E}$ which acts only on $\cH_{E}$. In other words, Eq.~\ref{eq:oqec} holds for a given operator $O$ and codespace $\cC$, if and only if,
\begin{equation}
\langle \tilde{\psi}| [O, X_{E}] |\tilde{\psi}\rangle = 0, \; \forall |\tilde{\psi}\rangle \in \cC. \label{eq:commutant}
\end{equation} 
It is useful to note that the set of operators that satisfy Eq.~\ref{eq:commutant} form a $*$-subalgebra of the operators on the codespace $\cC$.

\begin{figure}
   \centering
        \resizebox{0.5\textwidth}{!}{%
        \includegraphics{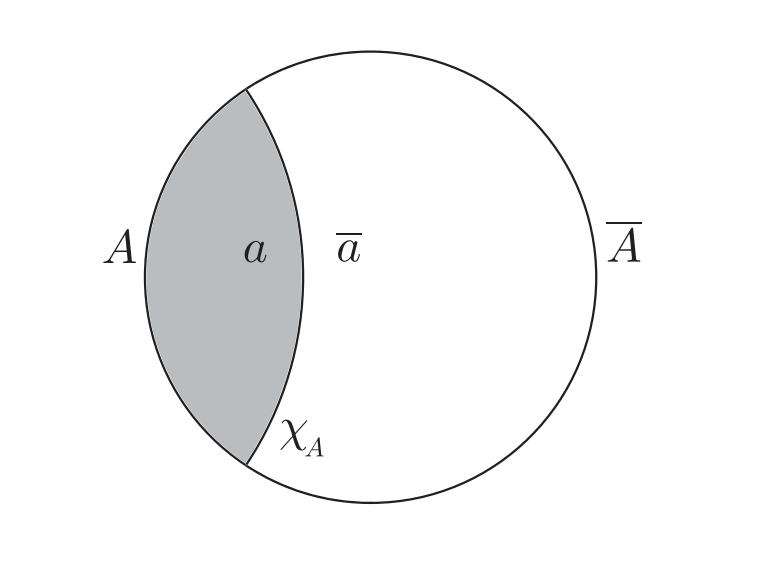}
        }
        \caption{Bulk-boundary factorization on a time slice and the entanglement wedge. The boundary CFT is factorized as $\cH_{A}\otimes\cH_{\bar{A}}$. The shaded region denotes the entanglement wedge $\cE_{A}$ corresponding to the boundary region $A$. $\cH_{a}$ is the Hilbert space of the bulk excitations in $\cE_{A}$. $\chi_{A}$ denotes the RT extremal surface. Figure taken from~\cite{dong_2016_2}.}
        \label{Fig:entWedge}       
    \end{figure}

\subsubsection{Proof of entanglement wedge reconstruction via operator QEC}\label{sec:recons-proof}

We proceed to prove entanglement wedge reconstruction using the framework of operator algebra error correction described above. The crucial input would be the JLMS result of the equivalence of bulk and boundary relative entropies~\cite{Jafferis_2016} discussed in section \ref{sec:modH_relEnt}. 

Formally, following the discussion in~\cite{dong_2016_2}, we may consider a factorization of the boundary CFT $\cH$ into $\cH_{A}\otimes\cH_{\bar{A}}$ as shown in Fig.~\ref{Fig:entWedge}. Let $\cE_{A}$ and $\cE_{\bar{A}}$ denote the associated entanglement wedges and let $\cH_{a}$ and $\cH_{\bar{a}}$ denote the Hilbert space of bulk excitations in $\cE_{A}$ and $\cE_{\bar{A}}$ respectively. The codespace is a suitably chosen subspace of the CFT, with a natural factorization of the form $\cC = \cH_{a}\otimes\cH_{\bar{a}}$. The JLMS proposal can then be stated as follows, for a pair of density operators $\rho_{A}, \sigma_{A} \in \cB(\cH_{A})$ and a pair of density operators $\rho_{a}, \sigma_{a}$ acting on the space $\cH_{a}$ corresponding to the bulk subregion $a$. 
\begin{equation}
    S(\rho_{A}\vert \sigma_{A}) = S(\rho_{a}\vert\sigma_{a}).  \label{eq:jlms}
\end{equation}
Since the relative entropy between a pair of operators $S(\rho\vert\sigma)$ vanishes if and only if $\rho = \sigma$, the above equality is identical to the statement that  $\rho_{A} = \sigma_{A}$ would imply $\rho_{a} = \sigma_{a}$ and vice-versa. 

The relative entropy equivalence in Eq.~\eqref{eq:jlms} in conjunction with the operator algebra QEC condition in Eq.~\eqref{eq:commutant} leads to the following reconstruction theorem~\cite{Dong_2016}.
\begin{theorem}[Bulk reconstruction]\label{thm:bulk_reconst}
Consider any code subspace $\cC \subset \cH$ of a finite-dimensional Hilbert space\footnote{The assumption that the Hilbert space $\cH$ is finite can be accomplished by imposing a UV cutoff in the CFT.} $\cH$ with the factorization $\cH = \cH_{A}\otimes\cH_{\bar{A}}$ and an operator $O$ that acts on $\cC$. Suppose there exists a factorization of the codespace into $\cC = \cH_{a}\otimes \cH_{\bar{a}}$ such that, 
\begin{itemize}
    \item[(i)] the operator $O$ acts only on $\cH_{a}$, and,
    \item[(ii)] for any pair of pure states $|\Psi\rangle, |\Phi\rangle \in \cC$, the reduced density operators $\rho_{\bar{A}} = \tr_{A}[|\Psi\rangle\langle\Psi|]$, $\sigma_{\bar{A}} = \tr_{A}[|\Phi\rangle\langle \Phi|]$ $\rho_{\bar{a}} = \tr_{a}[|\Psi\rangle\langle\Psi|]$ and $\sigma_{\bar{a}} = \tr_{a}[|\Phi\rangle\langle \Phi|]$ satisfy 
    \[ \rho_{\bar{a}} = \sigma_{\bar{a}} \Rightarrow \rho_{\bar{A}} = \sigma_{\bar{A}}  . \]
\end{itemize}   
Then, there exists an operator $O_{A}$ acting only on $\cH_{A}$ such that its action on the codespace is the same as that of the operator $O$. In other words,
\begin{equation}
    O_{A}|\Psi\rangle = O|\Psi\rangle, \, \forall \, |\Psi\rangle \in \cC. \label{eq:bulk_reconst}
\end{equation}
\end{theorem} 
We will now outline the proof strategy of Dong {\emph et al}~\cite{Dong_2016} here, for completeness. We first note that the operator algebra QEC condition in Thm.~\ref{thm:OAQEC} implies that the bulk reconstruction in Eq.~\eqref{eq:bulk_reconst} follows if we can establish that the action of the operator $O$ commutes with the action of any $X_{\bar{A}}$ (with support only on subsystem $\cH_{\bar{A}}$) on the codespace (see Eq.~\eqref{eq:commutant}). This can be shown easily for any Hermitian operator $O$, and then extended to general operators by linearity. For any real number $\lambda$ consider two states $|\Psi\rangle, |\Phi\rangle \in \cC$, such that,
\[|\Psi\rangle =  e^{i\lambda O} |\Phi\rangle . \]
If assumption (i) of the theorem holds, the operator $O$ acts only on the subsystem $\cH_{a}$. If we further assume that $O$ is Hermitian, the two states $|\Psi\rangle$ and $|\Phi\rangle$ are related by a unitary operator, so that,
\begin{eqnarray}
\rho_{\bar{a}} &=& \tr_{a}[|\Psi\rangle\langle\Psi |] \nonumber \\
&=& \tr_{a}[(e^{i\lambda O}\otimes I_{\bar{a}}) |\Phi\rangle\langle\Phi| (e^{-i\lambda O} \otimes I_{\bar{a}})] \nonumber \\
&=& \sigma_{\bar{a}}.
\end{eqnarray}
Now if we use assumption (ii) --  which is equivalent to the relative entropy condition in Eq.~\eqref{eq:jlms} -- we have, $\rho_{\bar{A}} = \sigma_{\bar{A}}$. This in turn implies that the expectation value of any operator $X_{\bar{A}}$ acting only on subsystem $\cH_{\bar{A}}$ is the same for the two states $|\Psi\rangle$ and $|\Phi\rangle$. Thus,
\begin{eqnarray}
\langle \Psi|X_{\bar{A}}|\Psi\rangle &=& \langle \Phi|X_{\bar{A}}|\Phi\rangle \nonumber \\
\Rightarrow \langle \Phi| e^{-i\lambda O} X_{\bar{A}} e^{i\lambda O} | \Phi\rangle &=& \langle \Phi|X_{\bar{A}}|\Phi\rangle. \label{eq:exp_value}
\end{eqnarray}
Expanding Eq.~\eqref{eq:exp_value} to order $\lambda$, we get the desired commutativity condition, namely,
\[ \langle \Phi|[O, X_{\bar{A}}] | \Phi\rangle = 0, \; \forall \; |\Phi\rangle \in \cC. \]

Theorem~\ref{thm:bulk_reconst} thus proves that bulk reconstruction is possible via the entanglement wedge prescription, provided there exists a codespace which can be factorized into the bulk subregion $a$ and its complement $\bar{a}$ in such a way that the corresponding density operators satisfy the relative entropy equality in Eq.~\eqref{eq:jlms}. We refer to~\cite{pastawski_preskill2017} for a more detailed exploration of the various connections between holography and operator algebra quantum error correction. Non-perturbative gravity corrections imply only an approximate recovery of the entanglement wedge is possible (with exponentially small errors and even in absence of horizons) and the generalization of the present discussion can be found in \cite{Gesteau:2021jzp}\footnote{Approximate state dependent recovery is a necessary starting point in the presence of horizons as discussed later.}.

One final aspect of holographic QEC that we would like to highlight here is the important link between the erasure correcting properties of a holographic code and the strong subadditivity inequality. Consider a tripartite quantum state $\rho_{ABC}\in \cB(\mathcal{H}_A \otimes \mathcal{H}_B\otimes\mathcal{H}_{C} )$. The strong subadditivity inequality (see \eqref{eq:SSA-def}) states that,
\begin{equation}
    S(\rho_{ABC}) + S(\rho_{B}) \leq S(\rho_{AB}) + S(\rho_{BC}).  \label{eq:ssa}
\end{equation}
This inequality can be simply understood as the positivity of conditional mutual information $I(A;C|B)$ between subsystems $A$ and $C$, given $B$. It turns out that the quantum states that saturate this inequality have an interesting structure, as shown in~\cite{hayden_SSA2004}. Equality of Eq.~\eqref{eq:ssa} implies that the Hilbert space $\cH_{B}$ can be decomposed as a direct sum of tensor products of the form $\cH_{B} = \sum_{i} \cH_{B^{1}_{i}}\otimes \cH_{B^{2}_i}$, and that the tripartite state then has the block diagonal form,
\begin{equation}
    \rho_{ABC} = \sum_{i} p_{i} \rho_{AB^{1}_{i}} \otimes \rho_{B^{2}_{i}C}. \label{eq:QMarkov}
\end{equation}
This structure essentially implies conditional independence of subsystems $A$ and $C$, given subsytem $B$. In other words the state $\rho_{ABC}$ can be thought of as a \emph{quantum Markov chain}, the quantum analogue of a classical Markov Chain $A\rightarrow B \rightarrow C$. Furthermore, it was shown that if subsystem $C$ is erased (or traced out) the tripartite state $\rho_{ABC}$ can be reconstructed from the marginal state $\rho_{AB}$ via the action of a recovery map $\cR:\cB(\cH_{B}) \rightarrow \cB(\cH_{B}\otimes \cH_{C})$ called the Petz map, described below in Sec.~\ref{sec:petz_aqec}. Note that this recovery map is completely independent of subsystem $A$.

\begin{figure}
   \centering
        \resizebox{0.5\textwidth}{!}{%
        \includegraphics{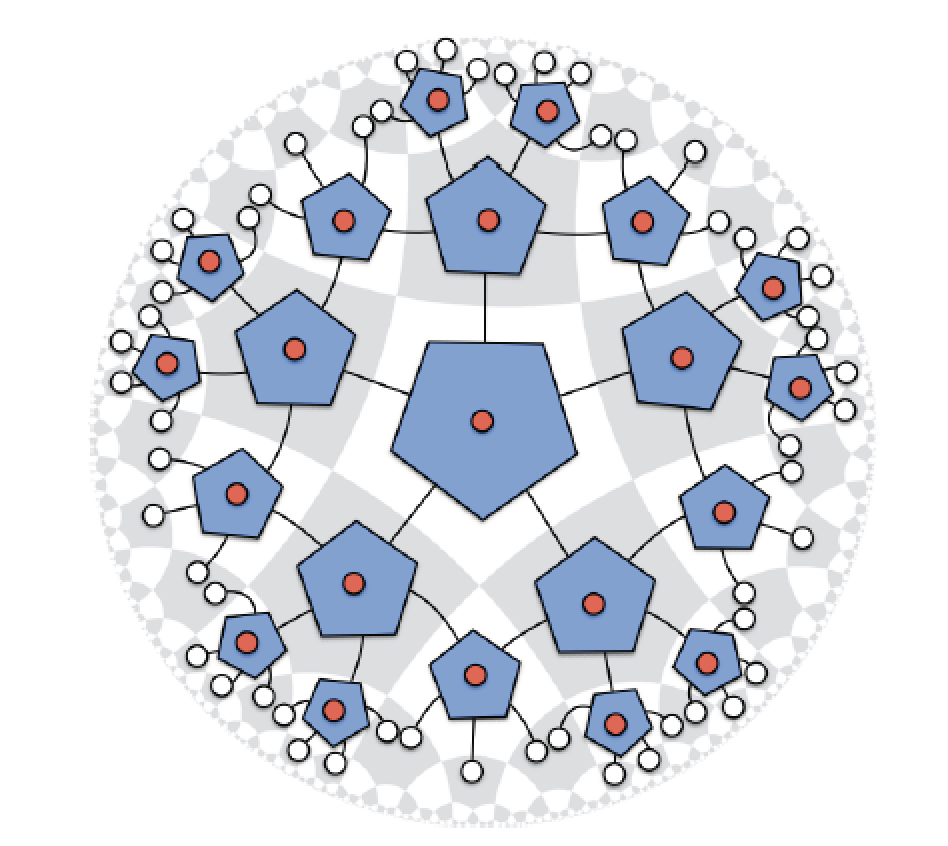}
        }
        \caption{The $5$-qubit code visualised as a tensor network~\cite{happy_2015}. The white dots represent uncontracted indices on the boundary and the red dots represent the contracted tensor legs in the bulk.}
        \label{Fig:5qubit}       
    \end{figure}

In the holographic setting, this Markov chain structure can be applied to get a stronger statement of bulk recovery, as follows. Let the boundary region $A$ in Fig.~\ref{Fig:entWedge} be made up of three disjoint regions $A_{1}\cup A_{2}\cup A_{3}$. Let $A' = A_{1}\cup A_{3}$ denote the union of two such unconnected, disjoint regions. The full boundary can be viewed as the union of the three regions $\bar{A}\cup A'\cup A_{2}$. The question is, does there exists a recovery map that can correct for erasure of the region $A_{2}$, without invoking the complementary region $\bar{A}$? The quantum Markov condition states that this is indeed possible if the boundary state is such that $A\rightarrow A' \rightarrow A_{3}$ form a quantum Markov chain, which will saturate the inequality in Eq.~\eqref{eq:ssa}. Then, there exists a recovery map $\cR: \cB(\cH_{A'}) \rightarrow \cB(\cH_{A})$ that recovers for the erasure of $\cH_{A_{2}}$, without involving the subsystem $\cH_{\bar{A}}$. In such a situation, the erasure is said to \emph{locally correctable}~\cite{pastawski_preskill2017}. This observation has potential implications for decoding of the black hole interior, as discussed in Sec.~\ref{sec:microstate}.

For a discussion on how such tensor network models can reproduce correlation functions and entropy of a three-dimensional black hole geometry see \cite{Bhattacharyya:2016hbx}.

\subsubsection{Tensor network toy model for bulk reconstruction}\label{sec:happy}
While the operator algebra QEC framework provided an abstract proof of existence of bulk reconstruction in the AdS/CFT correspondence, in this section we review the concrete tensor network based toy model of holography from~\cite{happy_2015}.  A tensor network can be visualised as a graph with a set of vertices $\{V_{x}, \; x =1,2, \ldots, N\}$, with a quantum state $|V_{x}\rangle \in \cH_{x}$ associated with each vertex. An isometric tensor is any linear map $T: \cH_{x} \rightarrow \cH_{y}$ such that $T^{\dagger} T = I_{x}$, the identity operator on $\cH_{x}$. Specifically, $T: |x_{i}\rangle \rightarrow \sum_{j}T_{ij} |y_{j}\rangle$, where $\{|x_{i}\rangle\}$ and $\{|y_{j}\rangle\}$ denote compete orthonormal bases for $\cH_{x}$ and $\cH_{y}$ respectively. The local Hilbert space $\cH_{x}$ at each vertex could admit a tensor product decomposition of the form $\cH_{x} = \otimes_{k=1}^{n_{x}} \cH_{k}$. The number of tensor indices depends on the factorization structure of the input and output spaces. For instance, the action of the isometric map $T : \cH_{1}\otimes\cH_{2} \rightarrow \cH_{y}$ on the basis states can be represented as 
\begin{equation}
    |i_{1} i_{2}\rangle \rightarrow \sum_{j} T_{j i_{2}i_{1}}|y_{j}\rangle. \label{eq:tensor1}
\end{equation} 
An interesting property of such isometric tensors is that it is possible to reinterpret an input factor as an output factor, upto a rescaling. Thus, the tensor map in Eq.~\eqref{eq:tensor1} can be recast as as a map $\tilde{T}: \cH_{A_{1}} \rightarrow \cH_{B}\otimes\cH_{A_{2}}$ as,
\[ |i_{1}\rangle \rightarrow \sum_{j, i_{2}} T_{j i_{2} i_{1}}|y_{j} i_{2}\rangle.  \]
In general, a tensor $T$ with $n$ indices, ranging over $d$ values represents a quantum state in an $n$-fold tensor product space of $d$-dimensional quantum systems.
\[|\psi\rangle = \sum_{i_{1}, i_{2}, \ldots, i_{n}} T_{i_{1}i_{2}\ldots i_{n}}|i_{1}i_{2}\ldots i_{n}\rangle. \]

A special class of tensors called \emph{perfect tensors} lead to \emph{encoding} isometries for quantum error correcting codes in the following sense. A perfect tensor with $2n$ indices describes a pure state of $2n$ quantum systems with the property that any subset of $n$ systems is maximally entangled with the complementary set of $n$ systems\footnote{Such states are called Absolutely Maximally Entangled (AME) states.}. Such a tensor can then be thought of as a linear map from $1$ ($d$-dimensional) system to $2n-1$ ($d$-dimensional) systems, encoding a single quantum system is such a way that it is protected against the erasure of any $n-1$ subsystems. In QEC terminology, a  perfect tensor with $2n$ indices corresponds to a $[[2n-1, 1, n]]$ code, as exemplified by the well known $[[5,1,3]]$ stabilizer code~\cite{5qubit_laflamme}.

In the context of holography, a tensor network can be interpreted as a map from the bulk to the boundary in the following sense. If we imagine the quantum states associated with the vertices of a graph to correspond to perfect tensors, the edges of the graph are associated with contractions of the tensor ``legs''. Contracted tensor indices can then be associated with bulk degrees of freedom and uncontracted vertices are associated with the boundary degrees of freedom, as depicted in Fig.~\ref{Fig:5qubit}.

The holographic pentagon code shown in Fig.~\ref{Fig:5qubit} is the simplest example of a tensor network based holographic code, and provides a nice demonstration of exact bulk reconstruction. The code geometry comprises of a uniform tiling of a hyperbolic disc by pentagons, with four pentagons adjacent at each vertex. A perfect tensor with six legs is placed at the center of each pentagon, so that each tensor has one uncontracted index, indicated by the red dot in Fig.~\ref{Fig:5qubit}. All other interior legs are contracted. The uncontracted leg in the interior can be interpreted as an encoded input in the bulk to the tensor isometry and the uncontracted legs at the boundary (the white dots at the boundary in Fig.~\ref{Fig:5qubit}) interpreted as the \emph{physical} outputs at the boundary. The entire system can thus be viewed as a tensor network that maps the input legs in the bulk to the output legs at the boundary. 

We refer to~\cite{happy_2015} for a detailed discussion of the error correction properties of the pentagon code.  Suffice it to note that the erasure-correcting properties of the underlying $5$-qubit code ensure that this toy model can achieve bulk reconstruction by accessing only a subregion of the boundary. A full description of the operator reconstruction involves some interesting techniques such as \emph{tensor pushing} and leads to a so-called greedy entanglement wedge reconstruction. Finally, we note that the $3$-qutrit code described in Eq.~\eqref{eq:qutrit} is also a perfect tensor and can be viewed as a triangular holographic code. 

Moving beyond holographic codes based on perfect tensors, toy models of holography have been proposed using random tensor networks~\cite{hayden2016_rtn}. Unlike perfect tensor codes which are based on fixed isometric tensors, these models involve projecting onto maximally entangled states via random projection operations. Such random tensor networks are known to successfully demonstrate several holographic properties~\cite{harlow2017_RT} and have been recently used to demonstrate approximate bulk reconstruction~\cite{jia2020_petz_RTN} via the universal recovery map described in Sec.~\ref{sec:bulk_petz} below. Infinite dimensional HaPPY codes have been discussed in \cite{Gesteau:2020hoz}. Here, it has been shown that the infinite dimensional code fails to reproduce long range correlations at the boundary, which are necessary ingredients of a dual CFT; thus indicating limitations of this model for AdS/CFT.

\subsection{Bulk-boundary reconstruction via quantum recovery maps}\label{sec:recovery}
The original bulk reconstruction proposal presented in Sec.~\ref{sec:BR_qec} relied on an exact equivalence of the bulk and boundary relative entropies, as stated in Eq.~\eqref{eq:entropy_eqv} (equivalently~\eqref{eq:jlms}). While such an exact equivalence of the bulk and boundary entropies maybe argued for in an asymptotic setting, in a finite regime, within the framework of bulk effective field theories, it is expected that these two entropies may only be \emph{approximately} equal. (This will also be a crucial issue in the context of establishing black hole interior reconstruction as an universal subsystem recovery map. It will be discussed in section \ref{sec:Python}.) The question then arises as to whether the QEC-based bulk reconstruction argument can be extended to this case of approximate equivalence of the bulk and boundary relative entropies. It turns out that the right framework to consider in this case if that of \emph{approximate quantum error correction} (AQEC), rather than the \emph{perfect} QEC situation considered thus far. In this section, we review some recent works~\cite{cotler2019_univR,chen2020_petz,jia2020_petz_RTN} that formulate the bulk reconstruction argument using ideas and techniques from approximate QEC. 

\subsubsection{Approximate QEC and the Petz map}\label{sec:petz_aqec}

As defined in Eq.~\eqref{eq:aqec} above, approximate QEC extends the framework of QEC to allow for situations where the state is not perfectly recovered, but recovered with high enough fidelity after the action of the noise map. In the case of perfect QEC, code constructions rely on decomposing the noise operators in terms of Pauli operators and then using the structure of the Pauli algebra to identify good code subspaces. The recovery operation is a two-step process comprising error detection, followed by application of the appropriate Pauli operators to correct for the errors~\cite{nielsen}. The problem of finding good approximate QEC codes is in general much harder, since it requires a search over both code spaces $\cC$ and recovery maps $\cR$, such that $(\cR\circ\cN)(\rho)$ is close in fidelity to the density operator $\rho\in \cB(\cC)$. 

An important tool that emerged in this context is the idea of a near-optimal universal recovery map, namely, the \emph{Petz map}~\cite{ohya_petz}, that can reverse the effect of the noise to a high degree of fidelity. Given a noise map $\cN$ with associated error operators $\{E_{i}\}$ and a code space $\cC$, the Petz map $\cR_{\rho, \cN}$ corresponding to any density operator $\rho$ with support on the codespace is defined in terms of its Kraus operators $\cR_{\rho,\cN} =\{(R_{\rho,\cN})_{i}\}$, as~\cite{barnum_Knill2002},
\begin{equation}
    (R_{\rho, \cN})_{i} = \rho^{1/2}E_{i}^{\dagger}(\cN(\rho))^{-1/2}, \label{eq:petz1}
\end{equation}
where the inverse is taken on the support of the positive operator $\cN(\rho) = \sum_{i}E_{i}\rho E_{i}^{\dagger}$. In other words, the action of the Petz map $\cR_{\rho,\cN}$ on an arbitrary density operator $\sigma \in \cB(\cC)$ can be written as,
\begin{eqnarray}
    \cR_{\rho, \cN}(\sigma) &=& \rho^{1/2}\left(\sum_{i}E_{i}^{\dagger}\cN(\rho)^{-1/2}\sigma\cN(\rho)^{-1/2}E_{i} \right) \rho^{1/2} \nonumber \\
    &=& \rho^{1/2} \cN^{\dagger}\left( \cN(\rho)^{-1/2}\sigma\cN(\rho)^{-1/2} \right) \rho^{1/2}.  \label{eq:petz2}
\end{eqnarray}
Here, $\cN^{\dagger}$ denotes the dual to the map $\cN$, with Kraus operators $\{E_{i}^{\dagger}\}$. Furthermore, it is easy to see that 
\[ \cR_{\rho, \cN}\circ\cN(\rho) = \rho.\] 
Thus, $\cR_{\rho, \cN}$ is the map that reverses \emph{perfectly}, the effect of the noise map $\cN$ on the state $\rho$.

In the context of approximate QEC, the Petz map defined in Eq.~\eqref{eq:petz1} was shown to be a universal, near-optimal recovery map, where optimality was characterized using the average entanglement fidelity~\cite{barnum_Knill2002}. Subsequently, a variant of the Petz map -- defined over a codespace $\cC$, rather than a specific state $\rho$ -- has been shown to be a universal, near-optimal recovery map in terms of the worst-case fidelity~\cite{approxQEC}. In what follows we will survey some of the recent works~\cite{cotler2019_univR,chen2020_petz,jia2020_petz_RTN} that use the Petz map construction to demonstrate a robust, universal recovery map for bulk reconstruction.

\subsubsection{Bulk reconstruction using the Petz map}\label{sec:bulk_petz}

The Petz map was originally conceived in the context of understanding the monotonicity of quantum relative entropy~\cite{petz_monotonicity2003,hayden_SSA2004}. Under the action of a noise map $\cN$, the relative entropy between two states $\rho, \sigma$ can never increase, that is,
\begin{equation}
S(\rho\vert \sigma) \geq S(\cN(\rho)\vert \cN(\sigma)).  \label{eq:monotonicity}    
\end{equation}
This is often referred to as Uhlmann's theorem~\cite{uhlmann1977relative} in quantum information theory. Since the relative entropy $S(\rho\vert\sigma)$ vanishes if and only if $\rho = \sigma$, it can be thought of as a measure of distance between quantum states. The difference between $S(\rho\vert\sigma)$ and $S(\cN(\rho)\vert\cN(\sigma))$ can thus be used to quantify the extent to which the noise map $\cN$ corrupts the quantum system. It was subsequently realised that the monotonicity inequality also captures the extent of \emph{recoverability} of the states $\rho$ and $\sigma$ under noise $\cN$. Suppose there exists a recovery map $\cR$ that recovers the states $\rho$ and $\sigma$ \emph{perfectly} from the effects of the noise $\cN$, namely, $(\cR\circ\cN)(\rho) = \rho$ and $(\cR\circ\cN)(\sigma)=\sigma$, then, the inequality in Eq.~\eqref{eq:monotonicity} is saturated. Interestingly, the converse is also true, and the specific form of the recovery map that saturates monotonicity is indeed given by the form of Petz map defined with respect to $\sigma$ and $\cN$~\cite{petz_monotonicity2003}:
\[\cR_{\sigma, \cN}(.) = \sigma^{1/2} \cN^{\dagger}\left( \cN(\sigma)^{-1/2} (.) \cN(\sigma)^{-1/2} \right) \sigma^{1/2}.\]
Note that this form is identical to the one in Eq.~\eqref{eq:petz2}, except that this is the Petz map that recovers the state $\sigma$ perfectly under the action of the noise $\cN$.

The fact that saturation of Eq.~\eqref{eq:monotonicity} is a necessary and sufficient condition for exact recoverability provides us with a nice information theoretic interpretation of the JLMS proposal for bulk construction. In essence, the JLMS condition in Eq.~\eqref{eq:jlms} can be thought of as a saturation of the monotonicity of the relative entropy between operators on the corresponding bulk and boundary subregions under the action of the erasure noise map on the complementary boundary region. This naturally begs the question of what happens when the monotonicity inequality is only \emph{approximately} saturated. In the holographic context, this would imply that the bulk boundary relative relative entropies are only approximately equal, perhaps to leading order. Does there exist a universal recovery map in this case, which can achieve approximate bulk reconstruction despite having access to only a certain subregion of the boundary? Remarkably, it turns out that the answer to this question is in the affirmative, and there exists more than one construction of such a universal recovery map for approximate bulk reconstruction, based on the Petz map.

\subsubsection{The twirled Petz map}\label{sec:twirled} 
One proposal for a universal recovery map comes from a time-averaged form of the Petz map, called the \emph{twirled Petz map}. The twirled form of the Petz map is motivated by a recent result in quantum information theory, which relates the difference in quantum relative entropies before and after the action of a noise map to the fidelity between the ideal and noisy states. Formally, for any two states $\rho, \sigma \in \cB(\cH)$ on some Hilbert space $\cH$ and any noise map $\cN$, there exists a recovery map $\tilde{\cR}_{\sigma,\cN}$ such that~\cite{Junge_2018}, 
\begin{equation}
    S(\rho\vert\sigma) - S(\cN(\rho)\vert\cN(\sigma)) \geq -2\log F(\rho, (\tilde{\cR}_{\sigma,\cN}\circ\cN)(\rho)).\label{eq:approx_mon}
\end{equation}
Here, $F(.)$ is the fidelity function defined in Eq.~\eqref{eq:fidelity} above. For a map $\cN$ that saturates monotonicity of relative entropy, both LHS and RHS identically vanish, for in this case the recovery map $\tilde{\cR}_{\sigma,\cN}$ is simply the Petz map which satisfies  $\cR_{\sigma,\cN}(\rho) = \rho$. For a map that does not saturate the monotonicity inequality, the LHS of Eq.~\eqref{eq:approx_mon} quantifies the deviation from saturating monotonicity, whereas the RHS quantifies the extent to which the recovery map $\tilde{\cR}_{\sigma,\cN}$ recovers the state $\rho$ after the action of the noise. In essence, the above inequality states that the fidelity with which a recovery map can correct for the action of the noise $\cN$ is bounded by how close the map $\cN$ comes to saturating the monotonicity inequality. 

Furthermore, an explicit form of the universal recovery map $\tilde{\cR}_{\sigma,\cN}$ for the case where monotonicity is not exactly saturated was also given in~\cite{Junge_2018}, as,
\begin{eqnarray}
&& \tilde{\cR}_{\sigma,\cN}(.) =  \int dt \beta(t) \nonumber \\
&& \sigma^{\frac{1-it}{2}}\cN^{\dagger}\left( [\cN(\sigma)]^{\frac{-1+it}{2}}(.)[\cN(\sigma)]^{\frac{-1-it}{2}} \right) \sigma^{\frac{1+it}{2}}, \label{eq:twirled}
\end{eqnarray}
where $\beta(t) =  (\pi/2)(\cosh (\pi t)+1)^{-1}$. This twirled form of the Petz map then provides a natural choice for a universal recovery map in holography, where $t$ is interpreted as the boundary modular time. This was formalised in the work of~\cite{cotler2019_univR}, which we briefly review here.

The starting point for approximate bulk reconstruction will be the approximate form of the relative entropy equivalence from~\cite{Jafferis_2016}.
\begin{equation}
S(\rho_{A}\vert\sigma_{A}) = S(\rho_{a}\vert\sigma_{a}) + \mathcal{O}\left(\frac{1}{N}\right), \label{eq:JLMS_approx}
\end{equation}
where $\rho_{A}$, $\sigma_{A}$ are density operators on the boundary subregion $A$ in Fig.~\ref{Fig:entWedge} and 
$\rho_{a}$, $\sigma_{a}$ are density operators on the corresponding entanglement wedge region $a$ in the bulk. In other words, the JLMS equality conditions holds upto leading order in the CFT gauge group rank $N$. Interpreting Eq.~\eqref{eq:approx_mon} as approximate saturation of the monotonicity of the relative entropy requires the existence of a mapping $\rho_{a} \rightarrow \rho_{A}$ of states from the entanglement wedge in the bulk to the boundary subregion $A$. The AdS/CFT correspondence can be described via an isometry $\cV: \cC \rightarrow \cH$ from the code subspace to the boundary CFT $\cH$. The map $\cN$ is simply the partial trace operation, which traces out the complementary boundary region $\bar{A}$. Thus, for any $\rho_{a} \in \cB(\cH_{a})$, and any fixed, full-rank state $\sigma_{\bar{a}} \in \cB(\cH_{\bar{a}})$, we have,
\begin{equation}
    \cN(\rho_{a}) = \tr_{\bar{A}}[\cV(\rho_{a}\otimes\sigma_{\bar{a}}) \cV^{\dagger} ] .\label{eq:erasure}
\end{equation}
Note that we only consider product density operators of the form $\rho_{a}\otimes\sigma_{\bar{a}}$ on the codespace, where $\sigma_{\bar{a}}$ is a fixed state on $\cH_{\bar{a}}$, so that the partial trace operation truly becomes a mapping of states $\rho_{a}$ on the entanglement wedge region (shaded region $a$ in Fig.~\ref{Fig:entWedge}) to states $\tr_{\bar{A}}[\cV (\rho_{a} \otimes \sigma_{\bar{a}}) \cV^{\dagger} ]$ on the boundary subregion $A$. 

Now, consider the twirled Petz map corresponding to the map $\cN$ defined in Eq.~\eqref{eq:erasure}, $\tilde{\cR}_{\sigma_{a}, \cN}$, defined using a full-rank state $\sigma_a \in \cB(\cH_{a})$. This is a map of the form $\tilde{\cR}_{\sigma_{a}, \cN} : \cB(\cH_{A}) \rightarrow \cB(\cH_{a})$. Finally, if we assume that the map $\cN$ defined in Eq.~\eqref{eq:erasure} approximately saturates the monotonicity inequality, Eq.~\eqref{eq:approx_mon} implies that the twirled Petz map $\tilde{\cR}_{\sigma_{a}, \cN}$ recovers any $\rho_{a} \in \cB(\cC)$ with a fidelity that is bounded by,
\begin{eqnarray}
&& -2\log F(\rho_{a},(\tilde{\cR}_{\sigma_{a},\cN}\circ\cN)(\rho_{a})) \nonumber \\
&\leq & S(\rho_{a}\vert\sigma_{a}) - S(\cN(\rho_{a})\vert\cN(\sigma_{a})).
\end{eqnarray}  
This argument was then extended in~\cite{cotler2019_univR} to show that the map $\tilde{\cR}_{\sigma_{a},\cN}$ can recover for \emph{all} states, extending the scope of this result beyond states that are factorised as $\rho = \rho_{a}\otimes\sigma_{\bar{a}}$. Furthermore, since $\tilde{\cR}_{\sigma_{a},\cN}$ recovers states on $\cH_{a}$ with high fidelity, it can be shown that the adjoint  $\tilde{\cR}^{\dagger}_{\sigma_{a},\cN}$ maps operators $\phi_{a}$ with support on the entanglement wedge region $\cH_{a}$ to boundary operators $\cO_{A}$ which are close in expectation values. Finally, it was shown in~\cite{cotler2019_univR} that an explicit formula for operators $\cO_{A}$ can be obtained by a specific choice of the fixed states $\sigma_{a}$ and $\sigma_{\bar{a}}$, namely, the maximally mixed states on $\cH_{a}$ and $\cH_{\bar{a}}$.  

We conclude this section by noting a few more recent results that demonstrate bulk reconstruction using variants of the Petz map. Moving away from the twirled Petz map which involves an averaging over the modular time, it was argued in~\cite{chen2020_petz} that the standard form of the Petz map in Eq.~\eqref{eq:petz1} suffices to achieve approximate bulk reconstruction. Their argument is based on the original result of Barnum and Knill~\cite{barnum_Knill2002}, which shows that the Petz map is the near-optimal universal recovery map in terms of the average entanglement fidelity between the ideal and noisy states. We restate the main result of~\cite{chen2020_petz} here, for completeness. Let $\cM_{a}$ be a subalgebra on the code space $\cC$ with dimension $d_{\rm code}$, and $\cN$ be any noise map. Suppose there exists an \emph{optimal} recovery map $\cR_{\rm opt}$ such that
\[ \parallel (\cR_{\rm opt}\circ\cN)(\rho) - \rho_{a} \parallel_{1} < \delta ,  \]
for all $\rho \in \cB(\cC)$ and its projection $\rho_{a}$ onto $\cM_{a}$. Then, the Petz map \[ \cR_{\tau, \cN}(.) \equiv \frac{1}{d_{\rm code}} \cN^{\dagger} \left[\cN(\tau)^{-1/2}(.)\cN(\tau)^{-1/2} \right], \] 
defined using the maximally mixed state $\tau$ on the codespace, satisfies,
\[\parallel (\cR_{\tau, \cN}) \circ \cN)(\rho) \vert_{a} - \rho_{a} \parallel_{1} \leq d_{\rm code}\sqrt{8\delta} , \]
where $\parallel (.)\parallel_{1}$ denotes the $1$-norm or the trace-distance and $(.)\vert_{a}$ denotes the projection onto the subalgebra $\cM_{a}$. In other words, so long as the error using the optimal recovery $\cR_{\rm op}$ is non-perturbatively small (in trace-norm), the error after the Petz map recovery will also be non-perturbatively small upto a factor $d_{\rm code}$. 

More recently, it has been shown that a Petz-like map can be used to demonstrate bulk reconstruction in random tensor network toy models~\cite{jia2020_petz_RTN,Penington:2019kki} of holography. From a dynamical point of view, there appears to be an interesting connection between the structure of the twirled Petz map and the action of the modular Hamiltonian, as hinted in~\cite{cotler2019_univR} and  also in sections \ref{subsec:modularhamexcited} and \ref{sec:BR_qec}. Exploring this connection further might lead to further insights on the problem of bulk reconstruction, and promises to be an exciting direction for future investigations. 

\subsection{A note on holography as a renormalization group flow}\label{sec:herg}
 Holographic renormalization scheme which defines the dictionary between boundary and bulk observables already makes it manifest that the radial direction in the holographic bulk is dual to the energy scale/scale of resolution in the dual field theory. Nevertheless, it does not in itself tell us how local bulk operators such as the metric can be reconstructed as coarse-grained operators in the dual field theory. One can say that there is a passive and an active point of view of bulk reconstruction. In the active point of view advocated in the HKLL and the more refined JLMS procedures, we \textit{extrapolate} the bulk operator to the boundary. On the other hand in the RG flow point of view, we should coarse-grain the boundary operator under a suitably defined RG flow such that it mimics the geometric radial flow, and then demonstrate the emergence of local bulk operators from these flowed operators. The coarse-graining is more specifically an evolution under a sequence of CP (completely positive) unital maps of the dual field theory operators.\footnote{A CP unital map is the dual of a CPTP (completely positive trace preserving) map which evolves density matrices to density matrices. Here dual implies the map which takes us to the Heisenberg picture from the Schrodinger picture. The unital map preserves the identity.} Interestingly, the real space RG has been already discussed explicitly from the quantum error correction perspective~\cite{furuya2020,ghodrati2021_modFlow} and the Petz map also naturally emerges in this context. The passive RG flow perspective also holds an enormous promise to reveal novel principles of bulk emergence as we discuss below. It is not a mere reinterpretation of the active (HKLL/JLMS) perspective since the latter is manifestly non-local while the RG flow tames non-locality in a controlled way. 
 
 Various proposals for the RG flow have been advocated \cite{Lee:2009ij,Lee:2013dln,Behr:2015aat,Behr:2015yna,Mandal:2016rmt,Sathiapalan:2017frk} based on fundamental insights developed in \cite{Heemskerk:2009pn,Heemskerk:2010hk}. Here we will focus on the \textit{highly efficient RG flow} construction developed in \cite{Behr:2015aat,Behr:2015yna} based on earlier works in the context of the fluid/gravity correspondence \cite{Kuperstein:2011fn,Kuperstein:2013hqa}. For a review see \cite{Mukhopadhyay:2016fre}. In this approach bulk locality is manifest especially through the Ward identities. We outline this approach here in the language of quantum error correction. 
 
 We first choose a code subspace. Let this be the space of all solutions of pure gravity with a negative cosmological constant. This class of states can be charactized by the expectation value of a specific single-trace operator, namely the energy-momentum tensor, i.e. $\langle t_{\mu\nu}\rangle$ since the expectation values of multi-trace operators factorize in the large $N$ limit and other single-trace operators have vanishing or fixed expectation values.\footnote{To be precise one needs to also specify intial conditions to describe the full geometry. However, $\langle t_{\mu\nu}\rangle$ is all we need to find the geometry in a radial tube in Fefferman-Graham or Eddington-Finkelstein coordinates when the boundary metric (dual to the physical metric in which the dual field theory lives) is also specified. A better way to implement this would be to invoke a Borel resummation of the derivative expansion at late time in $\langle t_{\mu\nu}\rangle$ \cite{Heller:2013fn}. Then the initial conditions are encoded in the Stokes parameters.}
 
 The aim is to define a sequence of CP unital maps parametrized by a coarse-graining scale $\Lambda$ under which
 $$t_{\mu\nu} \rightarrow t_{\mu\nu}(\Lambda).$$Since other single-trace operators do not play a role in this subspace, most generally we should obtain
 \begin{eqnarray}\label{Eq:RGFlowEvolve}
  t_{\mu\nu}(\Lambda) &=& t_{\mu\nu} + a_1 \frac{1}{\Lambda^2}\Box t_{\mu\nu} + \frac{1}{\Lambda^4}(a_2 t_\mu^{\,\,\rho}t_{\rho\nu} + a_3 \eta_{\mu\nu} t_{\alpha\beta} t^{\alpha\beta} +  a_4 \Box^2 t_{\mu\nu} + \cdots)\nonumber\\&& + \mathcal{O}\left(\frac{1}{\Lambda^6}\right).
 \end{eqnarray}
    Above $a_i$s are appropriate numerical constants which are determined by the specific CP unital map (the insight that single trace operators should mix with multi-trace operators under the RG flow even in the large $N$ limit is from \cite{Heemskerk:2009pn}). Essentially we put in all possible multi-trace operators on the right hand side except those which vanish due to the CFT Ward identities:
    \begin{equation}\label{Eq:WIs}
    \partial^\mu t_{\mu\nu} = 0, \quad \eta_{\alpha\beta}t^{\alpha\beta} =0.
    \end{equation}
Owing to large-$N$ factorization of multi-trace operators, the evolution \eqref{Eq:RGFlowEvolve} is effectively a classical equation. The point of the highly efficient RG flow is that the coarse-graining which generates these unital maps should be done with a very specific choice of the complementary subspaces which are traced out such that we can define a metric$$g_{\mu\nu}(\Lambda) = g_{\mu\nu}[t_{\alpha\beta}(\Lambda)] $$ in which we obtain a local Ward identity
\begin{equation}\label{Eq:WIprincipal1}
    \nabla_{(\Lambda)}^\mu t_{\mu\nu}(\Lambda) = 0
\end{equation}
at each scale $\Lambda$ with $\nabla_{(\Lambda)}$ being the covariant derivative constructed from $g_{\mu\nu}(\Lambda)$. The Ward identity is defined by considering $t^\mu_{\,\,\nu}(\Lambda)$ as the fundamental variable and we lower and raise indices using $g_{\mu\nu}(\Lambda)$ and its inverse respectively. In practice, we need to then restrict $a_i$s appearing in \eqref{Eq:RGFlowEvolve} such that we can obtain $b_i$s defining $g_{\mu\nu}(\Lambda)$ via the expansion
 \begin{eqnarray}\label{Eq:RGFlowEvolveMetric}
  g_{\mu\nu}(\Lambda) &=& \eta_{\mu\nu} + \frac{1}{\Lambda^4}t_{\mu\nu} + b_1 \frac{1}{\Lambda^6}\Box t_{\mu\nu} +\nonumber\\&& \frac{1}{\Lambda^8}(b_2 t_\mu^{\,\,\rho}t_{\rho\nu} + b_3 \eta_{\mu\nu} t_{\alpha\beta} t^{\alpha\beta} +  b_4 \Box^2 t_{\mu\nu} + \cdots) + \mathcal{O}\left(\frac{1}{\Lambda^{10}}\right)
 \end{eqnarray}
with which \eqref{Eq:WIprincipal1} is satisfied. This is possible only for specific choices of $a_i$s and hence the coarse-graining CP unital maps. The emergence of bulk spacetime follows by considering the $D+1$-dimensional metric in the Fefferman-Graham gauge after identifying $r$ with $\Lambda^{-1}$:
\begin{equation}\label{Eq:FGmetric}
    {\rm d}s^2 = \frac{ {\rm d}r^2 + g_{\mu\nu}(r,x){\rm d}x^\mu {\rm d}x^\nu  }{r^2}.
\end{equation} 
The metric above satisfies the $D+1$-dimensional Einstein's equations or appropriate classical gravity equations. We can always choose the $a_i$s in \eqref{Eq:RGFlowEvolve} such that we obtain $g_{\mu\nu}(\Lambda = r^{-1})$ with which \eqref{Eq:WIprincipal1} is satisfied and the metric \eqref{Eq:FGmetric} satisfies $D+1$-dimensional classical gravity equations. It was shown in \cite{Behr:2015aat} that \eqref{Eq:WIprincipal1} is sufficient to guarantee the emergence of a $D+1$-dimensional classical gravity because without the emergence of a gauge (diffeomorphism) symmetry \eqref{Eq:WIprincipal1} cannot hold along the RG flow. In fact the RG flow has an automorphism which is related to the residual gauge symmetry of the Fefferman-Graham gauge that maps to conformal transformations at the boundary. Via appropriate state-dependent conjugations one can obtain the RG flow that reproduces the bulk metric in other gauges. In the Fefferman-Graham gauge, the RG flow is manifestly state-independent (except for a subtlety which we describe below).

The key point of highly efficient RG flow is that one should choose the complementary subspace which is projected out such that the energy-momentum exchanges with the complement can be absorbed into a redefinition of the metric. One can generalize this to cases where other single-trace operators have non-trivial expectation values simply by introducing sources for these operators such that a general version of the local Ward identity \eqref{Eq:WIprincipal1} holds. These scale-dependent sources then are the dual bulk fields after we identify $r$ with $\Lambda^{-1}$. 

\begin{figure}
   \centering
        \resizebox{0.5\textwidth}{!}{%
        \includegraphics{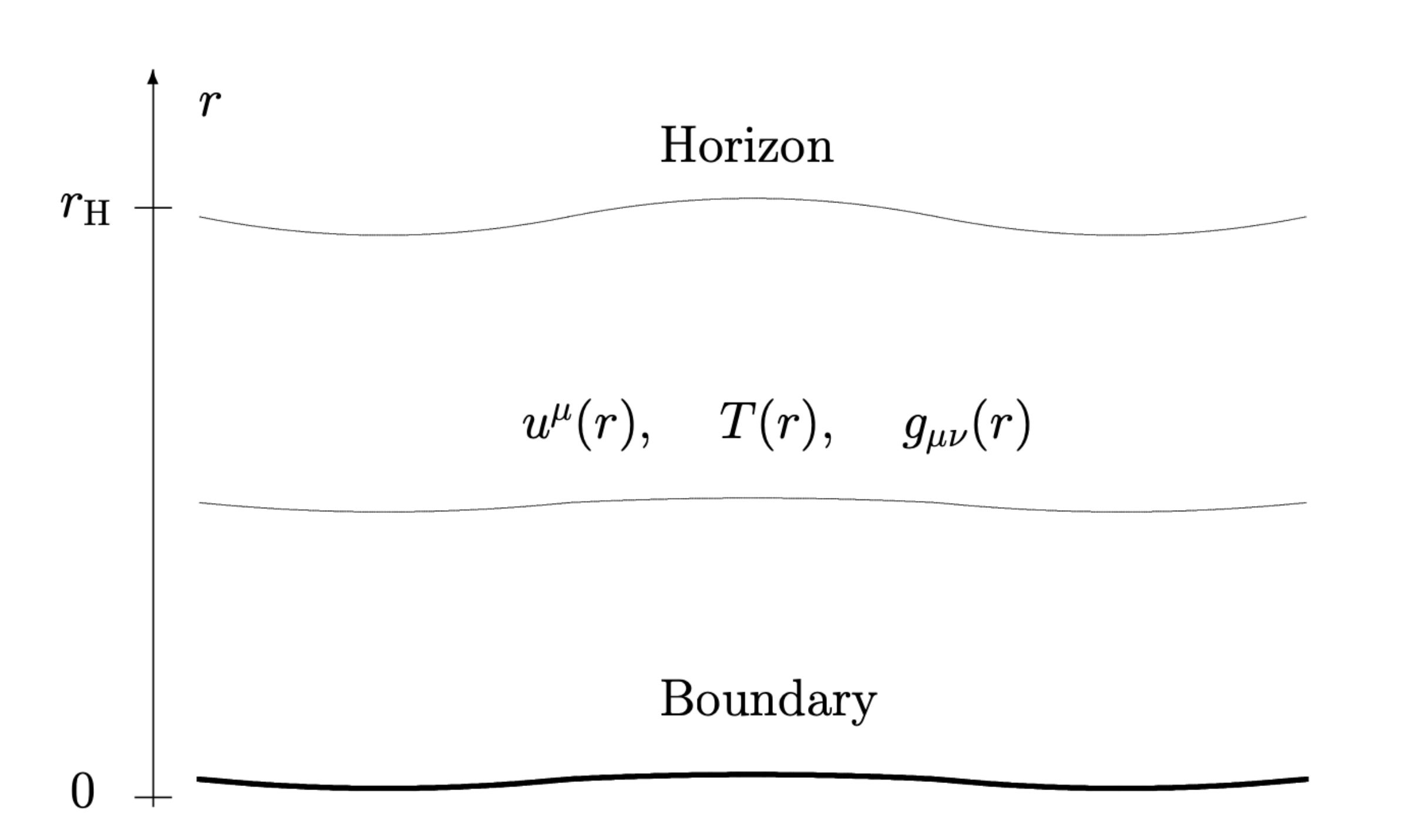}
        }
        \caption{The highly efficient RG flow in the hydrodynamic sector can be simply constructed as evolution of the velocity field $u^\mu$ and the temperature field $T$  with the coarse-graining length scale (identified with the bulk radial coordinate $r$) via the constitutive relation defining the stress-tensor. The effective background metric is evolved such that the stress tensor is locally conserved. This implies that the scale of coarse-graining $\Lambda$ becomes dependent on $u^\mu(r)$ and $T(r)$ in a very specific way so that the violation of the local energy-momentum conservation can be absorbed into a re-definition of the metric. The highly efficient RG flow simply reduces to the first order flow equations for the transport coefficients. These have unique solutions requiring that at the end-point corresponding to the location of the horizon we obtain an incompressible non-relativistic Navier-Stokes fixed point. This leads to values of transport coefficients at the boundary which give a dual spacetime without naked singularities at the future horizon. Figure from \cite{Kuperstein:2013hqa}. }
        \label{Fig:HERG}       
    \end{figure}

This coarse-graining was implemented in the hydrodynamic sector in \cite{Behr:2015yna}. To do this one assumes that $\langle t_{\mu\nu}\rangle$ is given in terms of constitutive relations via hydrodynamic variables which can be expanded in the derivative expansion. In this case, we need to sum to all orders in $\Lambda^{-1}$ at a fixed order in derivatives. The RG flow is implemented by coarse-graining of the hydrodynamic variables with the scale itself being a local functional of these variables (see Fig. \ref{Fig:HERG}). In this case, the RG flow \eqref{Eq:RGFlowEvolve} reduces simply to first order ODEs which evolve infinite number of transport coefficients. This reproduces Einstein's equations (or other classical gravity equations). To see this we simply need to follow the procedure in \cite{Kuperstein:2011fn,Kuperstein:2013hqa} and rewrite the gravity equations (in the derivative expansion) as a first order flow of transport coefficients.
This raises a profound question. In the fluid/gravity correspondence \cite{Bhattacharyya:2007vjd,Baier:2007ix}, one explicitly solves the equations of gravity in derivative expansion and obtains the transport coefficients in the UV by requiring that the dual solution has a regular future horizon. However in the RG flow we do not see the metric directly but only $t_{\mu\nu}(\Lambda)$. How do we then determine the transport coefficients? This was solved in \cite{Kuperstein:2013hqa}. It was shown that there is a particular scale $\Lambda_h$ where there is naively a singularity in the RG flow with the pressure and most transport coefficients blowing up -- this scale corresponds to the horizon in the dual geometry. However, if one redefines time and the scale in a universal way (analogous to taking the near-horizon limit), then the singularity actually transforms to a fixed point described by non-relativistic incompressible Navier-Stokes equations with a single parameter, namely the shear-viscosity. This is however possible only if the transport coefficients do not blow up faster than certain bounds as $\Lambda$ approaches $\Lambda_h$. These conditions then fix the integration constants in the ODEs describing the RG flow of the transport coefficients, and we precisely obtain those values of the transport coefficients (shear-viscosity at first order and five other transport coefficients at second order etc) at the boundary computed in \cite{Policastro:2001yc,Bhattacharyya:2007vjd,Baier:2007ix} which lead to a regular future horizon. Essentially the regularity of spacetime is simply a consequence of the regularity of the highly efficient RG flow which in the hydrodynamic limit is simply the condition that it ends at the scale of the mean-free path at a fixed point described by the non-relativistic incompressible Navier-Stokes equations (governed just by an effective shear viscosity).\footnote{In \cite{Kuperstein:2013hqa}, it was shown that the counterterms which remove the UV divergences are also fixed by the Navier-Stokes fixed point.}

The highly efficient RG flow can lead to a new way to understand bulk emergence. The RG flow can be naturally viewed as a encoding of infrared (coarse-grained) physics into microscopic degrees of freedom. The highly efficient RG flow principle of preservation of effective Ward identities \eqref{Eq:WIprincipal1} then leads to emergence of bulk fields. It might seem that at least in the Fefferman-Graham gauge this RG flow \eqref{Eq:RGFlowEvolve} is manifestly state-independent. However, this is not completely true because one needs to impose infrared boundary conditions which depends on the code subspace. In the case of the hydrodynamic sector this is specified by the Navier-Stokes fixed point. However, within the code subspace the RG flow can be cast in a state-independent form. This is precisely how the reconstruction of the black hole interior in the dual field theory should be state-dependnet as we will discuss in section \ref{sec:Python}. The boundary operator which will reconstruct the bulk operator in the black hole interior will depend on the choice of the code subspace but not on the specific state in this code subspace. This motivates a study of the highly efficient RG flow beyond the hydrodynamic sector.




\section{Decoding the black hole interior}\label{sec:bhinterior}
\subsection{How the islands emerge from replica wormholes}\label{sec:ReplicaWomhole}
The black hole interior poses the most formidable challenge in the understanding of bulk reconstruction. If the evolution of the black hole can be described by an unitary theory, we should expect that after the Page time (defined below), the information of the interior should begin to leak out into the Hawking radiation. This is a consequence of Page's theorem \cite{Page_1993,Page_2013} which states that for a typical pure state in a bipartite system with one system being much smaller than the other, the trace distance between the density matrix of the smaller subsystem and the microcanonical ensemble is of the order of $e^{- S/2}$ with $S$ being the number of qubits in the larger one. At initial stages, the black hole is much larger than its Hawking radiation, so the entanglement entropy of the latter should grow. The Page time occurs when the black hole and the emitted radiation have the same coarse grained entropy\footnote{The coarse-grained entropy is defined as follows. Consider expectations values of simple operators (averaged over certain time-scales) and construct all density matrices which reproduce them. The coarse-grained entropy is the entropy of such a density matrix which has maximal von-Neumann entropy. Since the argument of Page uses typicality, the Page time is appropriately defined via the coarse-grained entropy which is determined mostly by the size of the Hilbert space.}. After the Page time, the entanglement entropy of the radiation should decrease as it becomes the larger subsystem. The already emitted Hawking quanta should also be purified by the subsequently emitted Hawking quanta. Thus its total entanglement entropy should follow the Page curve -- initially it should grow linearly in time following Hawking's original computation but then it should fall back to zero after Page time as it gets purified.

The crucial question is that if the semi-classical description is valid for the effective field theory (EFT) observables, then what exactly goes wrong with the semi-classical computation of the entanglement entropy of the Hawking radiation after the Page time. The question is further sharpened by the Almheiri-Marolf-Polchinski-Sully (AMPS) \cite{Almheiri:2012rt} paradox which points that such assumptions would lead to violation of the monogamy of entanglement because the Hawking quanta are maximally entangled with their infalling counterparts according to the EFT computations, but they also have to be maximally entangled with the quanta emitted before Page time in order to purify it and validate the unitarity of the evolution. The strong sub-additivity property of the entanglement entropy would prevent simultaneous maximal entanglement of a system with two other systems \cite{PhysRevA.69.022309}. We will discuss more aspects of this paradox later.\footnote{The AMPS paradox builds on the discussions by Mathur \cite{Mathur:2009hf} which introduced the strong subadditivity originally to examine if small semi-classical corrections can restore unitarity in Hawking radiation. Many aspects of the AMPS paradox were discussed also in \cite{PhysRevLett.110.101301}.} 

It is to this question (paradox) that the AdS/CFT correspondence has recently produced some remarkable new insights along with a deeper understanding of which part of the black hole interior is encoded in a given subregion of the Hawking radiation system. The main lesson is that the EFT in the semi-classical black hole geometry is indeed unimpeachable in its usual domain of validity set by an energy scale, however there are other subleading Euclidean saddles in the path integral for R\`{e}nyi entropies which give sufficient contribution after Page time to restore unitarity. Furthermore, these saddles imply that the fine-grained entropy\footnote{The fine-grained entropy can be defined like the coarse-grained entropy above but taking into account more operators that probe the short distance structure of the density matrix. However, we should keep in mind that we are already doing an averaging which allows a tensor factorization between the black hole and radiation system to emerge. A very concrete discussion on this issue can be found in \cite{Ghosh:2021axl} within the context of the semi-classical approximation itself. We will have a more elaborate discussion on this in the context of microstate models in Section \ref{sec:microstate}.} of Hawking radiation (obtained from the R\`{e}nyi entropies) would involve contributions from \textit{islands} (which are regions of bulk spacetime diconnected from the asymptotic boundary and including portions of the black hole interior as already introduced in the context of explicit models in section \ref{sec:islands}) after Page time as these are connected to the already emitted quanta \textit{via wormholes}. This implies the ER= EPR mechanism \cite{Maldacena_2013} of resolution of AMPS paradox with wormholes implying that after Page time operators in the interior also affect this \textit{far away radiation}, and therefore the early radiation and the interior are not separable systems. We will discuss later that if one has to resolve this issue by explicitly following the state in real time we will need self-averaging and complexity. The crucial inputs in both cases would be the island reconstruction for which the framework of operator error correction would play an important role.

\begin{figure}
   \centering
        \resizebox{1.0\textwidth}{!}{%
        \includegraphics{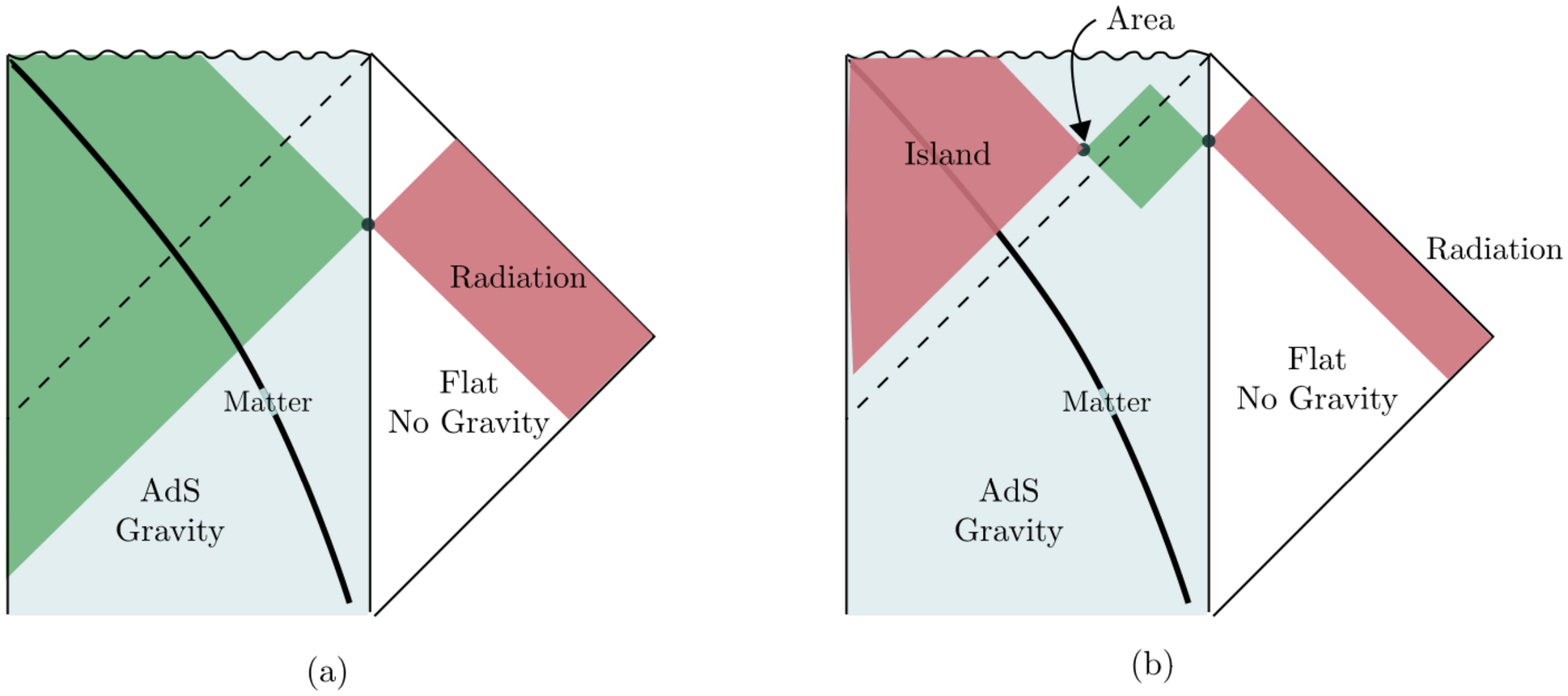}
        }
        \caption{This figure is from \cite{Almheiri2020Islands}. It depicts the construction where a holographic system $B$ is coupled to a large system $R$ without gravity. The dynamics of $B$ is described by the evaporating black hole with a constant curvature which is glued to flat Minkowski space in which $R$ lives. $R$ collects the Hawking quanta of the evaporating black hole. The red regions are the entanglement wedges of the interval(s) which contribute to the von-Neumann entropy of the Hawking quanta in $R$ while the green regions are those that contribute to $B$. (a) At early time, the entropy of the matter fields in whole of the Cauchy slice contributes to $B$. (b) After Page time, the interval exterior to the quantum extremal surface, whose entanglement wedge is the island, contributes to the von-Neumann entropy of the Hawking quanta in $R$. }
        \label{Fig:Island}       
    \end{figure}

The anti-de Sitter black hole cannot evaporate unless the boundary conditions allow the quanta of bulk matter to escape out of the asymptotic region. The understanding of how the black hole interior is encoded into these Hawking quanta especially after the Page time can be achieved by coupling a holographic system $B$ with another larger system $R$ living in one higher dimension, and which may or may not be holographic.  The role of $R$ is to collect the Hawking quanta of the evaporating black hole in the dual geometry that depicts the  evolving system $B$ holographically. For tractability, we also need the bulk matter to have large central charge $c$ so that we can ignore quantum gravity fluctuations. What has been shown to be crucial is that, after the Page time, the entanglement wedge of the quantum extremal surface (QES) of the asymptotic boundary is encoded in $B$, while the wedge including the exterior of the QES called the \textit{island} and containing parts of the black hole interior is encoded into $R$ as illustrated in Fig. \ref{Fig:Island} for $0+1$-dimensional $B$ coupled to a $1+1$-dimensional $R$. To see this, we need to consider the (pure) state of the quantum matter fields on a Cauchy slice of the black hole geometry and $R$, and consider contributions of saddles that contribute to the R\`{e}nyi entropy of the Hawking radiation in $R$ other than the semi-classical black hole itself. These saddles connect the replicas of the R\`{e}nyi entropy computation via wormholes, and are therefore called replica wormholes. These wormholes connect the island $I$ including parts of the black hole interior to the Hawking quanta in $R$, thus invalidating the tri-partition in the AMPS paradox and resolving it.
\begin{figure}
   \centering
        \resizebox{1.0\textwidth}{!}{%
        \includegraphics{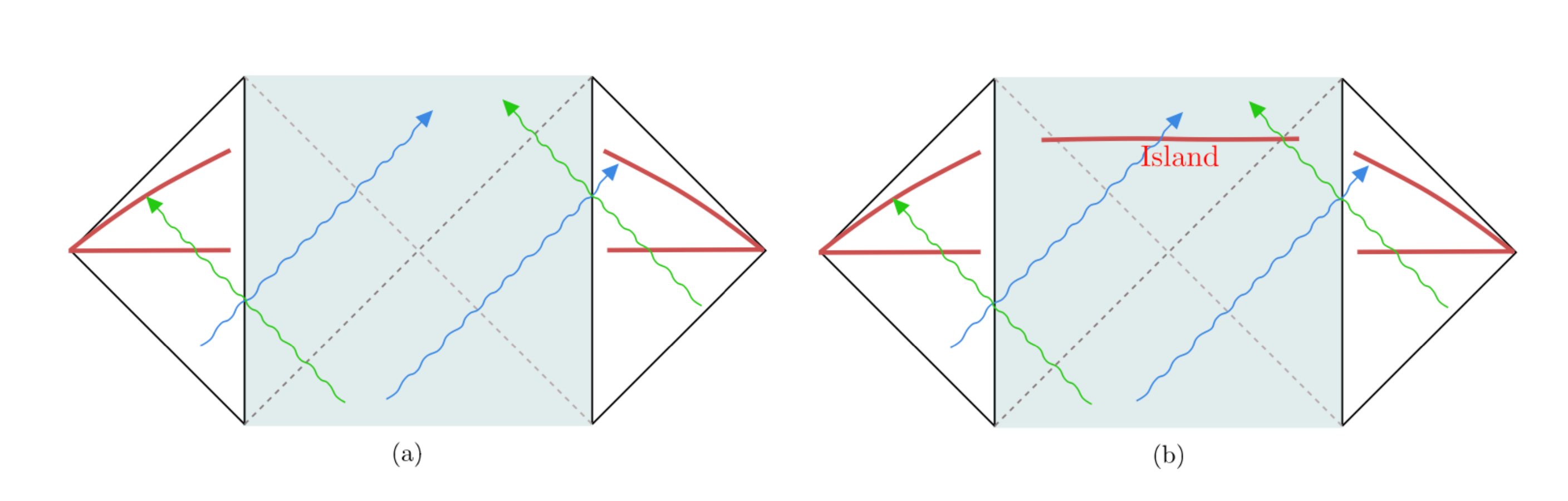}
        }
        \caption{A thermofield double version of the setup described in Fig. \ref{Fig:Island} taken from \cite{Almheiri2020Islands}. The particles with the same color (blue/green) are entangled according to effective field theory computations. Here we consider the entanglement entropy of the Hawking quanta in the red intervals at time $t=0$ and at $t= t_0$ with $t_0$ exceeding the Page time. In (a) we see that if we do not include the island, then the von-Neumann entropy of the red intervals ($R$ )should grow linearly with time and with the slope determined by the rate of Hawking pair production. We see in (b) that if one includes the island ($I$), then the matter contribution to the von-Neumann entropy in $I\cup R$ cannot grow as the entangled pairs get collected in this combined region.}
        \label{Fig:ThermoDouble}       
    \end{figure}

We need to explain why such saddles which give exponentially small contributions in the semi-classical limit to the R\`{e}nyi entropy should be important at all. The leading saddle is the usual $\widetilde{\mathcal{M}}_n/Z_n$ geometry discussed in Section \ref{sec:QES} denoting $n$ copies of the semi-classical black hole geometry quotiented by the cyclic permutation symmetry. In the presence of gravity, we need to take into account the other replica wormhole saddles too. The point is that the contribution from the $\widetilde{\mathcal{M}}_n/Z_n$ saddle decreases exponentially with time (consistent with the linear growth of the von-Neumann entropy in the $R$ region) while the contribution from the other replica wormhole saddles grow with time essentially due to the growth of the island. Around the Page time, these effects compete with each other, and after sufficiently long time the replica wormholes give the leading contribution to the R\`{e}nyi entropies, and arrest the growth of the von-Neumann entropy of $R$ as illustrated in Fig. \ref{Fig:ThermoDouble}.\footnote{Note in the thermofield double setup the entanglement entropy of the radiation should not vanish but rather saturate to be consistent with unitarity since not all radiation escapes to the null infinities in $R$. The Page curve then involves a transition from linear growth to saturation.} 
\begin{figure}
   \centering
        \resizebox{1.0\textwidth}{!}{%
        \includegraphics{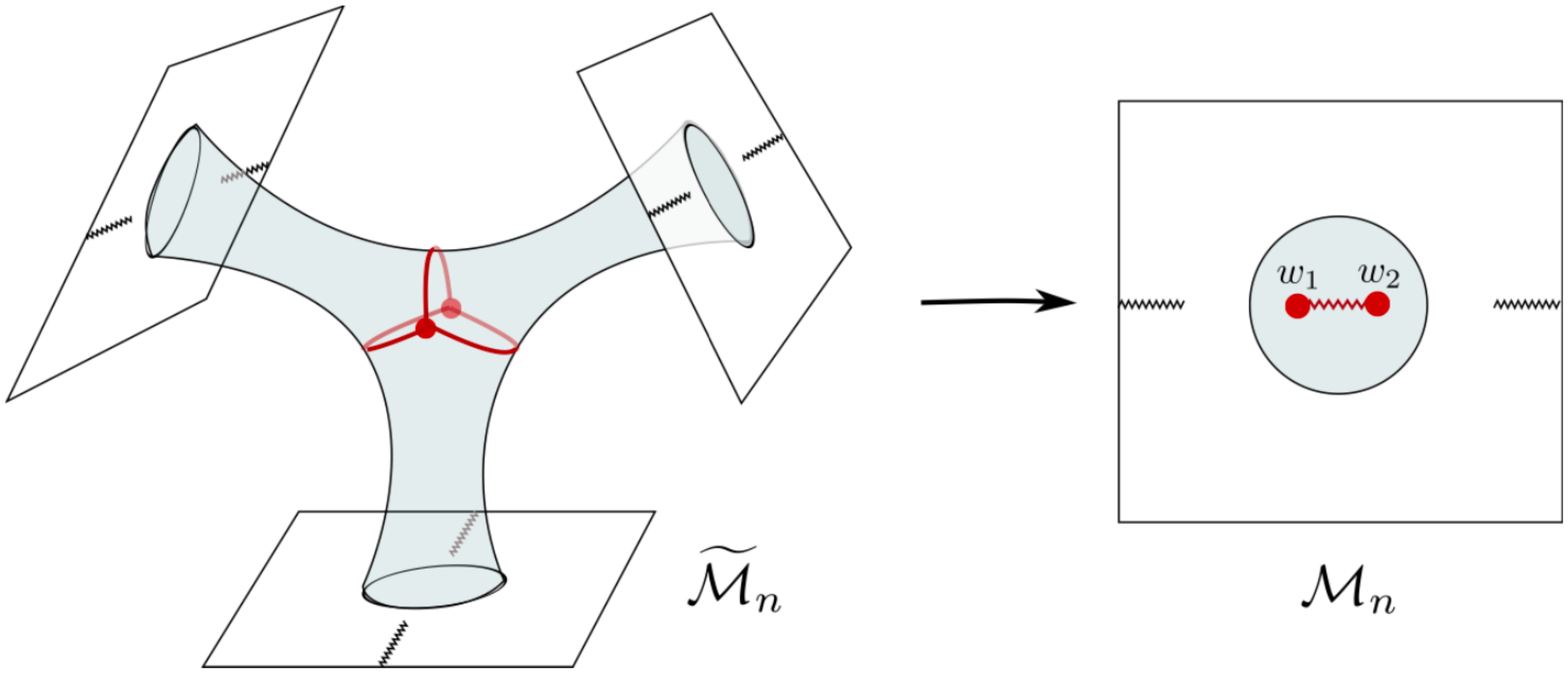}
        }
        \caption{An example of a replica wormhole from \cite{Almheiri2020Islands} for the computation of the $3$rd order R\`{e}nyi entropy of the Hawking quanta in $R$ in the setup described in Fig. \ref{Fig:ThermoDouble}. Firstly, we consider three copies of the Euclidean version of the diagrams in Fig. \ref{Fig:ThermoDouble}. In each copy, the shaded blue region is the Poincar\'{e} disk denoting Euclidean $AdS_2$ black hole with an appropriate time-reparametrization at the boundary of the disk. The white regions have cuts corresponding to the red intervals in Fig. \ref{Fig:ThermoDouble}. In (a) we denote the covering manifold $\widetilde{\mathcal{M}}_3$ and in (b) the actual replica wormhole solution which is  $\mathcal{M}_3 =\widetilde{\mathcal{M}}_3/Z_3$ which arises as a solution to the Euclidean gravity coupled to matter. The fixed points of the $Z_3$ action, namely $w_1$ and $w_2$, appear as conical singularities in the gravitational solution where the bulk twist operators are inserted. Note their positions are determined by the equations of motion also. }
        \label{Fig:ReplicaWormhole}       
    \end{figure}

These replica wormholes are essentially codimension two cosmic branes on the gravity side (giving rise to conical singularities in two-dimensional gravity) where the twist operators of the bulk matter theory must be inserted. See Fig. \ref{Fig:ReplicaWormhole} for an illustration. The positions of the these branes should be obtained by solving the gravitational equations of motion. For the semi-infinite interval in $R$ containing the Hawking quanta in Fig. \ref{Fig:Island}, it turns out that there is one such brane, and in the limit $n\rightarrow 1$ with $n$ denoting the number of replicas, the position of the cosmic brane is exactly that of the quantum extremal surface which computes the entropy of $B$. In case of the double interval in Fig. \ref{Fig:ThermoDouble}, the cosmic branes coincide with the two quantum extremal surfaces in the limit $n\rightarrow 1$ as well. For a recent pedagogical review on wormholes in semiclassical gravity including the explicit construction of these replica wormholes see \cite{Arnab_review}.

Specifically, the island rule  \eqref{Eq:IslandRule} which is now validated by the gravitational path integral accounting for the replica wormhole saddles, implies that
\begin{equation}\label{Eq:island-rule-specific}
S(\rho_R) = S(\rho_{RI}^{(g)}) + \frac{A({\rm QES})}{4 G}.
\end{equation}
Above $\rho_R$ denotes the density matrix of the interval(s) in $R$, $\rho_{RI}^{(g)}$ denotes the density matrix of the quantum fields (responsible for the Hawking quanta) in $R\cup I$ in the semiclassical geometry of the black hole glued to Minkowski space(s). Note that $A({\rm QES})$ above refers simply to the value of the dilaton for the JT theory since the QES is a point (or two points in the case of the thermofield double).  We should understand the above as a semi-classical statement only -- the island is defined with respect to the semi-classical geometry itself. On the other hand, if $D$ is the interval in this geometry which is the complement of $R\cup I$ and connects the QES to the boundary of the $AdS_2$ region, then the generalization of the HRT formula given by \eqref{Eq:GenEntropyJT} will imply that
\begin{equation}
S(\rho_B) = S(\rho_{D}^{(g)}) + \frac{A({\rm QES})}{4 G} 
\end{equation}
Above by $B$ we mean its two copies in the thermofield double case. Since, the quantum fields in the full Cauchy slice $I\cup D\cup R$ in the semiclassical description should be in a pure state, it implies that $S(\rho_{RI}^{(g)}) = S(\rho_{D}^{(g)})$, and therefore $S(\rho_R) = S(\rho_B)$. This is then consistent with unitarity of the evolution of the full (two copies of the) \textit{dual} $B\cup R$ system.

We can also then claim that the bulk effective field theory operators acting in the island can be reconstructed in $R$ as an extension of the entanglement wedge reconstruction described before in section \ref{sec:bulkreconstruction}. The code subspace is the span of states in the full (two copies of) $R\cup B$ which can be constructed in the gravitational description via operator insertions in the same semi-classical geometry containing the same island $I_i$ at leading order in perturbative EFT corresponding to an interval $R_i$ in $R$. The code subspace is assumed to admit a factorization of the form $\mathcal{H}_{code} =\mathcal{H}_{R_i}\otimes\mathcal{H}_{\overline{R}_i}$ where $R_i$ here denotes the interval(s) in $R$, and $\overline{R}_i$ the complement of $R_i$ that includes (two copies of) $B$. The EFT of bulk fields in the semi-classical gravity description including the Minkowski space region(s) imply that $\mathcal{H}_{code} = \mathcal{H}_{R_i}\otimes \mathcal{H}_{D_i}\otimes \mathcal{H}_{I_i}$ where $D_i$ is the complement of the island and $R_i$. Then \eqref{Eq:island-rule-specific} holds for any $\rho_{R_i}^k$ obtained from any state $\vert k\rangle \in \mathcal{H}_{code}$ by tracing out $\overline{R}_i$, and with corresponding $\rho_{R_iI_i}^{(g)k}$ on the bulk side obtained by tracing out $D_i$. An extension of the JLMS argument would then imply that the relative entropies of the semi-classical bulk states and the corresponding states in the full dual description are identical, i.e.
\begin{equation}
    S(\rho_{R_i}^m\vert \rho_{R_i}^n) = S(\rho_{R_iI_i}^{(g)m}\vert \rho_{R_iI_i}^{(g)n})
\end{equation}
for two states $\vert m \rangle$ and $\vert n \rangle$ belonging to $\mathcal{H}_{code}$. A repeat of the theorem of Dong, Harlow and Wall described in section \ref{subsec:OEC_BR} then implies that for any bulk EFT operator $\mathcal{O}_{I_i}$ localized in the island $I_i$ and any $\vert n \rangle \in \mathcal{H}_{code}$, we should have
\begin{enumerate}
\item  For any any $X$ localized in $\overline{R}_i$
\begin{equation}
    \langle n \vert[X, O_{I_i}]\vert n\rangle = 0,
\end{equation} and
\item there exists an operator $\mathcal{O}_{R_i}$ localized in $R_i$ such that its action on the code subspace is identical, i.e.
\begin{equation}
    \mathcal{O}_{R_i}\vert n\rangle =\mathcal{O}_{I_i} \vert n\rangle, \quad \mathcal{O}_{R_i}^\dagger\vert n\rangle =\mathcal{O}_{I_i}^\dagger  \vert n\rangle.
\end{equation}
\end{enumerate}

\begin{figure}
   \centering
        \resizebox{1.0\textwidth}{!}{%
        \includegraphics{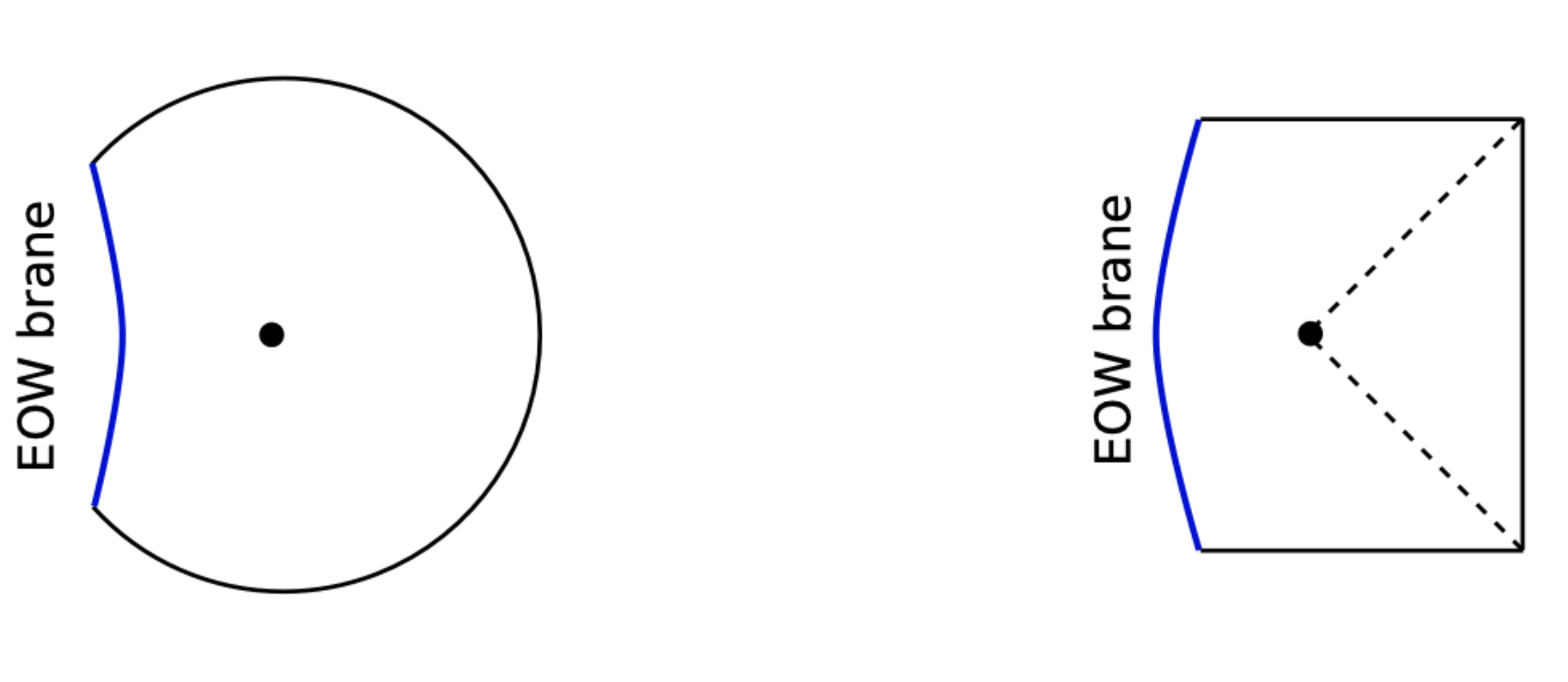}
        }
        \caption{The simplified PSSY model with an end of the world brane carrying a degree of freedom. It has zero extrinsic curvature and the Neumann boundary condition for the JT gravity dilaton is imposed on it. The left and right figures correspond to the Euclidean and Lorentzian scenarios. Figure from \cite{Penington:2019kki}}
        \label{Fig:EOW}       
    \end{figure}

Although the replica wormhole saddles produce results which are compatible with unitarity, there are crucial subtleties which are explicitly brought out by the Euclidean (Penington, Shenker, Stanford and Yang) PSSY toy model \cite{Penington:2019kki} in which we can sum over the planar replica wormhole saddles explicitly (via a Schwinger-Dyson type equation for the resolvent matrix $1/ (\lambda I - \rho_R)$). It also gives a simple way to see how the replica wormhole saddles give rise to the island which we will review below.  In this Euclidean model, we represent the state of the $B\cup R$ system in the maximally entangled form
\begin{equation}\label{Eq:PSSY-state}
    \vert \psi \rangle = \frac{1}{\sqrt{k}}\sum_{i=1}^k \vert \psi_i\rangle_B \vert i\rangle_R
\end{equation}
which should be a good approximation to the full state of the Hawking quanta especially after Page time. The state of the holographic bulk $B$ is captured by an end of the world (EOW) brane state $i$ as shown in Fig. \ref{Fig:EOW}. The evolution of time is captured by simply dialling $k$. To capture the dynamics post Page time then we need $k \gg e^{S_{BH}}$ where $S_{BH}$ is the entropy of the bulk black hole. The bulk theory is pure JT gravity with a dilaton and no bulk matter, while the EOW brane has vanishing extrinsic curvature and the Neumann boundary condition for the bulk dilaton is also imposed at its location. Explicitly, the Euclidean action for the metric $g$ and the dilation $\phi$ is
\begin{eqnarray}\label{Eq:PSSYaction}
S &=& - \frac{S_0}{4\pi}\left( \int_{\mathcal{M}} \sqrt{g}R + 2 \int_{\partial \mathcal{M}}\sqrt{h}K\right)\nonumber\\&&
-\frac{1}{2}\left( \int_{\mathcal{M}} \sqrt{g}\phi(R+2) + 2 \int_{\partial \mathcal{M}}\sqrt{h}\phi K\right)\nonumber\\&&
+\mu\int_{EOW-brane} {\rm d}s,
\end{eqnarray}
with $h$ denoting the induced metric on the boundary and $\mu \geq 0$ the mass of the EOW brane. The mentioned boundary conditions imply that on the EOW brane we should impose
\begin{equation}
    K =0, \quad \partial_n \phi = \mu
\end{equation}
with $\partial_n$ denoting the normal derivative. The standard asymptotic boundary conditions imply that at the boundary
\begin{equation}
    h = \frac{1}{\epsilon^2}, \quad \phi = \frac{1}{\epsilon}, \quad \epsilon \rightarrow 0.
\end{equation}
The limit $\epsilon \rightarrow 0$ implies implementing the standard procedure of holographic renormalization which extracts  physical observables of the $B$ system from the regularized action of the classical gravity solution. The solutions of the gravitational theory are simply the black hole solutions. It is easy to note that the computations of the replica wormhole saddles simplify in absence of bulk matter too. The results of the computations are the same. When $k \gg e^{S_{BH}}$, the replica wormhole saddles produce a dominant contribution as a result of which the QES in the limit $n\rightarrow 1$ is close to the bifurcate horizon opening up an island which is the entanglement wedge of the $R$ system.

\begin{figure}
   \centering
        \resizebox{1.0\textwidth}{!}{%
        \includegraphics{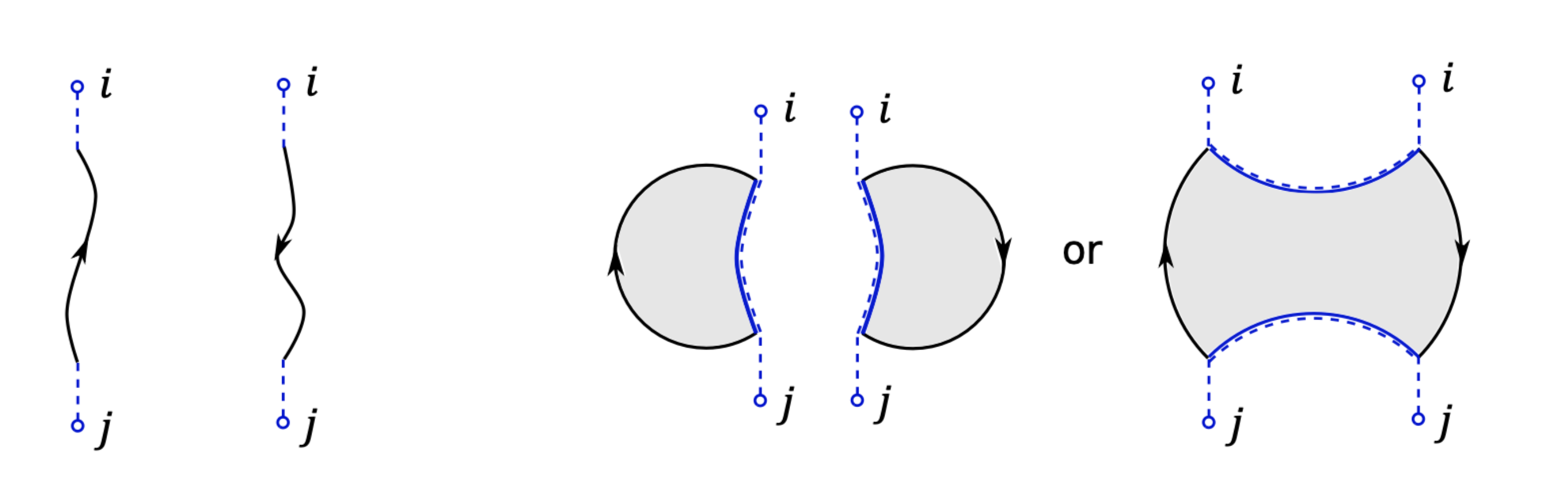}
        }
        \caption{The computation of ${\rm Tr}(\rho_R^2)$ involves two saddles. The first one is the product of two disconnected disks and the second one is a two-sided replica wormhole. To compute the trace we must join the lines dashed lines. The disconnected disks then produce a single loop implying a factor of $k$ while the double sided wormhole produces two loops implying a factor of $k^2$. Thus we obtain the numerator in Eq. \eqref{Eq:rho2}. Figure from \cite{Penington:2019kki}}
        \label{Fig:EOWreplica}       
    \end{figure}

\begin{figure}
   \centering
        \resizebox{1.0\textwidth}{!}{%
        \includegraphics{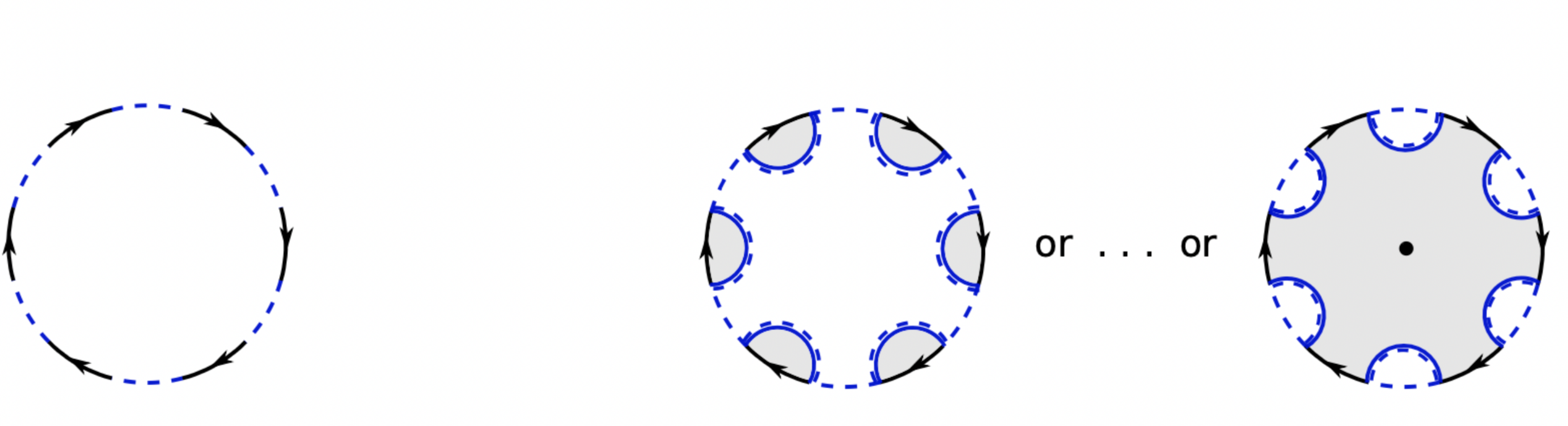}
        }
        \caption{The computation of ${\rm Tr}(\rho_R^6)$ involves several saddles of which the fully disconnected and the fully connected ones are explicitly shown above. The first one dominates when $k \ll e^{S_{BH}}$ while the second one dominates when $k \gg e^{S_{BH}}$. The latter has a $Z_6$ rotational symmetry whose action has a fixed point. Figure from \cite{Penington:2019kki}.}
        \label{Fig:Replica6}       
    \end{figure}

\begin{figure}
   \centering
        \resizebox{1.0\textwidth}{!}{%
        \includegraphics{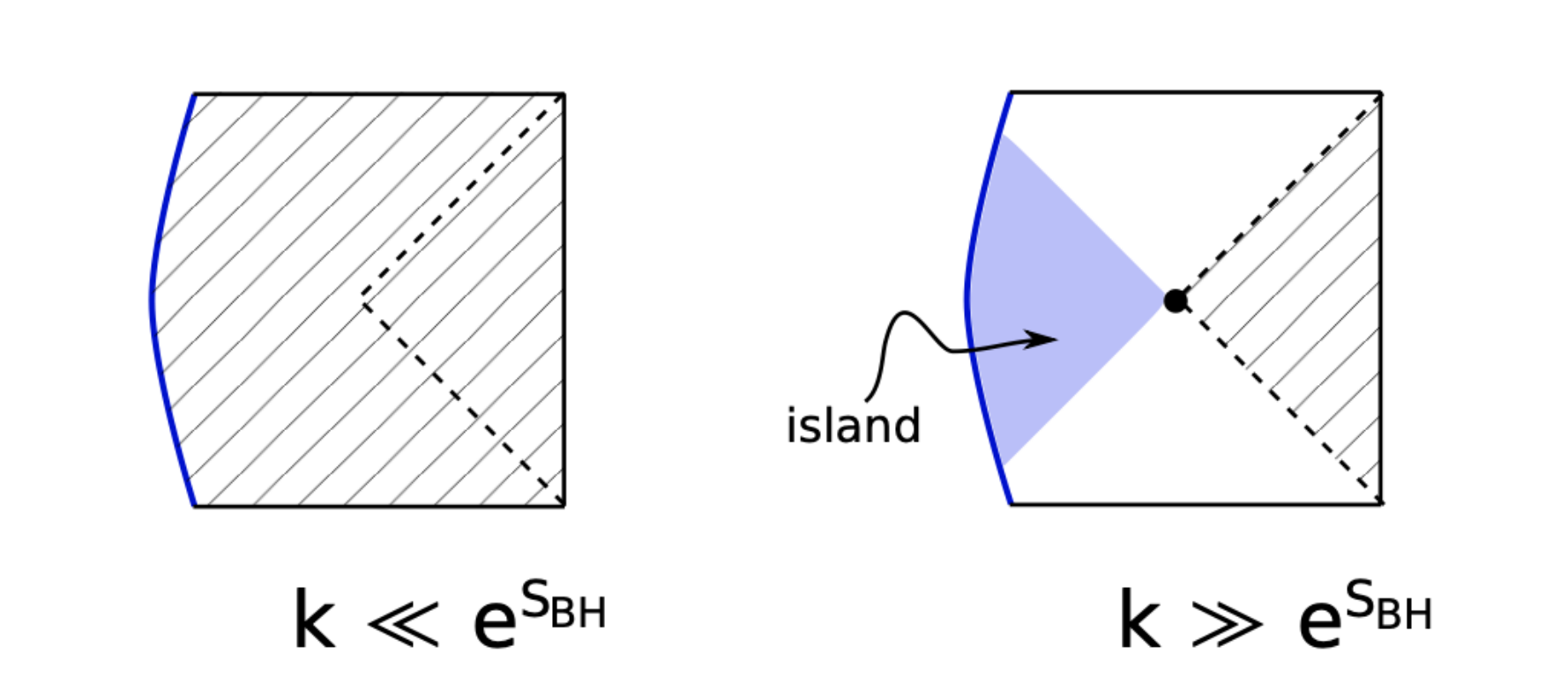}
        }
        \caption{As described in the text, the fully disconnected replica wormholes dominate in the limit $k \ll e^{S_{BH}}$ implying that ${\rm Tr}(\rho_R^n) = 1/k^{n-1}$. This is obtained by tracing over the entire black hole spacetime which thus forms the entanglement wedge of $B$. In the limit $k \ll e^{S_{BH}}$, the fully connected replica wormhole geometry (one single disc with $n$ asymptotic boundaries) dominates. So we obtain ${\rm Tr}(\rho_R) = S_{BH}$ when $n\rightarrow 1$. This implies that the entanglement wedge of $R$ is the interior of the horizon which forms the island, while the black hole exterior forms the entanglement wedge of $B$. Figure from \cite{Penington:2019kki}. }
        \label{Fig:PSSYisland}       
    \end{figure}

The subtlety is that we would naively think that if the EOW brane states $\vert i \rangle$ are orthogonal, then the reduced of the $R$ system which follows from \eqref{Eq:PSSY-state} is 
    \begin{equation}\label{Eq:denmatrixPSSY}
        \rho_R = \frac{1}{k} \sum_{i=1}^k \vert i\rangle \langle i\vert_R.
    \end{equation}
Furthermore, ${\rm Tr}(\rho_R^2) = 1/k$. The latter is however not correct if we take into account the replica wormhole saddle as illustrated in Fig. \ref{Fig:EOWreplica}. Explicitly we obtain
\begin{equation}\label{Eq:rho2}
    {\rm Tr}(\rho_R^2) = \frac{k Z_{(1)}^2 + k^2 Z_{(2)}}{(k Z_{(1)})^2} =\frac{1}{k} + \frac{Z_{(2)}}{Z_{(1)}^2}
\end{equation}
where $Z_{(2)}$ is the partition function of the two-sided wormhole and $Z_{(1)}$ is the disk partition function (the denominator comes from the appropriate normalization of the path integral). This is clearly incompatible with \eqref{Eq:denmatrixPSSY}.
The reconciliation is achieved if we interpret that the replica wormhole saddles of JT gravity are essentially computing the path integrals for an ensemble average of Hamiltonians describing the $B$ system, so that
\begin{equation}
    \rho_R = \frac{1}{k} \sum_{i,j=1}^k \vert i\rangle \langle j\vert_R \overline{\langle \Psi_j\vert\Psi_i\rangle}_B
\end{equation}
and 
\begin{equation}\label{Eq:Trrh02}
   {\rm Tr} (\rho_R^2) = \frac{1}{k^2} \sum_{i,j=1}^k \overline{\vert\langle \Psi_j\vert\Psi_i\rangle\vert^2}_B.
\end{equation}
Supposing that 
\begin{equation}
    \langle \Psi_j\vert\Psi_i\rangle_B = \delta_{ij} + e^{-S_0/2} R_{ij}
\end{equation}
where $S_0$ is the entropy of the JT gravity black hole at zero temperature and $R_{ij}$ are random phases, we should obtain
\begin{equation}
\overline{\langle \Psi_j\vert\Psi_i\rangle}_B = \delta_{ij}, \quad \overline{\vert\langle \Psi_j\vert\Psi_i\rangle\vert^2}_B = \delta_{ij} + \mathcal{O}(e^{-S_0})
\end{equation}
where the bar on top denotes averaging over theories. Note that $e^{-S_0}$ is exactly the order of magnitude of $Z_{(2)}/Z_{(1)}^2$ because of the respective topologies (two discs vs one with each disk being $\mathcal{O}(e^{S_0})$\footnote{The first term in the action \eqref{Eq:PSSYaction} is topological and accounts for $e^{S_0}$ weighting of the disk.}). Note it is easy to see from \eqref{Eq:rho2} that when $k \gg e^{S_0}$, then the replica wormhole gives the dominant contribution $Z_{(2)}/Z_{(1)}^2$. This in fact generalizes, and we can readily see that
\begin{equation}
    {\rm Tr}(\rho_R^n) \approx \frac{k^n Z_{(n)}}{(k Z_{(1)})^n} = \frac{Z_{(n)}}{Z_{(1)}^n} = \mathcal{O}(e^{-(n-1)S_0})
\end{equation}
so that the contribution is entirely from the simply connected replica wormhole with $n$ asymptotic boundaries (see Fig. \ref{Fig:Replica6} for an illustration in the case of $n=6$). The $Z_n$ quotient of these wormholes $Z_{(n)}$ can be readily analytically continued (the $Z_n$ symmetry is simply a rotational symmetry and quotienting produces a fixed point which is exactly the horizon in the limit $n\rightarrow 1$ in which the geometry reduces to the original unreplicated one). It is also then easy to see that in the limit $n\rightarrow 1$, the von Neumann entropy is just the generalized entropy of the QES that is located at the horizon, i.e.
$${\rm Tr}(\rho_R) = S_0 + 2\pi \phi_h = S_{BH},$$ 
where $\phi_h$ is the value of the dilaton at the horizon and $S_{BH}$ is the thermodynamic entropy of the black hole. The latter equality follows from the action \eqref{Eq:PSSYaction}) and note that there is no bulk matter in this model. Thus the $k\rightarrow\infty$ limit indeed implies that the QES is at the bifurcate horizon in the Lorentzian picture and the island is the full interior of the black hole. We explicitly see that the replica wormholes reproduce the correct QES. In the opposite limit $k \rightarrow \epsilon$ with fixed $S_0$, the fully disconnected geometry dominates (see Fig. \ref{Fig:Replica6} for an illustration in the case of $n=6$) so that
\begin{equation}
    {\rm Tr}(\rho_R^n) \approx \frac{k Z_{(1)}^n}{(k Z_{(1)})^n} = \frac{1}{k^{n-1}}.
\end{equation}
The von-Neumann entropy is then $${\rm Tr}(\rho_R) =\log k$$ in this limit signifying the absence of any island and that we should trace over the entire black hole spacetime which thus forms the entanglement wedge of $B$. This toy model then reproduces features of the growth of the islands in the full Lorentzian computation with the bulk matter fields described earlier. The island vanishes at early time, and as the black hole evaporates it forms and eventually encompass the original interior. The flow of time is indeed represented in the toy model as the growth of $k$ as illustrated in Fig. \ref{Fig:PSSYisland}.

The toy model also allows us to derive a version of the Petz map via path integrals with replica wormhole saddles for reconstruction of the operators acting on the island in the $R$ system. This makes sense if the wormholes perform an ensemble averaging over theories describing the $B$ system. The replica wormholes dominate when $k \gg d_{\rm code} e^{S_{0}}$ where $d_{\rm code}$ is the dimension of the code subspace, and in this limit the mentioned Petz map produces a perfect reconstruction as expected. Of course when $k \ll d_{\rm code} e^{S_{0}}$, there is no island and the $R$ system should have no knowledge of the black hole interior.

Recently, Bousso and Shahbazi-Moghaddam have revisited the holographic entropy bounds discussed in the introduction and have examined them in the context of islands \cite{Bousso:2021sji}. In particular, they have been able to find some general conditions for the existence of islands and how holographic entropy bounds should be revised in the context of island rule. They have argued that such arguments could guarantee behavior consistent with unitarity in more general contexts. This also explains absence of islands \cite{Manu:2020tty} found in the context of AdS-Kasner spacetimes coupled to non-gravitating reservoirs. 

The Page curve for other entanglement measures such as entanglement negativity (which is a more suitable measure than the von Neumann entropy in the case of mixed states) of the Hawking radiation leaking to a bath has also been recently reproduced in these two-dimensional setups \cite{KumarBasak:2020ams,KumarBasak:2021rrx} developing on prescriptions obtained in \cite{Kudler-Flam:2018qjo,Kusuki:2019zsp,KumarBasak:2020ams} and are consistent with results obtained from random matrix theory \cite{Shapourian:2020mkc}. The entanglement negativity has also been studied in the context of a generalized version of the PSSY model with a bipartite (non-gravitating) reservoir in \cite{Kudler-Flam:2021efr,Dong:2021oad}. Interestingly, saddles which break replica symmetry appear. The Page curve in doubly holographic setups with the (non-gravitating) reservoir subjected to relevant deformations have been studied in \cite{Caceres:2021fuw}. In this work, it has been shown that the coarse-graining of the reservoir generated by the RG flow leads to an increase in the Page time.  

Finally, the emergence of islands should be understood from modular flow which is the fundamental element of explicit entanglement wedge reconstruction as discussed in section \ref{sec:bulkreconstruction}. See \cite{Chen:2019iro} for advances made in this direction.

The fact that the replica wormhole saddles of the Euclidean path integrals including the gravitating regions perform some kind of averaging is a generic feature which should hold for higher dimensional setups also. In a full unitary quantum gravity computation in real time in which we should be able to take into account the microstates of the black hole (such as fuzzballs) explicitly, the averaging should emerge dynamically via quantum ergodicity. The late-time self-averaging in many-body dynamics which justifies the Euclidean computation of R\`{e}nyi entropies leading to an universal approximation has been discussed in \cite{PRXQuantum.2.010344}\footnote{See also \cite{Vardhan:2021npf} for a related discussion on how such \textit{equilibrium approximation} of chaotic many-body dynamics makes novel predictions for entanglement in Hawking radiation before Page time.}. Nevertheless, it has to be understood from the bulk point of view.\footnote{In fact JT gravity has a full non-perturbative dual description in terms of an ensemble of random matrices \cite{Saad:2019lba,Stanford:2019vob}. That such randomness eg random couplings should be described by wormholes has been already discussed by \cite{Giddings:1988wv,Polchinski:1994zs}. In an unitary description, such averaging should emerge from ergodicity and then the question is to understand the dual bulk mechanism from explicit microstate models.}

It is already remarkable that the replica wormholes produce results for the Page curve which are compatible with unitarity. Nevertheless, many crucial features of the encoding of the black hole interior into the Hawking quanta and other related phenomena such as quantum information mirroring cannot be addressed by such an approach. We will examine very soon if we can construct tractable microstate models to address these fundamental issues.

\subsection{State-dependence in microstate reconstruction: Alpha bits and Python's lunch}\label{sec:Python}
The natural question to ask of course is how do we reconstruct (microstate of) a non-evaporating black hole which is not in contact with an auxiliary reservoir.\footnote{By non-evaporating we mean with reflecting asymptotic boundary conditions. Note Hawking radiation is present but the black hole cannot lose mass.} This should be formulated in the general context of entanglement wedge reconstruction. This question was originally discussed in the context of the resolution of the AMPS paradox. It was argued by Papadodimas and Raju \cite{Papadodimas:2012aq,Papadodimas:2013jku,Raju:2020smc} that the reconstruction of the black hole interior in the dual boundary theory has to be necessarily state-dependent and this can resolve the AMPS and related information paradoxes. The operators needed for the reconstruction of the interior should themselves be complicated and would be also microstate dependent. Its consistency with the framework of quantum mechanics has been debated, see \cite{Harlow:2014yoa} for an instance. We will review these issues pertaining to the AMPS paradox in the next section. In this section, we will present developments related to state-dependence of black hole interior reconstruction which has been reformulated in the framework of quantum error correction with crucial inputs from the extremal surfaces in \cite{Hayden:2018khn,Akers:2019wxj,penington2020entanglement} which is manifestly consistent with the basic postulates of quantum mechanics. We will also discuss how this formulation leads to the quantification of the complexity of the reconstruction of the interior \cite{Brown:2019rox,Engelhardt:2021qjs} that is necessary for the resolution of the AMPS and related paradoxes (more on this later). Note this discussion relates to the properties of the entanglement wedge and is also applicable to evaporating black holes.

\begin{figure}
   \centering
        \resizebox{0.6\textwidth}{!}{%
        \includegraphics{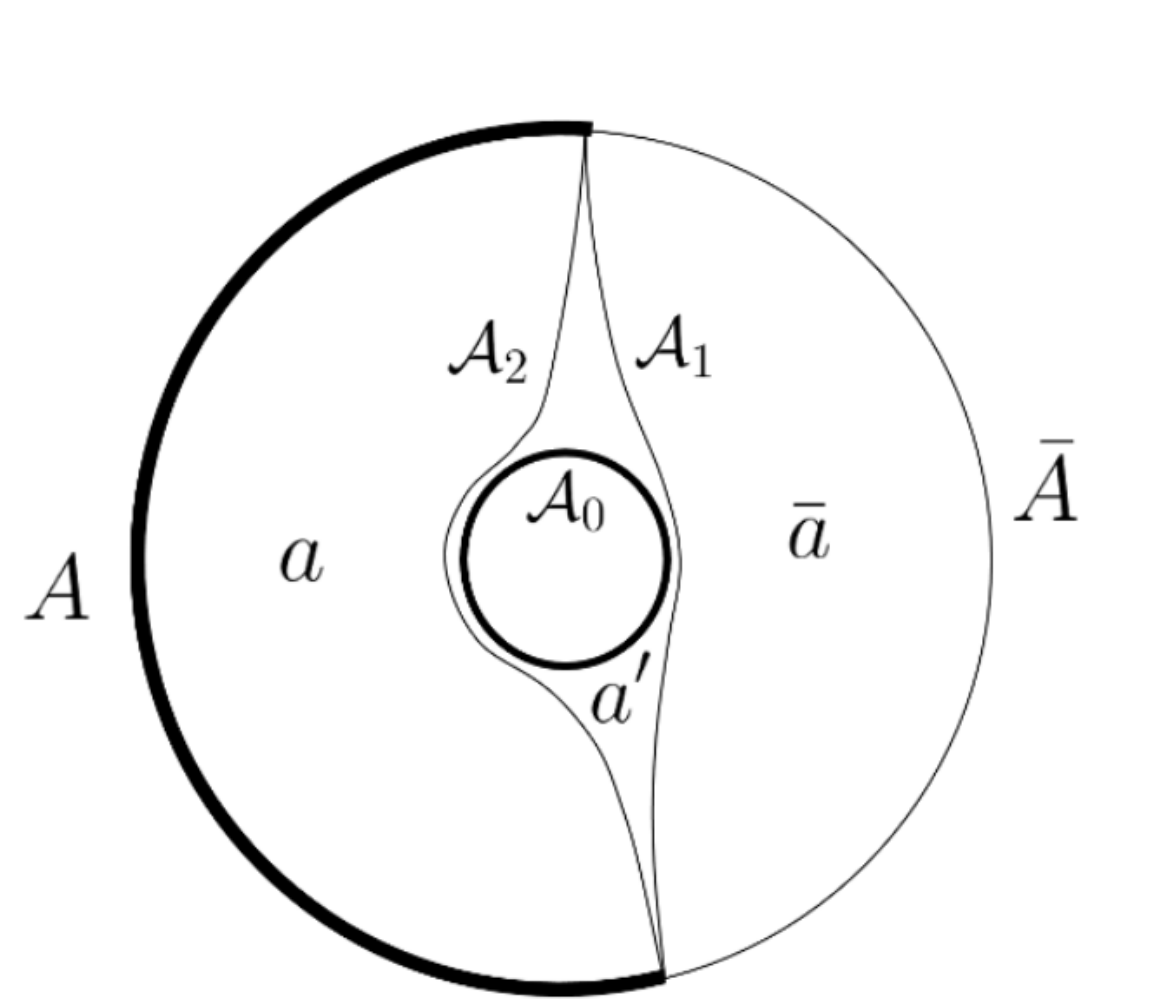}
        }
      \caption{In presence of a horizon, there are multiple extremal surfaces attached to the boundary $\partial A$ of a boundary subregion. As shown above, there are two with areas $\mathcal{A}_2$ and $\mathcal{A}_1$. If the boundary subregion inlcudes more than half of the full boundary, then $\mathcal{A}_2 > \mathcal{A}_1$. However, $\mathcal{A}_1$ is homologous to $\overline{A}$ and $\mathcal{A}_2$ is homologous to $A$. The homology constraint implies that the entanglement wedge of $A$ is $a$, the bulk region bounded by $A$ and $\mathcal{A}_2$ (and that of $\overline{A}$ is similarly $\overline{a}$) in the thermal state dual to a black hole. The bulk region $a'$ bounded by the two minimal surfaces, which includes portion of the exterior of the black hole, cannot be reconstructed at the boundary. As discussed in text, the situation is different in case of a microstate. Figure from \cite{Hayden:2018khn}.}
        
        \label{Fig:A1A2}       
    \end{figure}

The crucial point in \cite{Hayden:2018khn} is that state-dependence in the (approximate) reconstruction of the interior occurs when we consider a boundary sub-region instead of the whole dual system. Furthermore, state dependence actually implies dependence on the code subspace and not dependence on the specific bulk density matrix with support in this code subspace. Geometrically it originates from the dependence of the entanglement wedge on the choice of the code subspace and it is simply determined by the maximally mixed state in this code subspace. To illustrate this form of state-dependence in the reconstruction of the interior, consider a region $A$ at the boundary as shown in Fig. \ref{Fig:A1A2}. There are two extremal RT surfaces with areas $\mathcal{A}_1$ and $\mathcal{A}_2$ respectively. Here $\mathcal{A}_2$ is homologous to $A$ and $\mathcal{A}_1$ to its complement $\overline{A}$. If $A$ is larger than half the boundary, then  $\mathcal{A}_2 > \mathcal{A}_1$. The bulk is separated into three regions $a$, $a'$ and $\overline{a}$ where $a(\overline{a})$ is the region bounded by $\mathcal{A}_2(\mathcal{A}_1)$ and $A(\overline{A})$, while $a'$ is the region between the two minimal surfaces containing the black hole horizon with area $\mathcal{A}_0$ and portions of the black hole exterior also. Furthermore, $\mathcal{A}_2 - \mathcal{A}_1 < \mathcal{A}_0$ and so we can define
$$\alpha = \frac{\mathcal{A}_2 - \mathcal{A}_1 }{\mathcal{A}_0}$$so that $0\leq\alpha\leq 1$. The code subspace is 
$$\mathcal{H}_{\rm code}=\mathcal{H}_a  \otimes \mathcal{H}_{a'} \otimes \mathcal{H}_{\overline{a}}.$$In the semi-classical limit $G\rightarrow0$, the dimension of the full code subspace is $$ e^{S_{\rm BH}}= e^{\frac{\mathcal{A}_0}{4 G}}.$$

Consider the thermal state dual to the actual black hole geometry. Then the homology constraint implies that the entanglement wedge of $A$ is $a$ and that of $\overline{A}$ is $\overline{a}$. Any bulk operator $\mathcal{O}_{a'}$ localized in $\mathcal{H}_{a'}$ whether it is in the interior of the black hole or in the exterior cannot be reconstructed either in $\mathcal{H}_A$ or in $\mathcal{H}_{\overline{A}}$. This is not surprising at all. Firstly note that only $\mathcal{H}_{a'}$ which contains the horizon region will have dimension $\mathcal{O}(e^{1/G})$ whereas $\mathcal{H}_a$ and $\mathcal{H}_{\overline{a}}$ will have dimension $\mathcal{O}(1)$ in the limit $G \rightarrow0$. This implies that none of the microstates can be resolved as should be the case in the thermal ensemble. Only in the case of a microstate, we should be able to access (reconstruct) states in $\mathcal{H}_{a'}$ at the boundary.

To be precise, we assume that for any typical microstate we can consider the same semi-classical geometry up to where the surfaces $\mathcal{A}_1$ and $\mathcal{A}_2$ are located (indeed valid for fuzzballs which differ from the black hole significantly only at the scale of the horizon \cite{Mathur:2005zp}). This assumption is however not crucial for what follows -- it is enough if the location of $\mathcal{A}_1$ and $\mathcal{A}_2$ have sub-leading state dependence. Crucially, in a microstate (with a smooth geometry) there will be no homology constraint. For reasons which will be clear later, we should consider a mixed state in the bulk. In order to do this, let us entangle the black hole with a reference system $R$ (note this does not imply that they are in physical contact or $R$ is acting as a reservoir of Hawking quanta). Let $\vert \psi\rangle$ be a pure state in $\mathcal{H}_{\rm code}\otimes \mathcal{H}_R$. The question is then can we decode an operator $\mathcal{O}_{a'}$, localized in $\mathcal{H}_{a'}$, in $A$. It will be possible if the entanglement wedge of $A\cup R$ is $a\cup a'\cup R$. This would be so if $S_{gen}(\mathcal{A}_2)>S_{gen}(\mathcal{A}_1)$ implying that the quantum extremal surface corresponding to $A$ should be $\mathcal{A}_1$ instead of $\mathcal{A}_2$ for \textit{all} states $\vert \psi\rangle\in \mathcal{H}_{\rm code}\otimes \mathcal{H}_R$. Therefore, we need (using $S(aa')_\psi = S(\overline{a}R)_\psi$ for a pure state $\vert\psi\rangle$):
\begin{equation}\label{Eq:IneqEW}
   S(a)_\psi + \frac{\mathcal{A}_2}{4G}>S(aa')_\psi + \frac{\mathcal{A}_1}{4G} =S(\overline{a}R)_\psi + \frac{\mathcal{A}_1}{4G}.
\end{equation}
As discussed above, in the limit $G\rightarrow 0$, $S(a)_\psi$ is $\mathcal{O}(1)$. Also the triangle inequality implies $$\vert S(R)_\psi - S(\overline{a}R)_\psi\vert \leq S(\overline{a})_\psi = \mathcal{O}(1). $$So \eqref{Eq:IneqEW} reduces to the condition that if
\begin{equation}\label{Eq:IneqEW1}
    4 G S(R)_\psi < \mathcal{A}_2- \mathcal{A}_1,
\end{equation}
then the entanglement wedge of $A$ is $a\cup a'$. This requires that $\vert\psi\rangle$ belongs to a code subspace $\mathcal{H}_\mathcal{S}\subseteq \mathcal{H}_{\rm code}$ of dimension $d_S = d_R$ such that the above inequality is satisfied for any state in this subspace. The reconstruction criterion then arises from the maximally mixed state in the subspace for which $S(R)_\psi = \log d_S$ implying that
\begin{equation}
    d_S < e^{\frac{\mathcal{A}_2- \mathcal{A}_1}{4G}} = e^{\alpha S_{\rm BH}}.
\end{equation}
Irrespective of whichever state we choose in this subspace $\mathcal{H}_\mathcal{S}$, any bulk operator $\mathcal{O}_{a'}$ localized in $\mathcal{H}_{a'}$ can be reconstructed via the same operator $\mathcal{O}_A$ in $\mathcal{H}_A$. Nevertheless, this operator $\mathcal{O}_A$ will depend on the choice of the code subspace -- outside this subspace there exists states (eg. the fully entangled state of the black hole interior and $R$) the action of $\mathcal{O}_{a'}$ on which cannot be simulated by  $\mathcal{O}_A$ since the entanglement wedge of $A$ does not contain $a'$ for such states. The state-dependence of $\mathcal{O}_A$ is via the choice of this appropriate code subspace alone. As the size of $A$ grows, clearly $\alpha \rightarrow 1$. Therefore, the code subspace involves all typical states in the full Hilbert space and there is no state dependence in the reconstructed operators $\mathcal{O}_A$ which have the same action as $\mathcal{O}_{a'}$ acting in the interior of the black hole.

The quantum error correcting protocol which reproduces the above desired behavior is the universal subsystem recovery channel first considered in \cite{Hayden:2017xed}. Suppose there is a quantum channel that applies a Haar-random unitary $U$ to $n$ qubits, throws away a fraction of them and transmits the rest. Let the input Hilbert space be $\mathcal{H}_{\rm in}$. If one retains a fraction $(1+\alpha)/2$ of the qubits, then one will be able to decode (recover) any subspace in the input Hilbert space which has $\alpha n$ qubits (this subspace therefore has dimension $2^{\alpha n}$.) The universal subspace recovery map however can only approximately reverse the channel. Sometimes, there has to be an error $\epsilon > e^{-\eta n}$ with $\eta >0$. In the context of reconstruction of microstates this implies error
$$\epsilon > e^{-\eta/G} $$which should be typically non-perturbatively small in $G$. The explicit universal recovery map is closely related to the twirled Petz map \cite{Hayden:2018khn} which has tantalizing connections with the modular Hamiltonian as discussed before. The bulk region $a'$ between the minimal surfaces contains the $\alpha$-bits in the language of universal subspace error correction. 

Such issues in bulk reconstruction arise not only in the case of black holes but also when such competing extremal surfaces enclose bulk matter with high entropy -- we refer the reader to \cite{Akers:2019wxj,Akers:2020pmf} for extensive discussions. In particular \cite{Akers:2020pmf} discusses how the quantum extremal surface prescriptions should be refined in such generic circumstances with information-theoretic interpretations.

Intuitively it should be hard to reconstruct operators lying in $a'$ between the two minimal surfaces in Fig \ref{Fig:A1A2} from $A$ even within the code subspace essentially due to state-dependence. This is similar to the Hayden-Harlow conjecture of exponential complexity of reconstruction of operators in the interior of the black hole restricted to Hawking radiation to be discussed in the following subsection. 
A remarkable geometric way of capturing this complexity was proposed in \cite{Brown:2019rox} based on the generic existence of non-minimal extremal surfaces forming a Python-lunch geometry (sandwiched between minimal extremal surfaces or behind one of them) and interpreting this geometry in terms of tensor network constructions. This proposal further develops earlier proposals for the holographic dictionary for complexity \cite{Susskind:2014rva,Brown:2015bva,Brown:2015lvg}. For a recent review on quantum complexity and its holographic description see \cite{Shira_review}.

\begin{figure}
   \centering
        \resizebox{0.6\textwidth}{!}{%
        \includegraphics{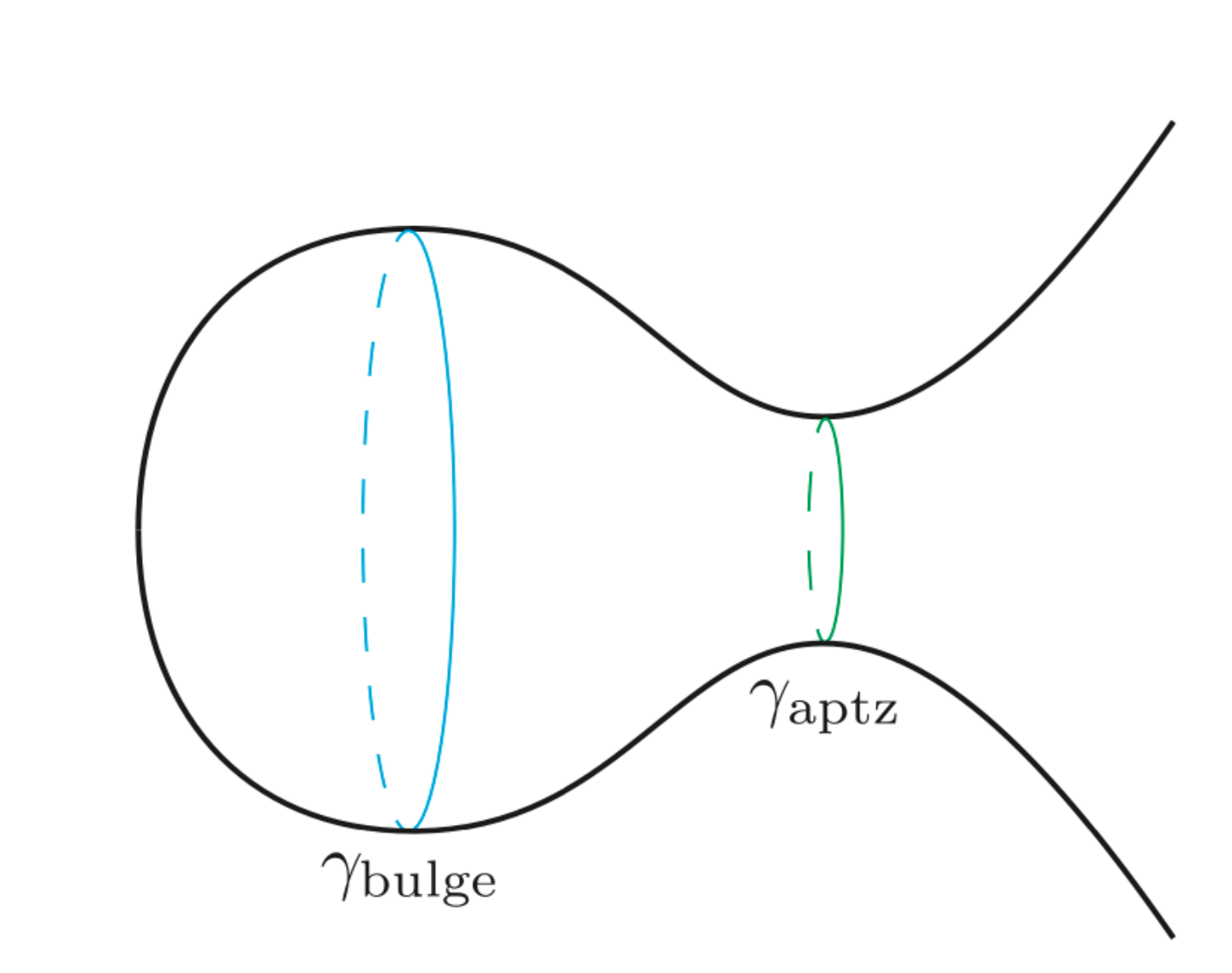}
        }
      \caption{In order to consider excited modes in the black hole interior, one first needs to restrict to a code subspace where the geometry will have a Python's lunch -- a non-minimal extremal surface $\gamma_{\rm bulge}$ behind the outermost minimal extremal surface $\gamma_{\rm aptz}$. The throat refers to the asymptotic region. There can be other locally minimal surfaces behind $\gamma_{\rm bulge}$ (not shown in figure). Figure from \cite{Engelhardt:2021qjs}.}
        
        \label{Fig:Python}       
    \end{figure}

The mechanism for nucleation for such Python lunches was proposed in \cite{Engelhardt:2021qjs}. Essentially the argument in the case of the black hole setup of Fig \ref{Fig:A1A2} is that we should first consider the appropriate code subspace where the interior can be decoded in $A$. In the maximally mixed state in this code-subspace the Hawking pairs will experience disentanglement and the bulk entropy gradients will be larger than the Hartle-Hawking state. Using results from \cite{Wall:2012uf,Marolf:2019bgj,Akers:2019lzs}, then it has been argued that due to blueshift of the entropy gradients when extrapolated in the past there would be nucleations of non-minimal (highly non-classical) extremal surfaces (since it can compete with the leading order $1/G$ term) behind the horizon ensuring that they do exist also at late time (this is somewhat similar to the existence of the transition in the extremal surfaces we have seen before in the context of reproduction of the Page curve). Even if one considers the full boundary, one needs to consider a code subspace for discussing the interior outgoing modes and then there will be such a bulge surface $\gamma_{\rm bulge}$ behind the outermost extremal surface $\gamma_{\rm aptz}$ (the appetizer of the Python's lunch) -- see Fig. \ref{Fig:Python}.  The complexity of the decoding (defined below) within this code subspace will then be half the difference in the generalized entropies of $\gamma_{\rm bulge}$ and $\gamma_{\rm aptz}$.

\begin{figure}
   \centering
        \resizebox{1.0\textwidth}{!}{%
        \includegraphics{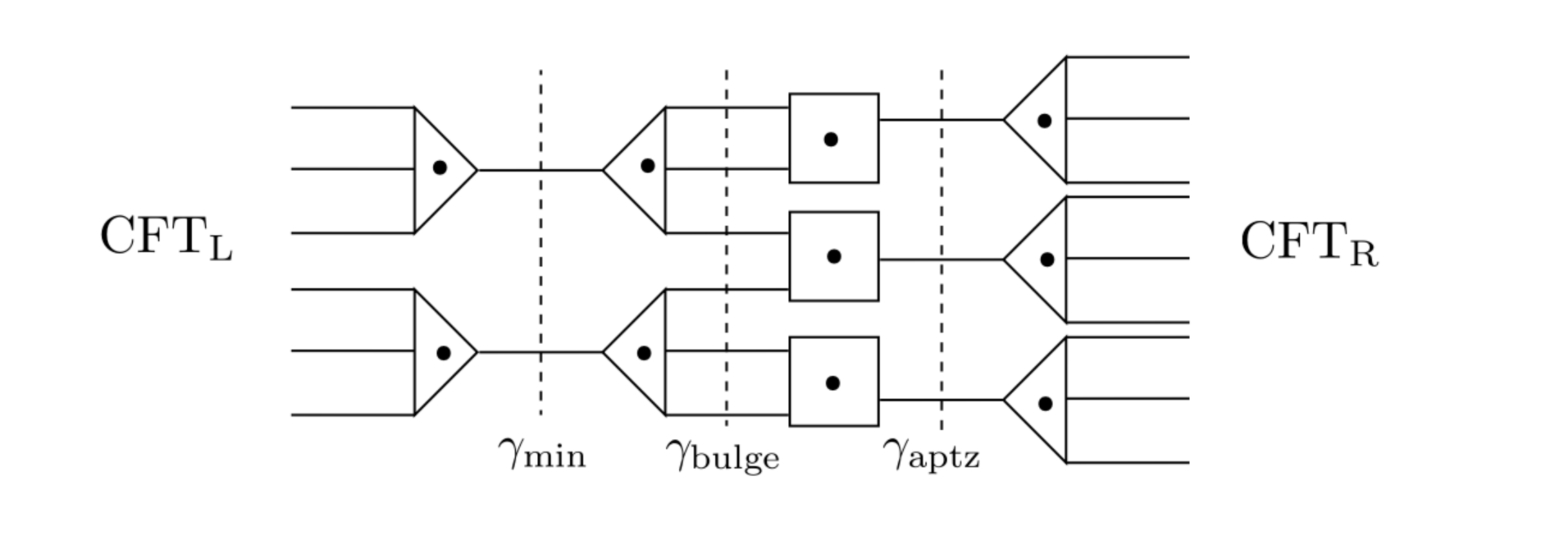}
        }
      \caption{A tensor network dual to a Python's lunch geometry such that the map from the input state in the left Hilbert space and the black dots representing bulk legs to the right Hilbert space is an approximate isometry. It has two locally minimal cuts $\gamma_{\rm aptz}$ and $\gamma_{\rm min}$ with $\gamma_{\rm aptz}> \gamma_{\rm min}$. Furthermore, there is a maximal cut $\gamma_{\rm bulge}$ between these two cuts. The triangles denote isometries whereas the squares have one or more of their legs projected to $\vert 0 \rangle$ state. Figure from \cite{Engelhardt:2021qjs}.}
        
        \label{Fig:TNPython}       
    \end{figure}

For the moment, let us assume that such a Python lunch geometry is generic when one considers appropriate code subspaces where interior modes are excited and can be defined in terms of the macroscopic geometry corresponding to the maximally mixed state in this code subspace. In this case, one can justify the quantification of the complexity of decoding the interior from tensor network models. Such models were also used to derive the holographic complexity conjectures \cite{Susskind:2014rva,Brown:2015bva,Brown:2015lvg} which need to be modified in the presence of the Python lunches. For an illustration consider such a wormhole with a Python lunch in the two-sided thermofield double geometry represented in the form of a tensor network in Fig. \ref{Fig:TNPython}. We discuss the circuit complexity of decoding following \cite{Engelhardt:2021qjs}.

The tensor network represents the map from the bulk represented as black dots in Fig \ref{Fig:TNPython} to the tensor product of the left and right CFT Hilbert spaces. Triangles represent isometries  and the squares involve \textit{postselection} -- an isometry where one (or more) of the legs is projected to $\vert 0\rangle$. In a generic tensor network of such type, there will be two locally minimal cuts $\gamma_{\rm min}$ (global minimum) and  $\gamma_{\rm aptz}$ (representing the locally minimal extremal surfaces) and a $\gamma_{\rm bulge}$ locally maximal cut in the middle. When we view the figure from the left to right it represents an approximate isometry from the tensor product of bulk legs and left CFT Hilbert space to the right CFT Hilbert space. For this to be the case, we need to assume that the bonds cut in $\gamma_{\rm aptz}$ have larger dimension than those in the $\gamma_{\rm min}$ cut plus the dimensions of the bulk legs (denoted as black dots) between these two cuts.  However, the problem is in the middle because the map from the $\gamma_{\rm bulge}$ cut plus the bulk legs in between the $\gamma_{\rm bulge}$ and the $\gamma_{\rm aptz}$ cut to the  $\gamma_{\rm aptz}$ cut is not an isometry as the Hilbert space dimension of the latter is smaller. To make it work in practice we need to somehow implement the postselection (which is not unitary as it involves projection) in a different way via unitary circuits. This can be done via Grover search algorithm in the manner discussed first in \cite{Yoshida:2017non} -- we need to do sequential unitary transformations such that we bring the qubits that are supposed to be postselected already in $\vert 0\rangle$ state. The complexity of the decoding of the Python lunch is essentially that of this Grover search algorithm. By general arguments this complexity is$$ C = O(\tilde{C} 2^{m/2})$$where $m$ is the number of qubits which should be postselected (equal to the difference between the bulge and appetizer cuts) and $\tilde{C}$ is related to the overall size of the network.\footnote{For a discussion on complexity utilizing the Petz map in the context of a subregion see \cite{Zhao:2020wgp}.}
    
Now in the gravitational analogue $\gamma_{\rm aptz}$ and $\gamma_{\rm min}$ are surfaces with locally minimal generalized entropies. By our earlier discussion the code subspace that can be reconstructed in the right Hilbert space has dimension $$ S_{gen}(\gamma_{\rm aptz})- S_{gen}(\gamma_{\rm min})$$with generalized entropies defined by considering the maximally mixed state in the code subspace. The condition $S_{gen}(\gamma_{\rm aptz})> S_{gen}(\gamma_{\rm min})$ is the reconstructibility criterion analogous to that for the tensor network to be an isometry to the right Hilbert space. Furthermore, $\gamma_{\rm bulge}$ has locally maximal generalized entropy and for the Python lunch geometry $S_{gen}(\gamma_{\rm bulge})> S_{gen}(\gamma_{\rm aptz})$. Then the tensor network analogy suggests complexity of decoding should be $$C= \mathcal{O}(\tilde{C}\exp\frac{1}{2}\left(S_{gen}(\gamma_{\rm bulge})- S_{gen}(\gamma_{\rm aptz}))\right).$$The exponential dependence in $G^{-1}$ comes crucially only from the difference of the generalized entropies while the dependence on $G^{-1}$ in $\tilde{C}$ (which is of same order in $G^{-1}$ as the generalized entropies) is only linear. So the former accounts for exponential complexity. A similar discussion can be repeated for a single sided microstate geometry shown in Fig \ref{Fig:Python} also. 
    
The understanding of the mechanism for generation of exponential complexity already alerts us towards the need for more explicit microstate geometries that can describe appropriate code subspaces. We will further motivate microstate models in the following subsection.

Interestingly a converse of the Python's lunch conjecture has been discussed in \cite{Engelhardt:2021mue}. It has been claimed that bulk operators between the boundary and the outermost extremal surface should have a simple reconstruction at the boundary in the sense that they can be recovered efficiently from a dual coarse-grained state with an effective local modular Hamiltonian. See also \cite{Levine:2020upy} for a related discussion. Recent discussions of the reconstruction of the experience of a bulk observer in the dual conformal field theory can be found in \cite{Jafferis:2020ora} via the use of modular flow and through emergent properties of von-Neumann algebras in \cite{Leutheusser:2021frk} .

\subsection{Decoding the interior in real time}
\subsubsection{Quantum information mirroring and the resolution of the AMPS paradox via complexity}\label{sec:AMPS}
Quantum information theory has been applied to understand how the black hole acts as a quantum channel with a motivation to resolve the AMPS paradox. Such analyses predict very non-trivial features of how the interior gets encoded into the outgoing Hawking quanta in real time. 

\begin{figure}
   \centering
        \resizebox{0.6\textwidth}{!}{%
        \includegraphics{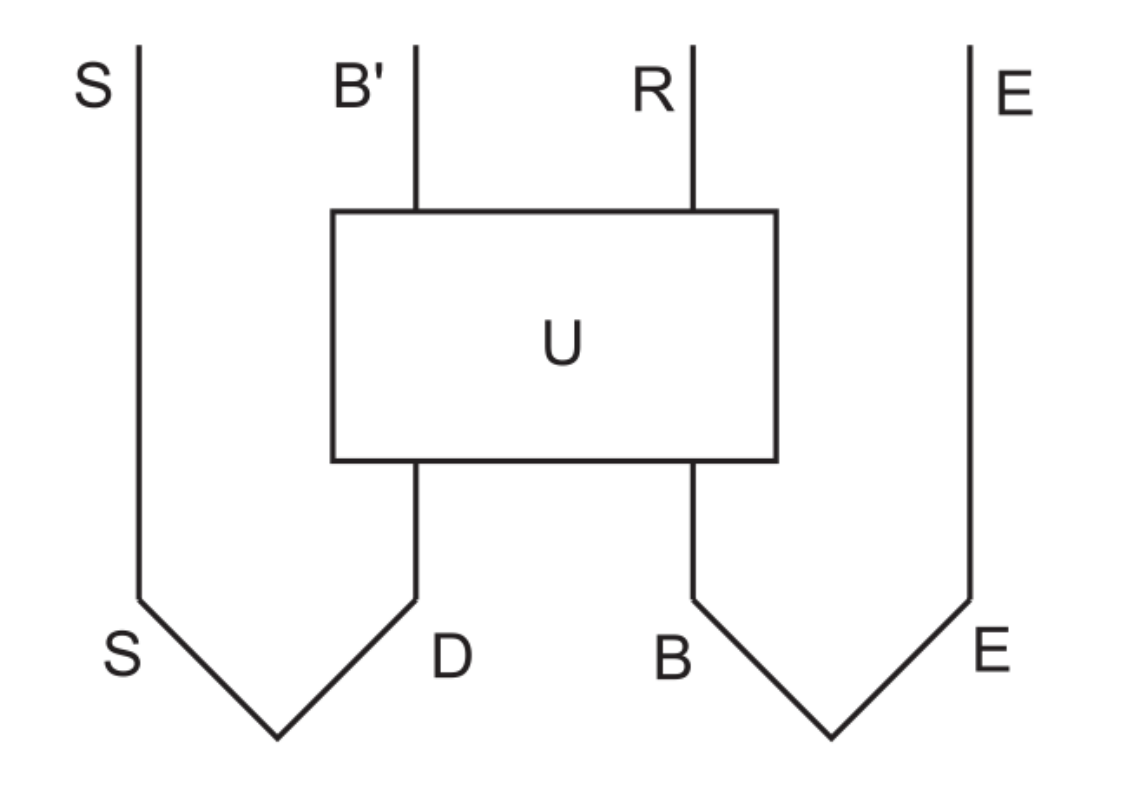}
        }
      \caption{An illustration of the Hayden-Preskill thought experiment which argues for quantum information mirroring in old black holes -- figure taken from \cite{Harlow:2014yka}. A quantum diary $D$ entangled with a reference state $S$ is thrown into an old black hole $B$ which is entangled with its radiation $E$. The diary $D$ and $B$ then gets scrambled by the unitary evolution operator $U$ after which we obtain the remaining black hole $B'$ and some more radiation $R$. If $R$ has only a few more qubits than $D$, then we can show that mutual information between $S$ and $B'$ should be negligible given that $U$ is random. The information in $D$ is then in the $E\cup R$ system after the scrambling time.}
        
        \label{Fig:Mirror}       
    \end{figure}

An extension of Page's argument implies that old black holes (past its Page time) would act as quantum information mirrors following a thought experiment and its analysis due to Hayden and Preskill \cite{Hayden_2007} as illustrated in Fig. \ref{Fig:Mirror}. A typical state of an old black hole would have Hawking quanta in the exterior ($E$) maximally entangled with the interior modes ($B$). If the black hole is a fast scrambler, then qubits ($D$) thrown into it would be maximally scrambled with $B$ (one could understand this as an action of a random unitary operator $U$\footnote{Hayden and Preskill showed that this unitary operator need not have exponentially large number of gates for a Page-like argument to work. It suffices to pick the unitary randomly from a unitary two-design, which can be achieved by a quantum circuit of depth $\mathcal{O}(\log n)$ where $n = S_{BH}$ is the total number of qubits at the horizon with $S$ being the black hole entropy. If each step in the circuit takes Plank time, then it should be redshifted to $r_s$ in the time of the asymptotic observer. It then reproduces the scrambling time as the circuit time which is $r_s \log S_{BH} $. This argument was refined in \cite{Sekino:2008he}}) resulting in a remnant $B'$ and newly radiated Hawking quanta $R$. For a formal statement, we consider that $D$ is maximally entangled with a reference system $S$, so that post-scrambling an extension of Page's arguments would imply that $S$ and $B'$ have no mutual information (entanglement). However, the information in $D$ would be transferred to the combined $R\cup E$ system of Hawking quanta implying information mirroring which would essentially happen at scrambling time. The latter has been computed based on growth of thermal commutators by Shenker and Stanford \cite{Shenker:2013pqa}. In the context of black holes, it is essentially the time (from the point of view of an asymptotic observer) it takes a light ray to reach Planck distance close to the stretched horizon, which is $\approx r_s \log S_{BH}$, with $S_{BH}$ being the entropy of the black hole. For a detailed review of the Hayden-Preskill protocol and progress on experimental realization of quantum simulators which achieves the scrambling needed to realize it see \cite{Lata_review}.

The natural question that arises is how easy it would be to decode the qubits in $D$ from $E\cup R$. Before considering this question we need to encounter first  the most pragmatic operational way to resolve the AMPS paradox that leads to an additional feature of the encoding of the interior into the Hawking quanta. In order to present this, it is useful to restate the AMPS paradox. Once again consider the old black hole for which the interior modes $B$ should be maximally entangled with $E$, the already radiated quanta. Unitarity of the time-evolution would imply that the newly emitted quanta $R$ that decouples from the black hole can be purified by a factor $E_R$ of $E$ so that $\rho_{RE_R}$ is a pure state. However, $R$ must also be maximally entangled with (a factor of) $B$ if semiclassical EFT holds at the horizon. A simple way to resolve this paradox is to invoke the \textit{philosophy} of black hole complementarity \cite{PhysRevD.48.3743,PhysRevD.50.2700} which postulates that the violation cannot be observed \textit{operationally} via a quantum complexity conjecture due to Harlow and Hayden \cite{Harlow:2013tf} (see \cite{Aaronson:2016vto} for a nice discussion). Essentially this argument states that in order to \textit{distill} the factor $E_R$ from $E$ needed to purify $R$, it would take time which is exponential in the entropy of the black hole (at the time when $R$ was emitted). This conjecture can be made more precise by evoking pseudorandom encoding of the interior into the outgoing Hawking quanta \cite{Kim:2020cds} (in microstate models we will be able to relate inherent pseudorandom/chaotic dynamics with Python lunches that can macroscopically amplify small excitations). Since the black hole evaporates in time that is polynomial in its entropy, the violation of monogamy of entanglement cannot be demonstrated \textit{operationally}.
 
 This discussion raises a few fundamental questions. Firstly, what exactly does an operational resolution mean? Does it mean we need to modify the framework of quantum mechanics actually to describe black holes although we cannot operationally test violation of its postulates? Or does it mean that somehow these Hilbert spaces, especially $B$ (the interior) and $E$ (the pre-Page time Hawking quanta) are actually not separable but only in an operational sense? Then in this operational framework would the AMPS paradox be resolved by complexity? The Euclidean replica wormhole saddles which connect $E$ and $B$ would support the latter point of view. However, could we actually understand how to make sense of such an operational framework in real time? The latter is challenging as we would also need to validate the usual semi-classical picture of the black hole (at least from the point of view of measurements of the EFT observables). In fact, the Page curve can be computed using only the semi-classical geometry as discussed before and therefore it should be valid for a large class of measurements.
 
 The second class of issues are related with the encoding of the information in Hawking radiation. If rapid mirroring of the information thrown in after Page time could happen together with the complex encoding of the Hawking interior, then which of these two possibilities could be true: (a) one can decode the qubits $D$ thrown into the black hole from the Hawking quanta immediately after scrambling time without explicit knowledge of the interior encoded in $E$ (decoding the latter would take time exponential in the entropy of the black hole), or (b) we can only decode it after we actually know the interior after an enormously long time. The discussion on islands in the form of entanglement wedge of $R$ would actually prefer the first possibility. In the setups discussed in the previous subsection, the newly emitted Hawking quanta would be in the entanglement wedge of the holographic system $B$ and not yet in the bath region. The physical separation between these regions would indicate that the mirrored information could be readily decoded soon after the new Hawking quanta emerges in the entanglement wedge of $B$. However, one still needs to demonstrate that the decoding is possible without the knowledge of the island (the black hole interior) that is encoded in this new radiation since the knowledge of how the island has been modified by the infalling bits should eventually leak into the bath. Therefore, one needs to identify such features of the encoding of the infalling qubits which co-exist with the complex encoding of the interior, and explain their physical origins. One could formulate these questions also more generally without the setups of the previous subsection by advocating an emergent infrared holographic theory that describes the near-horizon geometry of the black hole and which could play the role of $B$.
 
 In the next subsection, we will review a microstate model that could be promising for finding answers to such questions. 
 
 \subsubsection{Microstate dynamics: Towards understanding quantum black holes in real time}\label{sec:microstate}
\begin{figure}
   \centering
        \resizebox{1.0\textwidth}{!}{%
        \includegraphics{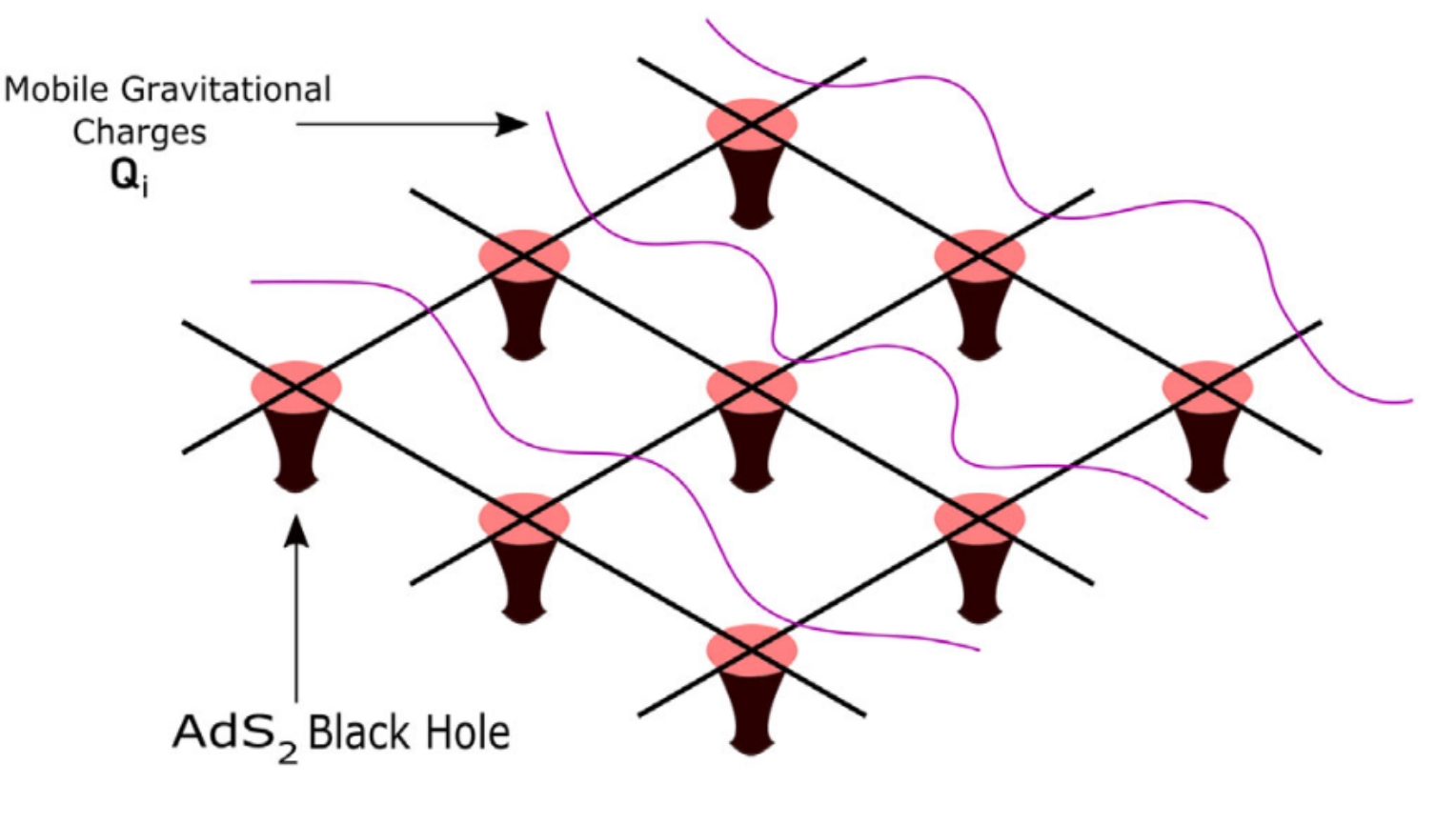}
        }
      \caption{An illustration of the microstate models comprising of a lattice of two-dimensional $AdS_2$ throats coupling to each other via mobile hair carrying gravitational charges. The lattice is a discretization of the horizon, a compact space. The $AdS_2$ throats may join in the interior forming complex networks. Hawking evaporation occurs due to appropriate asymptotic boundary conditions at the throats. This figure is from \cite{PhysRevD.102.086008}.}
        
        \label{Fig:Microstate}       
    \end{figure}

 It is useful to study tractable models of black hole microstate dynamics to gain insights into how the features of quantum information mirroring and complex encoding of interior appear in the Hawking radiation, and also for understanding how the principle of black hole complementarity and the averaging implied in replica wormholes could emerge operationally. The fuzzball program \cite{Mathur:2005zp}, if developed to its full potential, would be able to reveal these mysteries. However, at present, it will be very difficult to study the quantum dynamics of fuzzballs in a sufficiently detailed way. The same could be said about large $N$ BFSS matrix models \cite{Banks:1996vh}, etc. We will argue that certain simplified models could be promising for understanding many (if not all) aspects of these issues. 
 This class of models described in \cite{PhysRevD.102.086008} essentially simplify some of the otherwise untractable aspects of quantum gravity in a suitable way, and can be studied in the same spirit as the setups described earlier in this review. One key aspect of this class of models is to give prominence to hair degrees of freedom on the horizon which are essentially (approximately) conserved (non-)gravitational charges whose role in the information paradox have been emphasised in \cite{PhysRevLett.116.231301,Strominger:2017aeh}. The crucial and novel aspect of the hair that will be of interest to us is how it can enable an operational definition of separability of the interior and exterior Hilbert spaces while resolving AMPS type paradoxes that threaten black hole complementarity, and enable mechanisms for the complex encoding of interior, and quantum information mirroring where decoding could be possible without the knowledge of the encoding of the interior.  
 \newline \newline
\textbf{A class of models:}
The setup of these microstate models can be motivated from the fragmentation instability \cite{Maldacena:1998uz} of the near-horizon geometry of the near-extremal black hole mediated by the Brill instantons of semi-classical gravity. These imply the fragmentation of the near-horizon geometry into several two-dimensional throats of the type $AdS_2 \times X$, where $X$ is a compact space that has the topology of the horizon and $AdS_2$ is the two-dimensional anti-de Sitter space. At the boundaries of the instanton moduli space where the centers of some of the throats come within Planckian distance proximity to each other, there is a proliferation of soft modes. However, quantum gravity effects also become large, so the semi-classical computation becomes intractable. Heuristically, we can assume that these fragmented throats \textit{crystallise} into a stable configuration forming a lattice of $AdS_2$ spaces as shown in Fig. \ref{Fig:Microstate}. These throats should interact with each other via mobile hair which are the (non-)gravitational charges of the original unfragmented part of the geometry and other soft modes that live here. The lattice is simpy a discretization of the compact $X$ space that has the same topology of the horizon. Additionally, we can allow the $AdS_2$ spaces to join in their interiors and form complex networks. Here, we will restrict to the simplest version where the $AdS_2$ throats do not join with each other and are infinitely extended in the ingoing Eddinton-Finkelstein radial coordinate.

 For further tractabilibity and simplification, we consider that each $AdS_2$ throat in the lattice is described a by semiclassical JT-gravity theory with conformal bulk matter. The semi-classical picture is thus valid for the observer infalling at any throat, however non-local measurements will see struture at the horizon. We will soon discuss how the energy absorbing and relaxation dynamics of the semi-classical black hole emerges from this microstate model. We begin our discussion by first arresting Hawking radiation by enforcing the usual reflecting boundary conditions in each $AdS_2$ throat so that we can study the intrinsic properties of the microstates first. Note that we retain only a coarse-grained (effective infrared) description of each throat which is dual to a quantum dot in the Planckian lattice. Regardless, we should enforce total energy conservation of this combined lattice of quantum dots and mobile hair system in absence of Hawking radiation. Although, we do modify the effective description of horizon physics, usual statistical arguments would imply that the results of typical measurements would be almost the same as in a microcanonical ensemble which we will describe below. For simplicity, we will consider the horizon to be $S^1$, so the lattice would be a ring (chain) with periodic boundary conditions.
 
 The state of the $i$-th throat (dual to the $i$-th quantum dot) can then be described by $t_i(u)$ where $u$ is the time of the asymptotic observer. Essentially $t_i(u)$ is the time-reparametrization mode of the $i$-th throat which determines the correlation functions in the localized state in terms of the vacuum correlation functions as in the Sachdev-Ye-Kitaev model \cite{Sachdev_1993,Kitaev_2015} which could be the (UV complete) theory governing the quantum dots. In other words, $t_i(u)$ is the time determining the state of the $i$-th throat as a function of a common vacuum state time which is identified with the time of the vacuum observer. For ease of simulations, it is useful to define $\tau_i(u)$ via
 \begin{equation}
     t_i(u) = \tanh\left(\frac{\tau_i(u)}{2}\right)
 \end{equation}
 where $\tau_i(u)$ maps to the common time of a thermal state (instead of the vacuum state) with $\beta = 2\pi$. In absence of coupling between throats and sources for the bulk matter, each individual throat should have conserved $SL(2,R)$ charges which are:
 \begin{eqnarray}
 \mathcal{Q}_i^0 &=& \frac{\tau_i'''}{\tau_i'^2} - \frac{{\tau_i''}^2}{{\tau_i'}^3} - \tau_i',\nonumber\\
 \mathcal{Q}_i^+ &=& e^{\tau_i}\left(\frac{\tau_i'''}{\tau_i'^2} - \frac{{\tau_i''}^2}{{\tau_i'}^3} - \frac{\tau_i''}{\tau_i'}\right), \nonumber\\
  \mathcal{Q}_i^- &=& e^{-\tau_i}\left(\frac{\tau_i'''}{\tau_i'^2} - \frac{{\tau_i''}^2}{{\tau_i'}^3} + \frac{\tau_i''}{\tau_i'}\right).
 \end{eqnarray}
 We denote these collectively as $\vec{\mathcal{Q}}_i$. The Casimir of these $SL(2,R)$ charges is the Arnowitt-Deser-Misner (ADM) mass $M_i$ of the $AdS_2$ throat, i.e.
 \begin{equation}
     M_i= {\mathcal{Q}_i^0}^2 - \mathcal{Q}_i^+  \mathcal{Q}_i^- =- 2\, {\rm Sch}(\tau_i(u), u) + {\tau_i'}^2,
 \end{equation}
 where $\rm{Sch}$ denotes the Schwarzian derivative
 \begin{equation}
     {\rm Sch}(f(u), u) = \frac{f'''}{f'} - \frac{3}{2}\frac{f''^2}{f'^2}.
 \end{equation}
 On top of these lattice $SL(2,R)$ charges, we need to consider additionally mobile hair charges which we can take to be $SL(2,R)$ charges too representing the gravitational charges of the unfragmented geometry. These we denote as $\vec{\mathsf{Q}}_i$ which follow discretized Klein-Gordon type equation when decoupled from the lattice charges. We denote $\vec{A}\cdot \vec{B}$ as the $SL(2,R)$ invariant dot product of two $SL(2,R)$ vectors $\vec{A}$ and $\vec{B}$.\footnote{Explicitly, $$\vec{A}\cdot \vec{B} = A^0 B^0 - \frac{1}{2}(A^+ B^- + A^- B^+).$$} The simplest equations of motion of this system of gravitational lattice charges $\vec{\mathcal{Q}}_i(u)$ and mobile hair $\vec{\mathsf{Q}}_i(u)$, which reproduce desired phenomenological properties of a classical black hole, take the form:
 \begin{eqnarray}\label{Eq:LatticeEoms}
 M_i' &=&  - \lambda(\vec{\mathcal{Q}}_{i-1}+\vec{\mathcal{Q}}_{i+1}- 2\vec{\mathcal{Q}}_{i})\cdot \vec{\mathsf{Q}}_{i}', \nonumber\\
 \vec{\mathsf{Q}}_{i}'' &=& \frac{1}{\sigma^2}(\vec{\mathsf{Q}}_{i-1}+\vec{\mathsf{Q}}_{i+1}- 2 \vec{\mathsf{Q}}_{i})\nonumber\\
 && + \frac{1}{\lambda^2}(\vec{\mathcal{Q}}_{i-1}+\vec{\mathcal{Q}}_{i+1}- 2\vec{\mathcal{Q}}_{i}).
 \end{eqnarray}
 where $\lambda >0$ is the coupling between lattice charges and hair, and $\sigma$ determines the velocity of propagation of hair in the continuum limit. The first equation above gives the equations for evolution of $t_i(u)$ and the second equation determines the evolution of the hair. These equations have unique solutions provided we specify the initial lattice charges, and the hair charges and their time-derivatives (we assume initial synchronicity, i.e. $t_i(u= u_0) = u_0$ at initial time $u_0$ but this is not necessary) and can be solved numerically following \cite{PhysRevD.101.066001}. Generalizations of the above equations with higher derivative corrections and desired phenomenological features are possible but we do not discuss them here. These equations imply a specific form of null matter in the $AdS_2$ throats, but alternatively we can simply think of these as the equations determining the quantum dots and hair. Crucially, in the full interacting system we have only one global $SL(2,R)$ symmetry, namely that of the original unfragmented geometry.
 
 Note that we should treat the lattice charges $\vec{\mathcal{Q}}_{i}$ semi-classically invoking a large-$N$ limit in each throat, but the hair $\vec{\mathsf{Q}}_{i}$ should be understood as an open quantum system interacting with the lattice. However, for present purposes it will be sufficient to consider coherent states of the hair and treat it classically. The equations \eqref{Eq:LatticeEoms} imply that the full system has a conserved energy of the form:
 \begin{equation}\label{Eq:Energy}
 \mathcal{E} = \mathcal{E}_\mathcal{Q} + \mathcal{E}_\mathsf{Q}
 \end{equation}
 which is simply a sum of the energy in the lattice charges and that in the hair with
 \begin{eqnarray}\label{Eq:EnergyBreak}
 \mathcal{E}_\mathcal{Q} &=& \sum_i M_i = \sum_i \mathcal{Q}_i\cdot \mathcal{Q}_i,\nonumber\\
 \mathcal{E}_\mathsf{Q} &=& \frac{\lambda^3}{2} \sum_i \vec{\mathsf{Q}}_i'\cdot \vec{\mathsf{Q}}_i' \nonumber\\&&
 +\frac{\lambda^3}{2\sigma^2} \sum_i (\vec{\mathsf{Q}}_i-\vec{\mathsf{Q}}_{i-1})\cdot( \vec{\mathsf{Q}}_i-\vec{\mathsf{Q}}_{i-1}).
 \end{eqnarray}
Thus $\mathcal{E}_\mathcal{Q}$ is simply the sum of the ADM masses of the throats and $ \mathcal{E}_\mathsf{Q}$ is the (discretised) kinetic energy of the hair. Note that $\lambda > 0$ is necessary for the positivity of the average energy in the microcanonical ensemble as we will see below.
\newline \newline
\textbf{The microstates:} The microstates of the black hole can be identified with stationary solutions of \eqref{Eq:LatticeEoms}. One can readily prove that in such microstates, we should have
\begin{equation}
    \vec{\mathcal{Q}}_i = Q \vec{\xi} + \vec{\mathcal{Q}}_i^\perp, \quad {\rm with} \quad \vec{\mathcal{Q}}_i^\perp\cdot \xi = 0.
\end{equation}
The $SL(2,R)$ charge vector $\vec{\xi}$ thus spontaneously breaks the global $SL(2,R)$ symmetry. We can set this global frame $\vec\xi$ in the $0$-direction without loss of generality so that $\vec{\mathcal{Q}}_i^\perp$ have only $+$ and $-$ components. We also normalize $\vec\xi$ such that $\vec\xi\cdot\vec\xi=1$.  Furthermore, for stationarity we need the hair charges to have the following configuration:
\begin{equation}\label{Eq:hair-split}
    \vec{\mathsf{Q}}_i(u) = \vec{\mathsf{Q}}_i^{\rm loc} + \vec{\mathsf{Q}}_i^{\rm mon}(u) + q_i^{\rm rad}(u) \vec{\xi}
\end{equation}
 where
 \begin{equation}\label{Eq:HairLocked}
     \vec{\mathsf{Q}}_i^{\rm loc} = - \frac{\sigma^2}{\lambda^2} \vec{\mathcal{Q}}_i + \vec{\mathcal{K}},
 \end{equation}
 are \textit{locked} to the lattice charges,
 \begin{equation}
     \vec{\mathsf{Q}}_i^{\rm mon} = \alpha \vec{\xi} u
 \end{equation}
 is a monopole component with a homogeneous $\vec{\mathsf{Q}}_i' = \alpha \xi$ in the direction of the global frame, and $q_i^{\rm rad}(u)$ are oscillating hair components \textit{decoupled} from the lattice and satisfying the normal mode equations
 \begin{equation}\label{Eq:Normal-Modes}
     {q_i^{\rm rad}}'' = \frac{1}{\sigma^2}(q_{i-1}^{\rm rad}+q_{i+1}^{\rm rad}- 2 q_{i}^{\rm rad}).
 \end{equation}
 To avoid redundancy, we can set
 \begin{equation}
     \sum_i q_i^{\rm rad}=\sum_i {q_i^{\rm rad}}' = 0.
 \end{equation}
 Crucially the Fourier transform of $q_i^{\rm rad}(u)$ will have support only on the discrete normal modes with non-vanishing frequencies. 
 
 It follows from \eqref{Eq:LatticeEoms-2} that $\sum_i \vec{\mathsf{Q}_i'}$ is conserved, and therefore $\alpha$, the monopole charge, should not change when the microstate is perturbed.
 
 Finally, we can define the microcanonical ensemble as the collection of microstate solutions subject to the constraints, that (i) the total $M_i$ should add up to the ADM mass of the black hole, i.e. $\mathcal{E}_\mathcal{Q} = M$ , and (ii) $t_i(u)$ and hence $\tau_i(u)$ should be real, continuous and have continuous first and second derivatives at all lattice sites (necessary to define the mass $M_i$ that is proportional to the Schwarzian derivative). This implies two set of possibilities. Firstly let's set $\xi^0 = 1$ and $\xi^\pm = 0$ without loss of generality as mentioned before. Then the first set of possibilities which satisfy both constraints are those which satisfy $0 \leq M_i \leq Q^2$, $Q>0$, $\sum_i M_i = M$ and
 \begin{align}
 &\mathcal{Q}_i^0 = - Q, \quad \mathcal{Q}_i^+ = - \rho_i \sqrt{Q^2 - M}, \nonumber\\ &\mathcal{Q}_i^- = - \frac{1}{\rho_i} \sqrt{Q^2 - M}
 \end{align}
 with
 \begin{align}
     \sqrt{\frac{Q - \sqrt{M_i}}{Q + \sqrt{M_i}}}\leq \rho_i \leq \sqrt{\frac{Q + \sqrt{M_i}}{Q - \sqrt{M_i}}}.
 \end{align}
 Remarkably, these imply that $t_i' \geq 0$ (and hence $\tau_i' \geq 0$), i.e. the arrows of time of all the lattice sites should be aligned towards the future. Thus a global arrow of time is a consequence of the equations of motion! The second set of possibilities lead to similar inequalities which align the global arrow of time towards the past. We discard this set and define the microcanonical ensemble with the mentioned set of (in)equalities.
 
 In the microstate solutions, the total energy of the hair $\mathcal{E}_\mathsf{Q}$ further splits into three parts, i.e.
 \begin{equation}
     \mathcal{E}_\mathsf{Q} = \mathcal{E}_\mathsf{Q}^{\rm pot} + \mathcal{E}_\mathsf{Q}^{\rm mon} + \mathcal{E}_\mathsf{Q}^{\rm rad}.
 \end{equation}
 with (setting $\xi^0 = 1$ and $\xi^\pm = 0$ without loss of generality as mentioned before )
 \begin{eqnarray}
 \mathcal{E}_\mathsf{Q}^{\rm pot} &=& -\frac{\sigma^2}{2\lambda} \sum_i (\mathcal{Q}_i^+ -\mathcal{Q}_{i-1}^+) 
 (\mathcal{Q}_i^- - \mathcal{Q}_i^-),\nonumber\\
 \mathcal{E}_\mathsf{Q}^{\rm mon} &=&\frac{1}{2}\lambda^3 \alpha^2, \nonumber\\
 \mathcal{E}_\mathsf{Q}^{\rm rad} &=& \frac{\lambda^3}{2}\sum_i {{q_i^{\rm rad}}'}^2+\frac{\lambda^3}{2\sigma^2}\sum_i(q_i - q_{i-1})^2.
 \end{eqnarray}
 Clearly, if $\lambda > 0$, then both $\mathcal{E}_\mathsf{Q}^{\rm mon}$ and $\mathcal{E}_\mathsf{Q}^{\rm red}$ are positive. Although $\mathcal{E}_\mathsf{Q}^{\rm pot}$ need not be positive, its ensemble average is zero. Thus the average energy is positive if $\lambda > 0$.

\begin{figure}
\centering
\subfigure[Evolution of lattice charges and masses after a single shock]{\includegraphics[width = 3.2in]{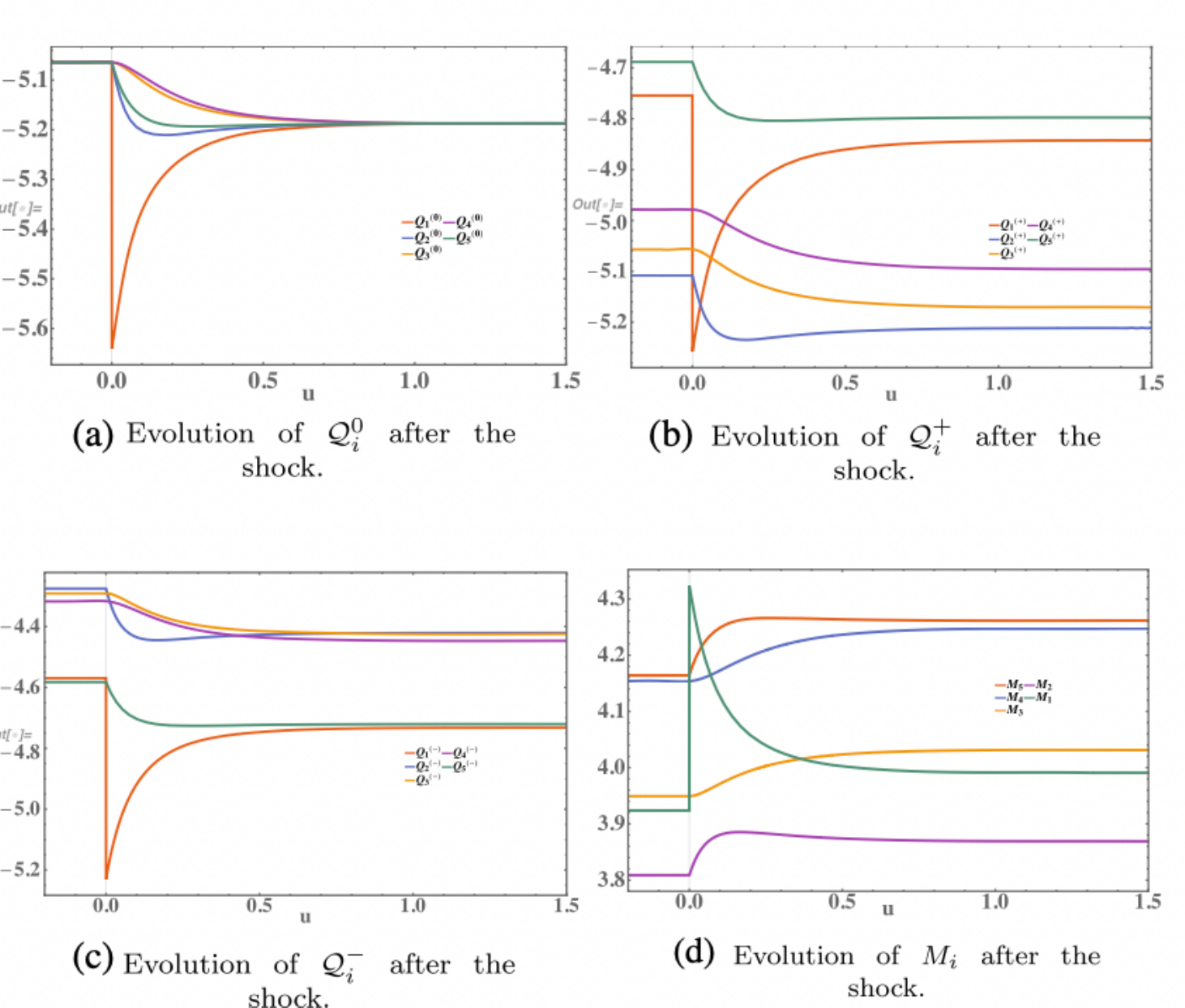}}\\
\subfigure[Evolution of the energies after a single shock]{\includegraphics[width = 2.5in]{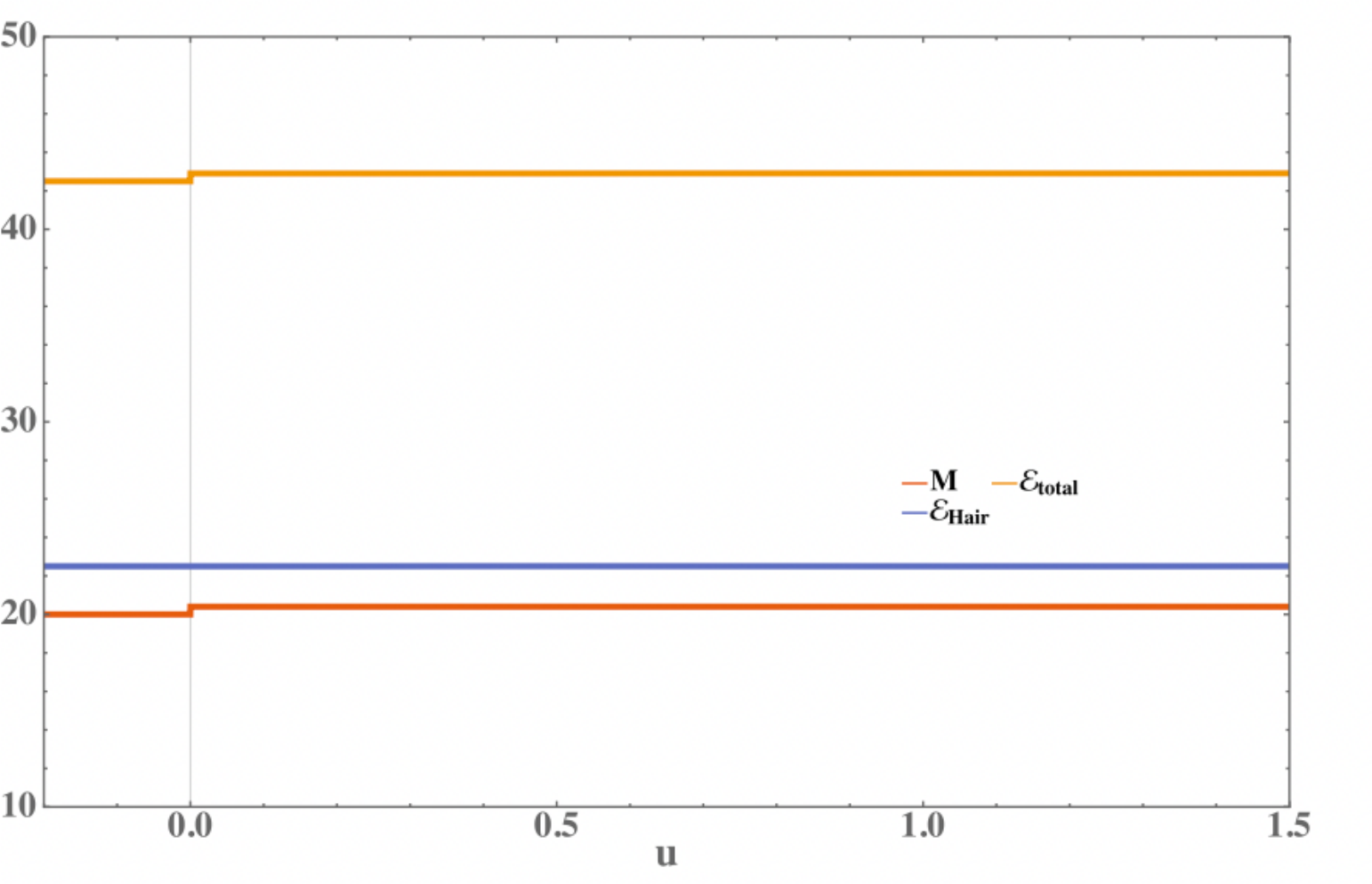}}
\caption{The evolution of a randomly chosen microstate after a single shock in a five site model is shown above. The first site is shocked with energy injection $e = 0.4$ at $u =0$. We set $\lambda =1$, $\alpha =1$ and $\sigma = 0.01$. The system relaxes to another microstate with a different distribution of masses and lattice charges. Almost the entire energy of the shock is absorbed by the total mass of the black hole $\mathcal{E}_\mathcal{Q}$. The energy in the hair $\mathcal{E}_\mathsf{Q}$ remains approximately constant. These figures are from \cite{PhysRevD.102.086008}. }\label{Fig:ShockedMicrostate}
\end{figure}

 Both classically and quantum-mechanically, the hair cannot be separated into interior and exterior components. However, in microstate solutions such a split operationally emerges since $\vec{\mathsf{Q}}_i^{\rm loc}$ is locked with the lattice $SL(2,R)$ charges which describe the configuration of the black hole interior, while both $\vec{\mathsf{Q}}_i^{\rm mon}$ and $\vec{\mathsf{Q}}_i^{\rm rad}$ do not affect the interior and are thus decoupled from it. The potential energy term $\mathcal{E}_\mathsf{Q}^{\rm pot}$ is determined solely by $\vec{\mathsf{Q}}_i^{\rm loc}$, while $\mathcal{E}_\mathsf{Q}^{\rm mon}$ and $\mathcal{E}_\mathsf{Q}^{\rm rad}$ are determined by $\vec{\mathsf{Q}}_i^{\rm mon}$ and $\vec{\mathsf{Q}}_i^{\rm rad}$ respectively. Therefore, the total energy also splits into an interior component which is the sum of $\mathcal{E}_\mathcal{Q} = M$ and $\mathcal{E}_\mathsf{Q}^{\rm pot}$, and the exterior component which is the sum of $\vec{\mathsf{Q}}_i^{\rm mon}$ and $\vec{\mathsf{Q}}_i^{\rm rad}$. Operationally, therefore, the Hilbert space of the hair has the structure
 $$\mathcal{H}_{\mathsf{Q}} = \bigoplus_\alpha\mathcal{H}_{\mathsf{Q}_\alpha}^{\rm int} \otimes \mathcal{H}_{\mathsf{Q}_\alpha}^{\rm ext}$$with $\alpha$ denoting microstates assuming that the evolution is adiabatic (we can employ an adiabatically evolving basis) with the off-diagonal terms being suppressed due to decoherence. Note that this would imply that the global frame $\vec\xi$ which is crucial to make the distinction between the interior and exterior would be evolving adiabatically too. We would come back to this in the context of the Hawking evaporation with asymptotic boundary conditions at the throats that allow the Hawking quanta to escape.
 \newline \newline
\textbf{Phenomenological viability:} The immediate question is that whether the ensemble of microstates in the model behave like a semi-classical black hole with relaxing and energy-absorbing properties. To see this, we once again return to the semi-classical limit in which Hawking radiation is absent, and perturb an arbitrary microstate solution in the ensemble by a sequence of shocks (injections of energies) $e_{i A}$ with $i$ denoting the $i$-th throat and $A$ referring to the instant $u_A$. The equations in \eqref{Eq:LatticeEoms} are then modified to
 \begin{eqnarray}\label{Eq:LatticeEoms-2}
 M_i' &=&  - \lambda(\vec{\mathcal{Q}}_{i-1}+\vec{\mathcal{Q}}_{i+1}- 2\vec{\mathcal{Q}}_{i})\cdot \vec{\mathsf{Q}}_{i}', \nonumber\\&& + \sum_{A} e_{iA}\delta(u - u_A),\nonumber\\
 \vec{\mathsf{Q}}_{i}'' &=& \frac{1}{\sigma^2}(\vec{\mathsf{Q}}_{i-1}+\vec{\mathsf{Q}}_{i+1}- 2 \vec{\mathsf{Q}}_{i})\nonumber\\
 && + \frac{1}{\lambda^2}(\vec{\mathcal{Q}}_{i-1}+\vec{\mathcal{Q}}_{i+1}- 2\vec{\mathcal{Q}}_{i}).
 \end{eqnarray}
 Note that the hair is not directly coupling to the shocks (this would imply the results such as mirroring to be more non-trivial). The shocks are localized on the lattice and cannot directly affect the delocalized gravitational charges comprising the hair. The shocks add a total energy of $$\Delta \mathcal{E} =\sum_{i,A} e_{i,A} $$ to the system. Starting from a random microstate with the (conserved) monopole charge $\alpha > 0$ and simulating the system via methods of \cite{PhysRevD.101.066001}, one finds that
 \begin{enumerate}
     \item Any microstate with or without decoupled hair oscillations rapidly settles down to another microstate \textit{with} decoupled hair oscillations after the sequence of shocks.
     \item The final microstate is determined by the initial microstate in a rather complex manner. Even for a single shock, one needs to take into account the initial states of all lattice sites and that of the hair indicating \textit{pseudorandom} dynamics.
     \item Almost all the energy in the shock is absorbed by the change in the total black hole mass, i.e. $\mathcal{E}_\mathcal{Q} = M$. This statement becomes better (for a typical initial microstate) if we increase the number of sites keeping the total initial black hole mass and total initial energy fixed. However, even for five lattice sites, less than $1$ percent of the energy in shocks is transferred to the hair.
 \end{enumerate}
 See Fig. \ref{Fig:ShockedMicrostate} for an illustration for the case of a five site lattice suffering a single shock. The above features imply that a typical microstate with a positive monopole charge behaves like a semi-classical black hole qualitatively as far its relaxation dynamics and energy absorption properties are concerned, and furthermore shows features of pseudorandom dynamics. 

As a consequence of these phenomenological features, we can argue that the effective split into interior and exterior emerges dynamically after the relaxation time (which goes to a finite value in the continuum limit as observed numerically).\footnote{ $\mathcal{H}_{\mathsf{Q}}^{\rm int} $ is supported at the inhomogeneous static fixed point configurations (or analogous adiabatic versions of these in the presence of Hawking radiation) while $\mathcal{H}_{\mathsf{Q}}^{\rm ext} $ is supported at the normal mode frequencies only (or analogous adiabatic versions of these in the presence of Hawking radiation). } In very general situations, we expect such a split to be operationally valid when the dynamics is coarse-grained over the relaxation time-scale, and can be demonstrated by employing a suitable coherent state basis. 
\newline \newline
{\textbf{Classical information mirroring in hair:}} The remarkable aspect of these microstate models is that it has the feature of information mirroring built into them intrinsically even though the Hawking radiation is arrested via boundary conditions at the throats. This can be demonstrated readily by studying the dynamics of the system in response to a sequence of shocks. As described above, a typical microstate relaxes to another microstate demonstrating pseudorandom dynamics. A part of the hair $\vec{\mathsf{Q}}_i^{\rm rad}$ decouples from the interior while another component gets locked to the interior lattice charges. The information of the shocks is mirrored in the $\vec{\mathsf{Q}}_i^{\rm rad}$ or equivalently in $q_i^{\rm rad}$ defined in \eqref{Eq:hair-split}. The crucial point is that there are features in $q_i^{\rm rad}$ which allows us to decode the information of the shocks without the knowledge of the interior of the initial or final microstate. Many aspects of the information encoded in the infalling shocks can thus be decoded by assuming that the interior is in the microcanocial ensemble. The only necessary information for the decoding is the global frame $\xi$ in \eqref{Eq:hair-split} which one can simply obtain from the (decoupled) monopole component, the analogue of the early radiation in the Hayden-Preskill thought experiment.

The Fourier transform of $q_i^{\rm rad}(u)$ will be supported only at the non-vanishing normal mode frequencies of \eqref{Eq:Normal-Modes}. It is sufficient to look into the phases and amplitudes of the positive normal mode frequencies since $q_i^{\rm rad}(u)$ is real. It turns out that only the differences of the phases of the two positive normal mode frequencies at the various sites are necessary to decode the information of the shocks. If we encode information into the time sequence of the location of sites which are shocked and the differences of energy injections into these shocks are not large, then the information of the sequence can be decoded from the ordering of the phase differences of the positive frequency normal modes of $q_i^{\rm rad}(u)$ at various lattice sites. All other features of $q_i^{\rm rad}(u)$ will be contaminated with the details of the initial microstate but this ordering of phase differences will be determined only by the specific time sequence of the shocks.

\begin{figure}
\centering
\subfigure[A sequence of two shocks into two non-neighboring sites. The dark blue site is shocked first and the light blue one later. The one shocked earlier has the highest and the one shocked later has the lowest phase differences between the positive normal modes.]{\includegraphics[width = 2.5in]{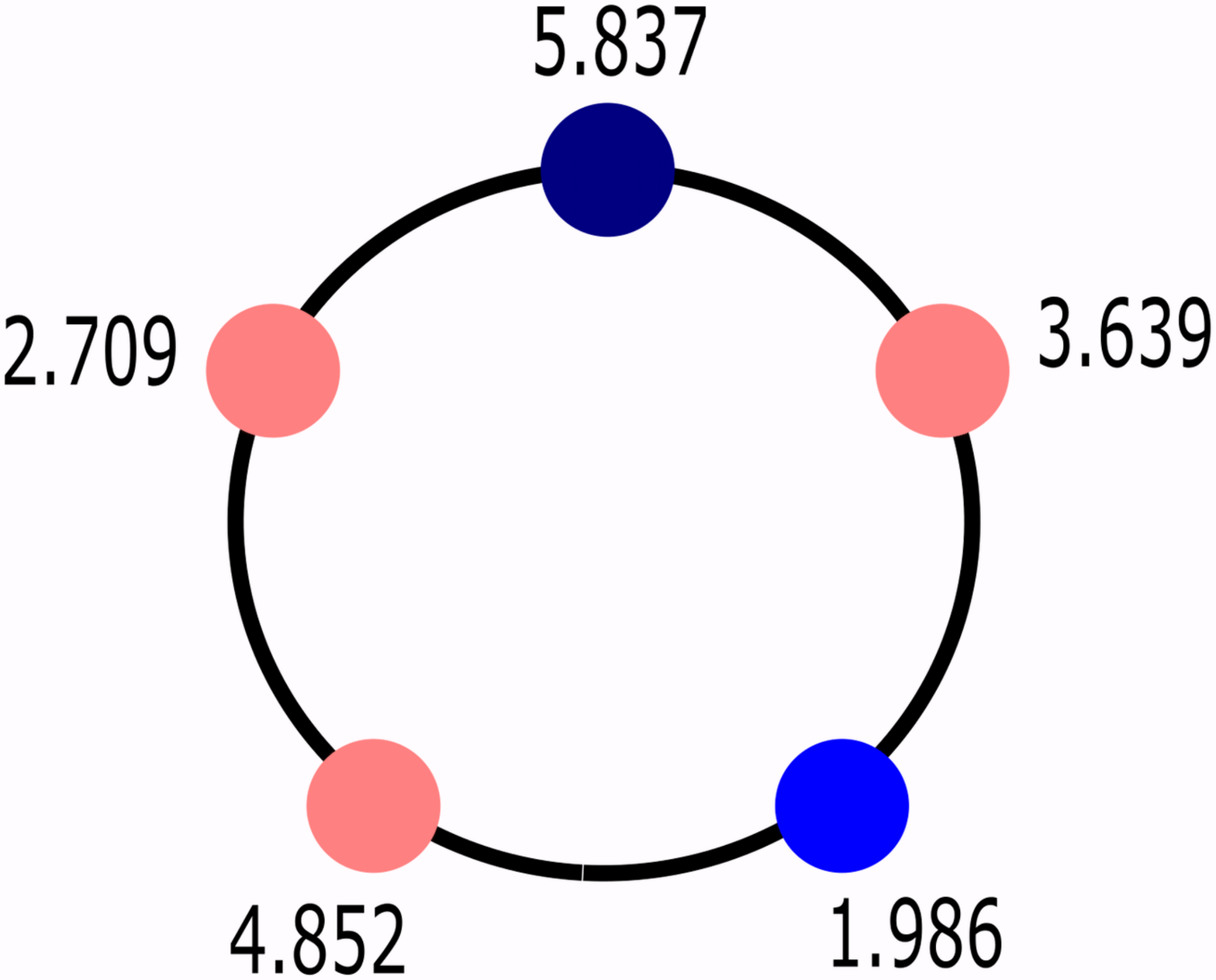}}\\
\subfigure[A sequence of two shocks into two neighboring sites. The dark blue site is shocked first and the light blue one later. The one shocked earlier has the lowest and the one shocked later has the highest phase differences between the positive normal modes.]{\includegraphics[width = 2.5in]{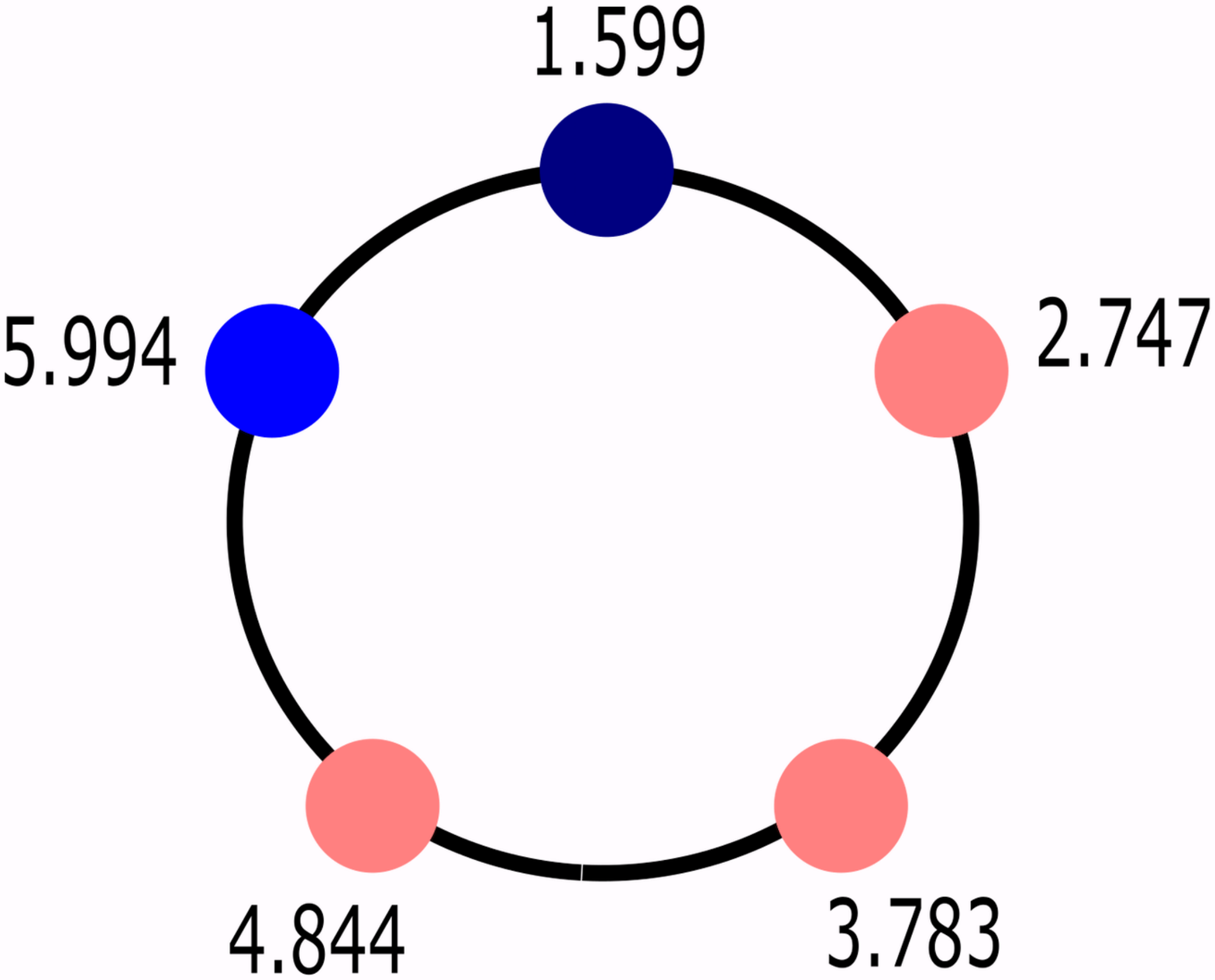}}\\
\subfigure[A sequence of two shocks into the same site. The shocked site has the highest phase differences between the positive normal modes. ]{\includegraphics[width = 2.5in]{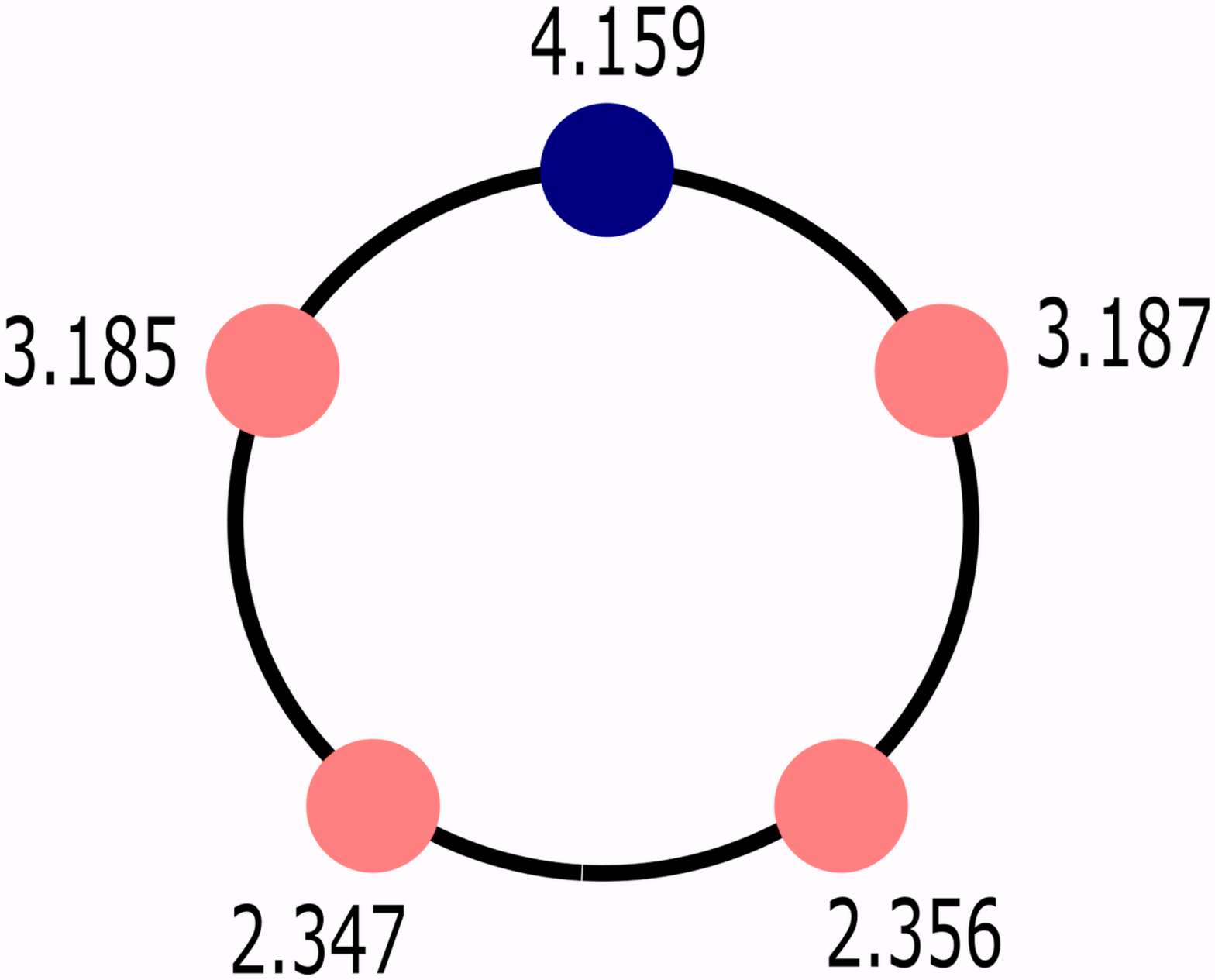}}
\caption{The decoding protocol for information mirroring when two sites are shocked in the five site microstate model via measurements of the phase differences between the two positive normal modes in the hair oscillations which decouple from the interior. Note that this protocol for decoding which sites have been shocked and in which sequence is the same for any initial microstate. Although the absolute values of the phase differences are somewhat sensitive to the initial microstate, the ordering is not.} \label{Fig:Mirroring}
\end{figure}

For a non-trivial example, consider the case of five sites and a sequence of two shocks into two sites. The decoding protocol for which sites have been shocked and in which sequence is illustrated in Fig. \ref{Fig:Mirroring}. This protocol is independent of the initial microstate.  

The information mirroring should carry over to the quantum regime if we quantize the hair. It would be interesting to find out if the information of the initial microstate prior to the shock is encoded in observables which are complementary, i.e. have large non-vanishing commutators with the phase differences in the normal modes. This should have consequences for understanding the black hole complementarity as we comment below.
\newline \newline
{\textbf{Encoding into the Hawking radiation:}} Hawking radiation can be readily incorporated into the microstate model described above by simply coupling each throat to a bath and implementing transparent boundary conditions following \cite{AEMM} as for instance. These boundary conditions choose a specific global null frame $\vec{\xi}$ (as for instance in \cite{AEMM} there is a null shock when the boundary conditions are altered). The first question to be asked is whether the information of the interior leaks out to the Hawking radiation completely. It is a non-trivial question because although the two dimensional black holes evaporate away at all the lattice sites, the $SL(2,R)$ directions of the lattice charges $\vec{\mathcal{Q}}_i$ may not homogenize and thus there would be a remnant information in the interior even after complete evaporation. One may need this remnant to decode the complete information. Firstly, we see that the $SL(2,R)$ directions of the lattice charges $\vec{\mathcal{Q}}_i$ align themselves with the null direction $\vec{\xi}$ set by the boundary conditions asymptotically at late time. Nevertheless, asymptotically $\vec{\mathcal{Q}}_i\rightarrow \mathcal{Q}_i \vec{\xi}$, so that the overall \textit{magnitude} $\mathcal{Q}_i$ is non-trivial although the Casimir (the mass of the black hole) vanishes at each site. However, the hair (being a smaller subsystem) will also dissipate its energy to the Hawking radiation that escapes away. The evolution equations \eqref{Eq:LatticeEoms} would imply that the hair charges should lock themselves with the lattice charges $\vec{\mathcal{Q}}_i\approx \mathcal{Q}_i \vec{\xi}$ following \eqref{Eq:HairLocked} so that the potential energy vanishes (since $\vec\xi$ is null) and thus the hair system reaches a fixed point. Furthermore, as discussed below the hair should be entangled with the Hawking radiation so the full decoding of the interior should be possible. 

The AMPS paradox, of course, is not about the final state after complete evaporation, but the system after Page time. Here, each individual throat would have its own Page time and therefore the Page time here refers to the average Page time of these throats (which will have small fluctuation about the average for a typical microstate of a large black hole). As argued above, the full system should have an effective split into interior and exterior because the hair can be split into $\mathcal{H}_{\mathsf{Q}} = \bigoplus_\alpha \mathcal{H}_{\mathsf{Q}_\alpha}^{\rm int} \otimes \mathcal{H}_{\mathsf{Q_\alpha}}^{\rm ext}$ when averaged over the (slowly evolving) relaxation time with $\alpha$ denoting different microstates. Also the outgoing Hawking radiation essentially forms a sequence of outgoing shocks. Therefore the information of coarse grained features of the outgoing Hawking radiation such as the overall energy outflow from the throats over a period commensurate with the relaxation time (which are insufficient to fully reconstruct the interior, especially the $SL(2,R)$ directions of the lattice charges) will be encoded into the radiation component of the hair which essentially forms $\mathcal{H}_{\mathsf{Q}}^{\rm ext}$ as happened in the case of mirroring of the information of the shocks. Crucially, these \textit{low energy/ coarse-grained} features will be encoded in special observables such as the phase differences of the normal modes in $\mathcal{H}_{\mathsf{Q}}^{\rm ext}$ from which decoding will be possible without the knowledge of the finer structure of the microstates. The detailed knowledge of the interior would be encoded in the complementary observables of $\mathcal{H}_{\mathsf{Q}}^{\rm ext}$ and also $\mathcal{H}_{\mathsf{Q}}^{\rm int}$\footnote{In \cite{Verlinde:2012cy}, one can find a general discussion on how effective field theory observables in a semiclassical but not necessarily smooth horizon geometry can decode the black hole interior. This discussion, cast in the formalism of quantum error correction, can be applied in the context of our microstate model. The crucial part of the construction of the recovery map from the local operators acting on the hair and the Hawking quanta at the horizon, is the knowledge of conditional transition matrices of the black hole interior. The resolution of the AMPS paradox in this context has been also discussed in \cite{Verlinde:2012cy}.}. 

Finally to understand encoding in the Hawking radiation, we should decipher which observables acting on the Hawking radiation will be strongly correlated with features of $\mathcal{H}_{\mathsf{Q}}^{\rm ext}$ that are responsible for mirroring the infalling qubits and encoding the coarse-grained information carried out by the outgoing Hawking radiation (these should have simple encoding in the sense that decoding can be possible by assuming that the interior is a microcanonical ensemble), and similarly which features of the Hawking radiation correlate strongly with $\mathcal{H}_{\mathsf{Q}}^{\rm int}$ and the fine grained features of the black hole interior (eg. the lattice $SL(2,R)$ charges). The encoding of the latter should be complex due to the underlying pseudorandom dynamics of the microstates (more on the generation of complexity via the Python lunch mechanism below). Since the split between the interior and the exterior is only an operational concept that emerges after averaging over an evolving relaxation timescale, there is no real contradiction that the Hawking radiation is in a way entangled maximally with both  $\mathcal{H}_{\mathsf{Q}}^{\rm ext}$ and $\mathcal{H}_{\mathsf{Q}}^{\rm int}$. Furthermore, the coarse-grained and finer information could be encoded with the best fidelity in observables that have large mutual commutators. This could be fundamental to understanding why the semiclassical geometry of the black hole emerges. Additionally, one possible way to realize the expected near saturation (but not violation) of the strong subadditivity property which is key to the AMPS paradox is that the tripartite interior, hair and outgoing Hawking radiation system has a quantum Markov chain like structure (see Section \ref{sec:recons-proof} for the definition of quantum Markov chain states).

The key to the detailed understanding would be to follow the evolution of the full tripartite system (after quantizing the hair) and analyze it using various tools of quantum information theory. This should reveal how black hole complementarity emerges in an operational sense free of paradoxes along with the self-averaging properties represented by the semi-classical black hole geometry and the Euclidean replica wormhole saddles.
\newline \newline
\textbf{Pseudorandom dynamics crucial for Python lunches:} The appearance of Python lunch geometries gives a natural quantification of encoding of the interior modes into outgoing Hawking radiation as shown in Sec. \ref{sec:Python}. As discussed there, instead of the vacuum of the bulk matter in each throat, we should consider first a small subspace of its Hilbert space which we can identify with the code subspace, and then study the microstate dynamics with the bulk matter in the throat in the maximally mixed state in this (small) code subspace (the importance of the maximally mixed state in the code subspace has been emphasized also). Via the arguments presented in Sec. \ref{sec:Python} we expect nucleation of a Python lunch or equivalently a (non-classical) locally maximal extremal surface in the throat. However, the inherent pseudorandom  dynamics would be needed to ensure a necessary amplification so that the Python's lunch manifests macroscopically across a significant fraction of throats (this would be needed in the higher dimensional setup) before the operational split of the hair Hilbert space into interior and exterior emerges, i.e. before black hole complementarity becomes operational. One can also similarly consider a code subspace of the interior  modes of the hair (modes with frequencies smaller than the first non-trivial normal mode) and study the Python lunch phenomenon. It will be also interesting to consider code subspaces of interior modes of bulk matter involving entanglement across several throats since it will give insights into understanding of the complexity of encoding of the entanglement of the interior. In all cases, the inherent chaos in microstate dynamics should connect the Python's lunch geometric mechanism of generating exponential complexity \cite{Brown:2019rox} in encoding with the pseudorandomness mechanism discussed in \cite{Kim:2020cds}.
\newline \newline
\textbf{Outlook:} These microstate models hold many promises for partial answers to how the encoding in Hawking radiation happens with desired features along with the validation of black hole complementarity, but further improvements would be necessary. Although we find information mirroring occurring at relaxation time in the microstates, it is not clear if they have shortest possible scrambling time for typical microstates. Furthermore, the bath region which collects the Hawking radiation is itself fragmented into two dimensional spaces. It would be necessary to glue these baths in a way we can form a connected flat asymptotic spacetime region. One way to construct more realistic models would be to figure out how simplified scenarios with characteristics of those discussed above can emerge from fuzzballs in string theory. The notion of fuzzball complementarity \cite{Mathur:2010kx,Avery:2012tf} which advocate emergent holographic descriptions for local measurements by an infalling observer could be useful.

\section{Discussion and Outlook}\label{sec:discussion}
One of the topics we have not been been able to discuss in this review is the fundamental relationship between tensor networks and gravity. Although tensor networks do model many aspects of holography, it is unclear if they reproduce all information theoretic aspects beyond the Ryu-Takanayagi type extremal surface which computes entanglement entropy. The latter was established first \cite{Swingle:2009bg} in the context of multiscale-entanglement-renormalization (MERA) tensor networks.\footnote{The correspondence between tensor networks and holography has also been explored via continuum versions of MERA \cite{Miyaji:2015yva}. See also \cite{Caputa:2017urj} for a path integral formulation in which the continuum version of the tensor network is reformulated as an optimization problem.} In \cite{Akers:2018fow} it has been argued that existing quantum error correcting tensor networks in the toy models discussed in section \ref{sec:happy} are area eigenstates of the bulk gravity theory rather than representatives of smooth bulk geometries. A better understanding of these issues could also come from constructing tensor networks which mimic the highly efficient RG flow described in section \ref{sec:herg} as it is designed to reproduce dynamical gravity. These issues are important to understand how we can simulate aspects of quantum gravity and also strongly interacting quantum field theories in an efficient qubit regularized way and in real time. 

Another interesting direction of research would be related to further elucidating the models of evaporating black holes in terms of quantum thermodynamics. These models described in this review involve coupling of a bath to a holographic system which has quantum matter in the bulk, and therefore usual ubiquitous outcomes like monotonic growth of entropy in classical gravity coupled to matter which satisfies the null energy condition, can be avoided. It could be possible to construct quantum engines in similar setups. Quantum thermodynamics is a rapidly emerging field which uses quantum information to generalize the laws of thermodynamics \cite{Goold_2016,PhysRevE.93.022126,Yunger_Halpern_2016,Guryanova_2016,RevModPhys.91.025001}. As for instance, bounds have been established for the one-shot work cost of creating a state and also for extractable work from the state in terms of the hypothesis testing relative entropy with respect to the Gibbs ensemble \cite{PhysRevE.93.022126}. Recently, even within the classical approximation in the bulk (with classical bulk matter) and assuming an infinite memoryless bath, it has been shown in \cite{Kibe:2021qjy} that for an instantaneous transition between thermal rotating states in holography, the quantum null energy condition \cite{Bousso_2016} of the boundary theory bounds the growth of thermodynamic entropy (temperature) for a fixed increase in temperature (entropy) from both above and below. Similarly, the rates of growth of entanglement can also be bounded from both above and below. Furthermore, one can recover the Landauer erasure principle and also understand how to construct erasure tolerant quantum memory \cite{Banerjee:2022dgv}. It would be interesting to pursue how one can construct various protocols by exploiting suitable quantum bulk matter in a semi-classical black hole geometry and interacting with a dynamical reservoir at the boundary. Also how the efficiency and work cost of such processes can be bounded via tools of quantum information theory especially the quantum null energy condition and its possible generalizations.

It is quite likely that the fundamental understanding bulk emergence and quantum black holes in holography will lead us to more fundamental and novel connections between quantum information and many-body dynamics.

Furthermore, we have restricted ourselves here to asymptotically anti-de Sitter space. For interesting discussions especially on how islands (entanglement wedges) may or may not generalize consistently for other asymptotic boundary conditions see \cite{Geng:2020fxl,Geng:2021hlu,Shaghoulian:2021cef} (an explicit discussion on de-Sitter space is in \cite{Shaghoulian:2021cef}). In the future, it would be of deep interest to understand bulk emergence in the context of both asymptotically flat and cosmological spacetimes. In the latter context, the approach of \cite{Bzowski:2012ih}, as for instance, may lead to promising results by extrapolating our existing understanding of anti-de Sitter spaces after incorporating extremal surfaces and consistent entanglement wedges.
\begin{acknowledgements}
The research of TK is supported by the Prime Minister's Research Fellowship (PMRF). PM and AM ackowledge support from the Institute of Eminence scheme of IIT Madras funded by the Ministry of Education of India. AM acknowledges the support of the Ramanujan Fellowship of the Science and Engineering Board of the Department of Science and Technology of India, the new faculty seed grant of IIT Madras and the CEFIPRA/IFCPAR grant no 6304-3.
\end{acknowledgements}

\bibliographystyle{spphys}
\bibliography{review}

\end{document}